\documentclass[a4paper]{iopart}
\usepackage{iopams}     
\usepackage{graphicx}
\usepackage{epsfig}
\usepackage{color}
\usepackage{url}
\usepackage{times}
\usepackage{bm}         

\usepackage[utf8]{inputenc}
\usepackage[english,american]{babel}
\usepackage{units}
\usepackage{latexsym}

\definecolor{darkred}{rgb}{0.5,0,0}
\definecolor{darkgreen}{rgb}{0,0.5,0}
\definecolor{darkblue}{rgb}{0,0,0.5}
\definecolor{black}{rgb}{0,0,0}

\newcommand{\Msun}{\,M_{\odot}}

\newcommand{\bea}{\begin{eqnarray}}
\newcommand{\eea}{\end{eqnarray}}
\newcommand{\beq}{\begin{equation}}
\newcommand{\eeq}{\end{equation}}

\newcommand{\cf}{\textit{cf.}~}
\newcommand{\ie}{\textit{i.e.}~}
\newcommand{\eg}{\textit{e.g.}~}
\newcommand{\mss}{{\rm ms}}
\newcommand{\km}{{\rm km}}

\newcommand{\mpc}{{\rm Mpc}}

\begin{document}

\title[Accurate evolutions of unequal-mass neutron-star
  binaries]{Accurate evolutions of unequal-mass neutron-star
  binaries: properties of the torus and short GRB engines}

\author{Luciano Rezzolla$^{1}$, Luca
Baiotti$^{2}$, Bruno Giacomazzo$^{3,4,1}$, David Link$^{1,5}$ and Jos\'e A. Font$^{6}$}

\address{$^{1}$ Max-Planck-Institut f\"ur Gravitationsphysik,
  Albert-Einstein-Institut, Potsdam-Golm, Germany }

\address{$^{2}$ Yukawa Institute for Theoretical Physics, University
of Kyoto, Kyoto, Japan }

\address{$^{3}$ Department of Astronomy, University of Maryland, College
Park, MD, USA}

\address{$^{4}$ Gravitational Astrophysics Laboratory, NASA Goddard
Space Flight Center, Greenbelt, MD, USA}

\address{$^{5}$ Institut f\"ur Physik, Humboldt-Universit\"at zu
  Berlin, Berlin, Germany }

\address{$^{6}$ Departamento de Astronom\'{\i}a y Astrof\'{\i}sica,
  Universitat de Val\`encia, Valencia, Spain }

\date{\today}

\begin{abstract}
We present new results from accurate and fully general-relativistic
simulations of the coalescence of unmagnetized binary neutron stars
with various mass ratios. The evolution of the stars is followed
through the inspiral phase, the merger and prompt collapse to a black
hole, up until the appearance of a thick accretion disk, which is
studied as it enters and remains in a regime of quasi-steady
accretion. Although a simple ideal-fluid equation of state with
$\Gamma=2$ is used, this work presents a systematic study within a
fully general relativistic framework of the properties of the
resulting black-hole--torus system produced by the merger of
unequal-mass binaries. More specifically, we show that: $(1)$ The mass
of the torus increases considerably with the mass asymmetry and
equal-mass binaries do not produce significant tori if they have a
total baryonic mass $M_{\rm tot} \gtrsim 3.7\, \Msun$; $(2)$ Tori with
masses $M_{\rm tor} \sim 0.2\,M_{\odot}$ are measured for binaries
with $M_{\rm tot} \sim 3.4\, \Msun$ and mass ratios $q \sim
0.75-0.85$; $(3)$ The mass of the torus can be estimated by the simple
expression $\widetilde M_{\rm tor}(q,M_{\mathrm{tot}}) = \left[c_1
  (1-q) + c_2\right](M_{\mathrm{max}}-M_{\mathrm{tot}})$, involving
the maximum mass for the binaries and coefficients constrained from
the simulations, and suggesting that the tori can have masses as large
as ${\widetilde M}_{\rm tor} \sim 0.35\,M_{\odot}$ for $M_{\rm tot}
\sim 2.8\, \Msun$ and $q \sim 0.75-0.85$; $(4)$ Using a novel
technique to analyze the evolution of the tori we find no evidence for
the onset of non-axisymmetric instabilities and that very little, if
any, of their mass is unbound; $(5)$ Finally, for all the binaries
considered we compute the complete gravitational waveforms and the
recoils imparted to the black holes, discussing the prospects of
detection of these sources for a number of present and future
detectors.
\end{abstract}



\section{Introduction}
\label{sec:Introduction}
  
The numerical investigation of the coalescence and merger of binary
neutron stars (NSs) within the framework of general relativity is
receiving increasing attention in recent years (see
\eg~\cite{Shibata06a,Anderson2007:shortal,Baiotti08,Anderson2008,Baiotti:2009gk,Etienne08,Giacomazzo:2009mp,Kiuchi2009}
and references therein). Drastic improvements in the simulation front
regarding mathematics (\eg formulation of the equations), physics (\eg
incorporation of equations of state (EOSs) from nuclear physics), and
numerical methods (\eg use of high-resolution methods and adaptive
mesh refinement), along with increased computational resources, have
allowed to extend the scope of the early simulations
(\eg~\cite{Shibata99d}).  Larger initial separations have recently
started being considered and some of the existing simulations have
expanded the range spanned by the models well beyond black-hole (BH)
formation~\cite{Baiotti08,Kiuchi2009,Giacomazzo:2009mp}. This is
allowing the computation of the entire gravitational waveform from the
early inspiral to the decaying tail of the late ringing of the formed
BH. The construction of such waveform templates is still one of the
driving motivations to perform binary NS simulations, as such events
are among the most promising sources of detectable gravitational
radiation for laser interferometric detectors.  The current estimate
for the detection rate relative to the first-generation
interferometric detectors is $\sim 1$ event per $\sim 40-300$ years,
increasing to an encouraging $\sim 10-100$ events per year for the
advanced detectors~\cite{Belczynski07}. The second major incentive
behind this type of simulations is establishing whether the
end-product of the merger can act as the underlying mechanism
operating at the central engine of short-hard gamma-ray bursts
(SGRBs)~\cite{Narayan92,Zhang04}. The consensus emerging from the
existing simulations suggests the formation, depending on the
suitability of the initial parameters of the simulated model, of a BH
of stellar mass surrounded by a hot disk. Driven by neutrino processes
and magnetic fields, such a compact system may be capable of launching a
relativistic fireball with an energy of $\sim 10^{48}$ erg on a
timescale of $0.1-1$ s~\cite{Piran:2004ba}.

This paper is dedicated in particular to investigating the late-time
dynamics of the torus formed after the merger of unequal-mass NS
binaries. As we describe below, all but one model of our initial sample
have mass ratio different from one. Previous simulations have shown that the key
parameter controlling the amount of mass left in the disk for a given
initial mass in the system and EOS is the NS mass
ratio~\cite{Shibata:2003ga,Shibata06a}. Broadly speaking the general
trend is simple: the larger the departure from equal-mass ratio, the
more important tidal effects become in the less massive star,
resulting in its tidal disruption. Because this takes place when the
separation is still comparatively large, the angular momentum of the
matter is still large and it results in larger-size and more massive
disks. Early and low-resolution simulations with an ideal-gas
EOS~\cite{Shibata:2003ga} have been shown to yield a disk mass of
several percents of the total mass of the system for a mass ratio of
$\sim 0.85$. Improved simulations by~\cite{Shibata06a} which adopted a
hybrid EOS to mimic realistic, stiff nuclear EOS, indicate that the
mass of the disk is $\sim 0.01\,\Msun$ or slightly larger when the
merger does not result in prompt collapse to a BH but in the formation
of a hypermassive NS of large ellipticity (which later
collapses to a BH following angular-momentum transport caused by emission
of gravitational radiation). Similar disk masses, as large as
$\sim0.02$ M$_{\odot}$, are also reported in the latest simulations
of~\cite{Kiuchi2009}, in which the initial orbital separation of the
two stars is larger than in previous works.

Observational data seems to indicate that the total gravitational
masses of the known galactic NS binary systems are in a narrow range
$\sim 2.65 - 2.85\,\Msun$~\cite{Stairs04} and there is also evidence
indicating that the masses of the two NSs are nearly equal, with the
baryonic mass ratio $q\equiv M_1/M_2$ being between $1$ and $\sim 0.7$
(Hereafter we will refer to $q$ simply as the ``mass ratio'' but
we report in table~\ref{tab:models} also the ratio in the ADM masses
$q_{_{\rm ADM}}$; $q$ and $q_{_{\rm ADM}}$ do not coincide because of
the nonlinear relation between baryonic and gravitational
mass). However, there is no theoretical reason to assume that
unequal-mass NS binaries are not produced in nature as often as
the seemingly prevailing almost exactly equal-mass systems, particularly for twin
giant progenitors~\cite{Pinsonneault06,Lee07}. There are, indeed,
recent computations and observations that contradict the predominance of
almost exactly equal-mass systems~\cite{Bulik04,Lee07}. On the one hand, binary population
synthesis computations performed by~\cite{Bulik04} show two peaks in
the observability-weighted distribution of double NSs. One of these
peaks is around $q\sim 1$ and appears when both masses are close to
$1.4\,\Msun$. The second peak is around considerably smaller mass
ratios and depends on the assumed maximum mass of a single NS (which is in
turn dependent on the EOS considered): the higher this mass (the
stiffer the EOS) the more significant the second peak~\cite{Bulik04,Rosinska:2005}. 
However, the crucial parameter
determining the shape of the distribution is the inclusion of
hypercritical accretion onto the compact object during the brief but
critical ``common-envelope'' evolution phase of the close
binary~\cite{Rosinska:2005}. Similarly, recent computations
by~\cite{Lee07} also accounting for the effects of hypercritical
accretion during the red-giant evolution of the less massive component
of the binary lead to a pattern of NS binaries consisting of pulsars
which are $\sim 50\%$ more massive than their companion NSs.

An additional issue which motivates our study has to do with the
investigation of the long-term stability and dynamics of the formed
accretion disks. It is well known that thick accretion disks orbiting
BHs may be subject to a number of instabilities, both axisymmetric,
such as the so-called ``runaway instability''~\cite{Abramowicz83}, or
non-axisymmetric, such as the ``Papaloizou-Pringle
instability''~\cite{Papaloizou84}. The first one, in particular, if
present, could destroy the torus on dynamical timescales, challenging
the viability of the BH--torus model for SGRBs. Early time-dependent,
axisymmetric, general-relativistic-hydrodynamical simulations of the
runaway instability of non-self-gravitating tori around BHs were
performed by~\cite{Font02a,Daigne04}. The distribution of specific
angular momentum in the disk, $\ell\equiv -u_{\phi}/u_{t}$, with
$u_{\phi}$ and $u_t$ being the corresponding components of the
4-velocity $u_{\mu}$, was the key parameter discriminating stable from
unstable models in those simulations. It was found that
$\ell-$constant models were runaway unstable while power-law
distributions $\ell=Kr^{\alpha}$ were stable for very small values of
the angular momentum slope $\alpha$ (much smaller than the Keplerian
value $\alpha=0.5$). Recent fully relativistic simulations
by~\cite{Montero09}, which for the first time have take into account 
the self-gravity of the disk, indicate that self-gravity does not favour the
appearance of the instability, irrespective of the angular-momentum
distribution. Under the effect of a perturbation, marginally stable
models show the presence of axisymmetric oscillations for several
dynamical timescales without the manifestation of the runaway
instability, as~\cite{Zanotti03} had previously found for the case of
non-self-gravitating tori. Indeed, the introduction of perturbations
triggers quasi-periodic oscillations (QPOs) lasting tens of orbital periods, with amplitudes that
are modified only slightly by the small loss of matter across the
cusp~\cite{Zanotti03,Montero07}. The spectral distribution of the
associated eigenfrequencies shows the presence of a fundamental $p$
mode and of a series of overtones in a harmonic ratio $2$:$3$, which
have been proposed to explain the QPOs observed in the X-ray
luminosity of low-mass X-ray binaries (LMXBs) containing a BH 
candidate with the QPOs of small tori near the
BH~\cite{Rezzolla_qpo_03a,Rezzolla_qpo_03b,Schnittman06}. In addition,
when sufficiently massive and compact, the oscillations of these tori
are responsible for an intense emission of gravitational
waves~\cite{Zanotti03,Zanotti05,Nagar2007,Montero07}. 

Overall, the {\it ab-initio} simulations reported here indicate that
large-scale tori with masses $M_{\rm tor} \sim 0.2\,M_{\odot}$ can be
produced as the result of the inspiral and merger of binary NSs with
unequal masses and that even larger masses can be predicted for
binaries with smaller total masses. These tori are typically of large
size, with quasi-Keplerian distribution of angular momenta, showing
quasi-stationary evolutions and the absence of dynamical
instabilities. As such, these results may provide additional
information relevant to all the above issues for BH--torus systems
formed in a fully consistent manner within the framework of general
relativity. Furthermore, the gravitational-wave emission computed here
reveals that the waveforms are sensitive to the mass ratio in the
binary, both during the inspiral and after the merger, and could be
used to extract important information on the structure and EOS of the
progenitor stars. Such observations, however, will most likely have to
rely on the advanced detectors, which will become operative in a few
years.

The paper is organised as follows: Section~\ref{sec:NumericalSetup}
describes the mathematical and numerical framework of our simulations.
Section~\ref{sec:CoalescenceDynamics} discusses the dynamics of the
coalescence and merger of our model sample. Next, in
Section~\ref{sec:TorusFormation} we focus on the analysis of the tori
formed after the merger and on their physical properties. The issue of
the gravitational-wave emission from unequal-mass NS mergers is
discussed in Section~\ref{sec:GravitationalWaves} and the main
conclusions of our investigation are presented in
Section~\ref{sec:Conclusions}. In addition Appendix A provides
quantitative measures of the accuracy of our numerical methods.  We
use a spacelike signature $(-,+,+,+)$ and a system of units in which
$c=G=\Msun=1$ (or in other units whenever more convenient). Greek
indices are taken to run from $0$ to $3$, Latin indices from $1$ to
$3$ and we adopt the standard convention for the summation over
repeated indices.

\section{Mathematical and Numerical Setup}
\label{sec:NumericalSetup}

All the details on the mathematical and numerical setup used for
producing the results presented here are discussed in depth
in~\cite{Pollney:2007ss,Baiotti08}. In what follows, we limit
ourselves to a brief overview.


\subsection{Einstein and Hydrodynamics equations}
\label{sec:Einsten_Hydro_eqs}

The evolution of the spacetime is obtained using the \texttt{CCATIE}
code, a three-dimensional finite-differencing code providing a
solution of a conformal traceless formulation of the Einstein
equations with a ``$1+\log$'' slicing condition and a ``Gamma-driver''
shift condition (the interested reader is addressed
to~\cite{Pollney:2007ss,Baiotti07} for a detailed discussion of the
equations and gauges used). The general-relativistic hydrodynamics equations are
instead solved using the \texttt{Whisky} code presented
in~\cite{Baiotti03a,Baiotti04,Baiotti07}, which adopts a
flux-conservative formulation of the equations as presented
in~\cite{Banyuls97} and high-resolution shock-capturing schemes. The
{\tt Whisky} code implements several reconstruction methods, such as
Total-Variation-Diminishing (TVD) methods, Essentially-Non-Oscillatory
(ENO) methods~\cite{Harten87}, and the Piecewise Parabolic Method
(PPM)~\cite{Colella84}. Also, a variety of approximate Riemann solvers
can be used, starting from the Harten-Lax-van Leer-Einfeldt (HLLE)
solver~\cite{Harten83}, over to the Roe solver~\cite{Roe81}, and the
Marquina flux formula~\cite{Aloy99b} (see~\cite{Baiotti03a,Baiotti04}
for a more detailed discussion). All the results reported hereafter
have been computed using the Marquina flux formula~\cite{Donat96} and
a PPM reconstruction. We stress again (as already done
in~\cite{Baiotti08,Giacomazzo:2009mp}) that the use of high-order
methods and high-resolution is \textit{essential} to be able to draw
robust conclusions on the inspiral and merger. Lower-order methods in
the reconstruction and low resolution may yield convergent and
apparently reasonable results which however contain a large truncation
error. Specific examples of this type of problem are presented in
Appendix A of~\cite{Baiotti08} and in Figure $4$
of~\cite{Giacomazzo:2009mp}. Also, a measure of our overall accuracy
is presented in Appendix A below and shows that by employing such
methods we are able to conserve energy and angular momentum to $\sim
1\%$ over a timescale of $\sim 140\,\mss$.

The system of hydrodynamics equations is closed by an EOS. As
discussed in detail in~\cite{Baiotti08}, the choice of the EOS plays a
fundamental role in the post-merger dynamics and significantly
influences the survival time, against gravitational collapse, of the
hyper-massive neutron star (HMNS) likely produced by the merger. It is
therefore important that special attention is paid to use EOSs that
are physically motivated, as done in~\cite{Oechslin07b} within a
conformally flat description of the fields and a simplified treatment
of the hydrodynamics. Because we are here mostly concerned with
drawing a first qualitative picture of the properties of the torus in
a space of parameters that is as vast as computationally affordable,
we have employed the commonly used ``ideal-fluid'' EOS, in which the
pressure $p$ is expressed as $p = \rho\, \epsilon(\Gamma-1) $, where
$\rho$ is the rest-mass density, $\epsilon$ is the specific internal
energy, and $\Gamma$ is the adiabatic exponent. Such an EOS, while
simple, provides a reasonable approximation and we expect that the use
of realistic EOSs would not change the main results of this
work. Furthermore, it was shown in~\cite{Bejger05} that the inspiral
of equal-mass binaries of NSs described by realistic EOSs can be
reproduced quite well by studying NSs with the same mass and radii but
constructed as polytropes with $\Gamma =2$.

As in~\cite{Baiotti08}, the gravitational-wave signal is extracted
using two methods. The first method uses the Newman-Penrose formalism
so that the gravitational-wave polarization amplitudes $h_+$ and
$h_\times$ are then related to $\Psi_4$ by simple time
integrals~\cite{Teukolsky73}
\begin{equation}
\ddot{h}_+ - {\rm i}\ddot{h}_{\times}=\Psi_4 \ ,
\label{eq:psi4_h}
\end{equation}
where the double overdot stands for the second-order time derivative 
and the curvature scalar
\begin{equation}
  \Psi_4 \equiv -C_{\alpha\beta\gamma\delta}
    n^\alpha \bar{m}^\beta n^\gamma \bar{m}^\delta
  \label{eq:psi4def}
\end{equation}
is defined as a particular component of the Weyl curvature tensor,
$C_{\alpha\beta\gamma\delta}$, projected onto a given null frame
$\{\boldsymbol{l}, \boldsymbol{n}, \boldsymbol{m},
\bar{\boldsymbol{m}}\}$ (see~\cite{Pollney:2007ss} for details). The
second and independent method is instead based on the measurements of
the non-spherical gauge-invariant perturbations of a Schwarzschild BH
(see refs.~\cite{Abrahams97a_shortlist,Rupright98,Rezzolla99a} for
some applications of this method to Cartesian-coordinate grids). In
particular, the gravitational-wave polarization amplitudes are in this
case expressed in terms of gauge-invariant metric
perturbations~\cite{Nagar05}
\begin{equation}
\label{eq:wave_gi}
h_+-{\rm i}h_{\times} =
  \frac{1}{\sqrt{2}r}\sum_{\ell,\,m}
  \Biggl( Q_{\ell m}^+ -{\rm i}\int_{-\infty}^t Q^\times_{\ell
          m}(t')dt' \Biggr)\,_{-2}Y^{\ell m}\ ,
\end{equation}
where $_{-2}Y^{\ell m}$ are the $s=-2$ spin-weighted spherical
harmonics and $Q_{\ell m}^\times$, $Q_{\ell m}^+$ the
(gauge-invariant) odd-parity (or {axial}) current multipoles and
even-parity (or polar) mass multipoles of the perturbed metric,
respectively. In practice, these multipoles are computed on a set of
$2$-spheres of fixed coordinate radius $r_{\rm iso}=200$ (\ie $\sim
300\, \km$). 

The two wave-extraction methods yield results whose
differences are smaller than $1\%$, and hereafter we will
concentrate only on the one using the gauge-invariant perturbations as
it reduces the number of integration constants to be determined when
computing the gravitational-wave strain.


\subsection{Adaptive Mesh Refinements and Grid setup}
\label{sec:GridSetup}

The grid hierarchy is handled by the \texttt{Carpet} mesh refinement
driver~\cite{Schnetter-etal-03b}. It implements vertex-centered mesh
refinement, also known as the box-in-box method, and allows for
regridding during the calculation as well as multiple grid
centers. With mesh refinement, a small number of grids with different resolutions, 
called refinement levels, overlay each other, and are nested in a
way that the coarsest grid has the largest extent and the finest grid
the smallest extent. While the refined grids in the interior allow for
an increased resolution where it is most desired, the outer boundary
can at the same time be kept at a large distance.

The timestep on each grid is set by the Courant condition (expressed
in terms of the speed of light) and so by the spatial grid resolution
for that level; the typical Courant coefficient is set to be $0.35$.
The time evolution is carried out using $4$th-order--accurate
Runge-Kutta integration algorithm. Boundary data for finer grids are
calculated with spatial prolongation operators employing $3$rd-order
polynomials and with prolongation in time employing $2$nd-order
polynomials. The latter allows a significant memory saving, requiring
only three timelevels to be stored, with little loss of accuracy due
to the long dynamical timescale relative to the typical grid timestep.

For the inspiral phase of the system of binary NSs, two grid centers
\ensuremath{\{r_{c,i}: i = 1, 2\}} are defined, with one grid center
located at the grid point where the rest-mass density reaches its
maximum \ensuremath{\rho_\mathrm{max} = \mathrm{max}(\rho)} and the
other grid center located at the $\pi$-symmetric point (\ie the grid
point correspondent to \ensuremath{\rho_\mathrm{max}} and rotated by
$180$ degrees around the \ensuremath{z}-axis). The grid hierarchy is
composed of six refinement levels and a $2:1$ refinement factor for
successive levels. Once the condition $\ensuremath{\rho_\mathrm{max} =
  \mathrm{max}(\rho_{\mathrm{max},i}) \geq
  1.2\,\rho_\mathrm{max,initial}}$ is satisfied, which is known from
experience to occur during the merger phase and well before collapse,
the grid hierarchy is reduced to a single grid center fixed at the
origin of the grid. At the initial time, the finest grids cover each
star completely. Later, during the merger phase, matter outflows cross
the boundary to the second finest grid and subsequently to the other
coarser refinement levels. The grid resolution varies from
$\ensuremath{\Delta_6 = 0.15}$ (\ie $\sim 221\,{\rm m}$)
for the finest level to $\ensuremath{\Delta_1 = 4.8}$ (\ie $\sim
7.1\,\km$) for the coarsest level, whose outer boundary is at $240$ in
our units (\ie $\sim 360\,\km$). Initially, the number of grid points 
across the linear dimension of a star is of the order of $100$. The torus
surrounding the BH after collapse is usually not contained within the
finest grid, but its high-density region is covered by the
second finest grid with resolution $\ensuremath{\Delta_2 = 0.3}$.

The whole grid is set up to be symmetric with respect to the
\ensuremath{(x,y)} plane for both unequal-mass binaries and equal-mass
binaries. The boundary conditions are chosen to be radiative for the
metric, in order to prevent gravitational waves from scattering back into the
grid, and static for the hydrodynamical variables. Note that the above
setup is identical to that adopted in~\cite{Baiotti08}.

\begin{table*}[ht]
  \caption{\label{tab:models}Properties of the binary NS initial
    data. From left to right the columns show: the name of the model
    (assembled from its rounded total baryonic mass preceded by the
    letter \texttt{M} and its mass ratio preceded by the letter
    \texttt{q}), the total baryonic mass
    \ensuremath{M_{{\mathrm{tot}}}} of the system, the total ADM mass
    \ensuremath{M_{_{\mathrm{ADM}}}} of the system, the ratio of the
    baryonic masses of the two stars \ensuremath{q = M_{2} / M_{1}},
    the ratio of the ADM masses of the two stars, the total angular
    momentum \ensuremath{J}, the initial orbital frequency
    \ensuremath{\nu_\mathrm{orb}}, the initial maximum rest-mass
    density \ensuremath{\rho_\mathrm{max}}, the mean radius
    \ensuremath{\bar r_i} of each star, the axis ratio
    \ensuremath{\bar A_i} of each star, the individual ADM mass 
    $M^{\infty}_i$ of each star as considered in isolation at infinity, 
    and the compactness ${\cal C}^{\infty}_{i}$ of each star as considered 
    in isolation at infinity. The mean radius is defined as
    \ensuremath{\bar r_i \equiv (r_\vdash + r_\dashv + r_\perp +
      r_\mathrm{pol}) / 4}, where \ensuremath{r_\vdash} and
    \ensuremath{r_\dashv} are the radii of the star parallel to the
    line connecting the stars, \ensuremath{r_\perp} is the radius in
    the equatorial plane perpendicular to that line, and
    \ensuremath{r_\mathrm{pol}} is the radius perpendicular to the
    equatorial plane. The axis ratio is defined as the ratio between
    the mean radius parallel to the line connecting the stars, and the
    mean radius in the plane perpendicular to that line, namely
    \ensuremath{\bar A_i \equiv (r_\perp + r_\mathrm{pol}) / (r_\vdash
      + r_\dashv)}. All values except \ensuremath{\rho_\mathrm{max}}
    are provided by the output of the \texttt{LORENE} code, and the
    accuracy of \ensuremath{M_{{\mathrm{tot}}}} and \ensuremath{J} is
    the one at which the \texttt{Whisky} code is able to reproduce
    them for the present setup.}  
\footnotesize{\begin{tabular}[t]{lccccccccc}
\br Model & {\tiny ${M_{{\mathrm{tot}}}}$, ${M_{_{\rm ADM}}}$} &
    {$q,\ q_{_{\rm ADM}}$} & {$J/10^{49}$} & ${\nu_\mathrm{orb}}$ &
    {\tiny ${\rho_\mathrm{max}}/10^{14}$} & ${\bar r_2},\ {\bar r_1}$ & ${\bar
      A_2},\ {\bar A_1}$ & $M^{\infty}_1$, $M^{\infty}_2$ &
    ${\cal C}^{\infty}_1,{\cal C}^{\infty}_2$ \\ 
    & \tiny{(${\Msun}$)} & & \tiny{(${\unit{g\,cm^2/s}}$}) & \tiny{(${\unit{Hz}}$)} &  \tiny{(${\rm g/cm}^3$)} & \tiny{(${\rm km}$)} & & \tiny{(${\Msun}$)} & \\
    \mr 
$\!\!\!\!\!$\texttt{M3.6q1.00} & $\!\!\!\!\!3.56, 3.23$ & $\!\!\!\!\!1.00, 1.00$ & $\!\!\!\!\!8.92$ & $\!\!\!\!\!303.32$ & $\!\!\!\!\!\!\!\!\!\!7.58$ & $\!\!\!\!\!\!\!\!\!\!12.0, 12.0$ & $\!\!\!\!\!0.95, 0.95$ & $\!\!\!\!\!1.643, 1.643$ &$\!\!\!\!\!0.130, 0.130$\\    
$\!\!\!\!\!$\texttt{M3.7q0.94} & $\!\!\!\!\!3.68, 3.33$ & $\!\!\!\!\!0.94, 0.94$ & $\!\!\!\!\!9.37$ & $\!\!\!\!\!306.56$ & $\!\!\!\!\!\!\!\!\!\!9.75$ & $\!\!\!\!\!\!\!\!\!\!12.0, 11.0$ & $\!\!\!\!\!0.95, 0.96$ & $\!\!\!\!\!1.643, 1.742$ &$\!\!\!\!\!0.130, 0.150$\\    
$\!\!\!\!\!$\texttt{M3.4q0.91} & $\!\!\!\!\!3.40, 3.11$ & $\!\!\!\!\!0.91, 0.92$ & $\!\!\!\!\!8.33$ & $\!\!\!\!\!299.06$ & $\!\!\!\!\!\!\!\!\!\!7.58$ & $\!\!\!\!\!\!\!\!\!\!13.1, 12.1$ & $\!\!\!\!\!0.93, 0.95$ & $\!\!\!\!\!1.512, 1.643$ &$\!\!\!\!\!0.111, 0.130$\\   
$\!\!\!\!\!$\texttt{M3.4q0.80} & $\!\!\!\!\!3.37, 3.08$ & $\!\!\!\!\!0.80, 0.81$ & $\!\!\!\!\!8.36$ & $\!\!\!\!\!303.62$ & $\!\!\!\!\!\!\!\!\!\!9.21$ & $\!\!\!\!\!\!\!\!\!\!13.8, 11.3$ & $\!\!\!\!\!0.90, 0.97$ & $\!\!\!\!\!1.400, 1.723$ &$\!\!\!\!\!0.097, 0.146$\\    
$\!\!\!\!\!$\texttt{M3.5q0.75} & $\!\!\!\!\!3.46, 3.14$ & $\!\!\!\!\!0.75, 0.77$ & $\!\!\!\!\!8.40$ & $\!\!\!\!\!300.84$ & $\!\!\!\!\!\!\!\!\!\!12.7$ & $\!\!\!\!\!\!\!\!\!\!13.0, 10.1$ & $\!\!\!\!\!0.89, 0.98$ & $\!\!\!\!\!1.390, 1.804$ &$\!\!\!\!\!0.096, 0.171$\\     
$\!\!\!\!\!$\texttt{M3.4q0.70} & $\!\!\!\!\!3.37, 3.07$ & $\!\!\!\!\!0.70, 0.72$ & $\!\!\!\!\!7.98$ & $\!\!\!\!\!298.47$ & $\!\!\!\!\!\!\!\!\!\!12.8$ & $\!\!\!\!\!\!\!\!\!\!14.6, 10.0$ & $\!\!\!\!\!0.85, 0.98$ & $\!\!\!\!\!1.304, 1.805$ &$\!\!\!\!\!0.087, 0.172$\\
\mr
\end{tabular}
}
\end{table*}


\subsection{Initial data}
\label{sec:InitialData}

We use quasi-equilibrium initial data generated with the multi-domain
spectral-method code \texttt{LORENE} developed at the Observatoire de
Paris-Meudon~\cite{Gourgoulhon01}. For more information on the code
and its methods, the reader is referred to the \texttt{LORENE} web
pages \cite{Lorene}. In particular, because the binaries are not
expected to be corotating~\cite{Bildsten92}, we use irrotational 
configurations, defined as having vanishing vorticity, and obtained 
under the additional assumption of a conformally flat spacetime
metric~\cite{Gourgoulhon01}.

Some of the models investigated in this paper are publicly available
on servers of the Meudon group~\cite{Lorene}. Others have been created
by us specifically for the unequal-mass simulations presented
here. The models of the lowest mass ratios have been kindly provided
by Dorota Gondek-Rosi\'nska). The EOS assumed for the initial data is
in all cases the polytropic EOS $p \equiv K\,\rho^\Gamma$ with an
adiabatic index ${\Gamma = 2}$ and a polytropic coefficient $K =
\unit[0.0332] \rho_{\rm nuc}\,c^2/n_{\rm nuc}^{\Gamma}=123.6$ (in
  units where $c=G=\Msun=1$), where $\rho_\mathrm{nuc}$ and
  ${n_\mathrm{nuc}}$ refer to the nuclear rest-mass and number
  densities, respectively. For this particular EOS, the allowed
  maximum baryonic mass for an individual stable NS is
  $\sim{\unit[2.00]{\Msun}}$. The initial coordinate separation of the
  stellar centers in all cases is $d = \unit[45]{km}$.

The models used as initial data include both equal-mass models and
most importantly unequal-mass models. As mentioned in the
Introduction, we here concentrate on the dynamics of the massive tori
resulting from the merger. The formation of the torus is enhanced for 
smaller mass ratios. Although the observational evidence
is not very firm~\cite{Stairs04,Chang07} there is also no theoretical
reason ruling out their possible
existence~\cite{Bulik04,Lee07}. A full list of all
considered models together with a selection of physical quantities
defining them, \eg baryon and ADM mass, orbital frequency and initial
angular momentum, etc., is given in table~\ref{tab:models}. To
distinguish simply the different binaries we adopt the following
naming convention: any initial data binary is indicated as
\texttt{M\%q\#}, with \texttt{\%} being replaced by the rounded total
baryonic mass ${M_{\mathrm{tot}}}$ of the binary neutron-star system
and \texttt{\#} by the mass ratio $q$. As an example,
\texttt{M3.4q0.80} is the binary with total baryonic mass
$M_{\mathrm{tot}}\simeq3.4\,\Msun$ and mass ratio $q=0.80$.

\begin{figure*}[ht] 
\begin{center}
{\label{fig:density-xy-em:1} 
\includegraphics[width=.45\textwidth]{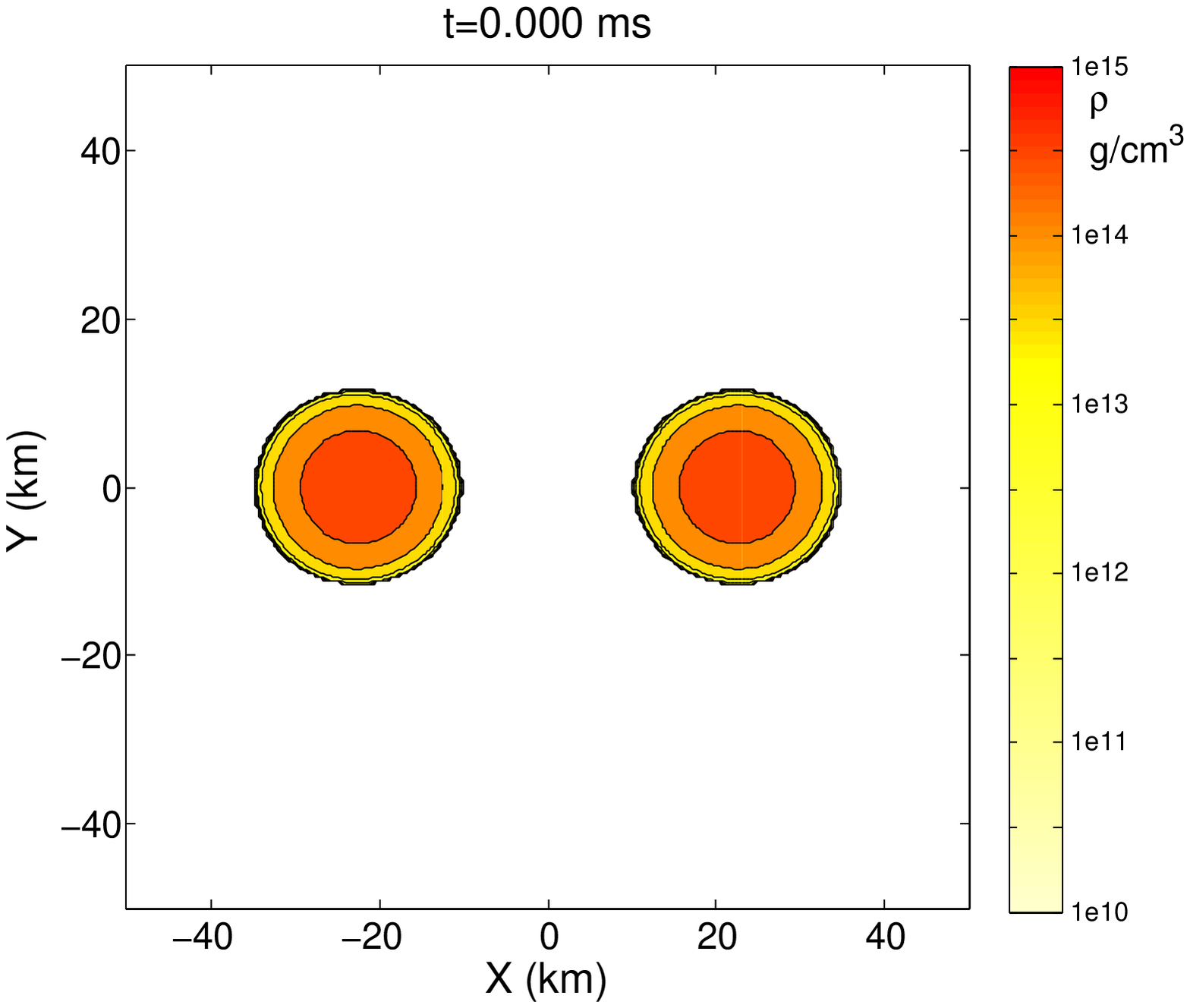}}
\hspace{.5 cm} 
{\label{fig:density-xy-em:2} 
\includegraphics[width=.45\textwidth]{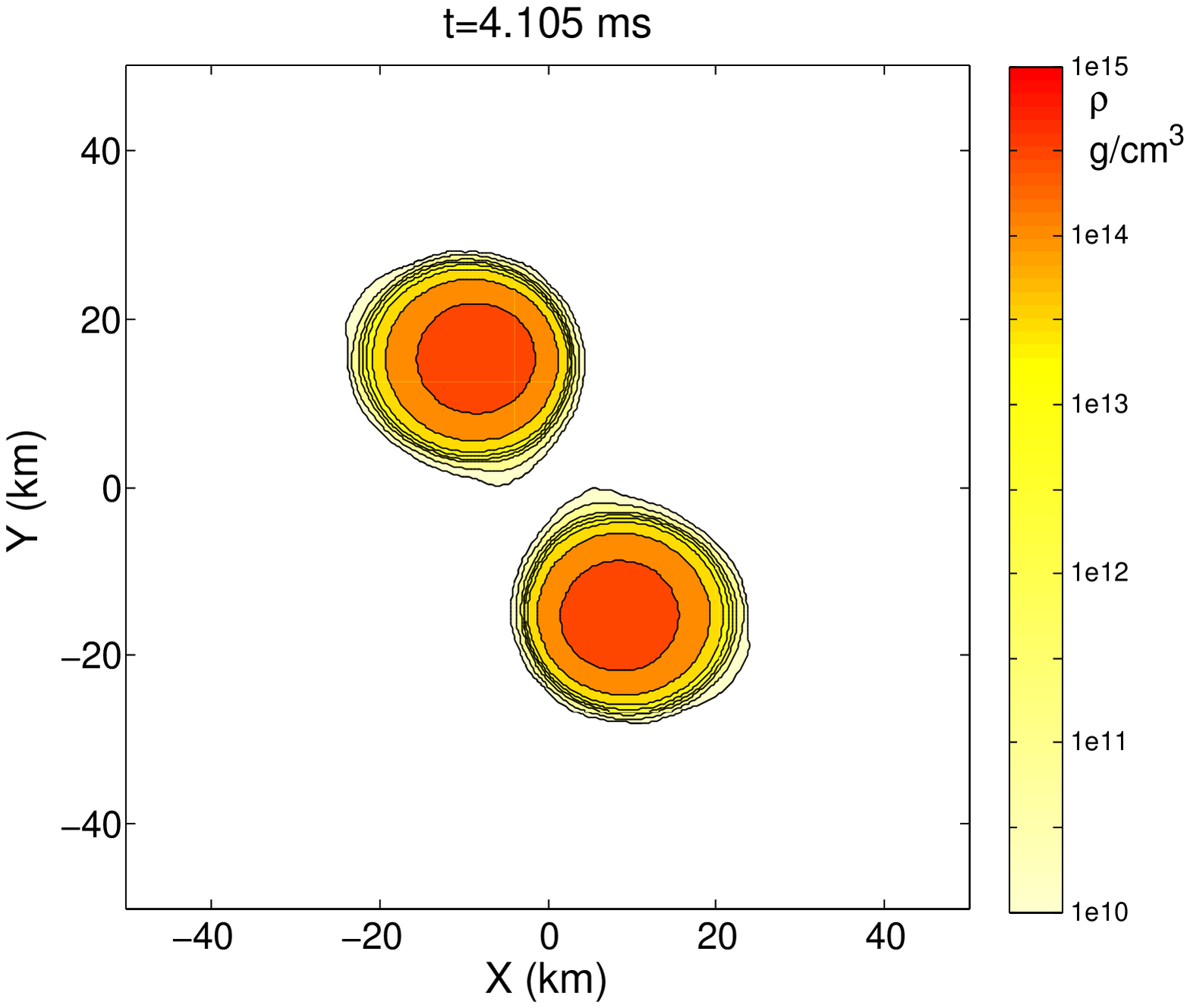}}
\vspace{.2 cm}
{\label{fig:density-xy-em:3} 
\includegraphics[width=.45\textwidth]{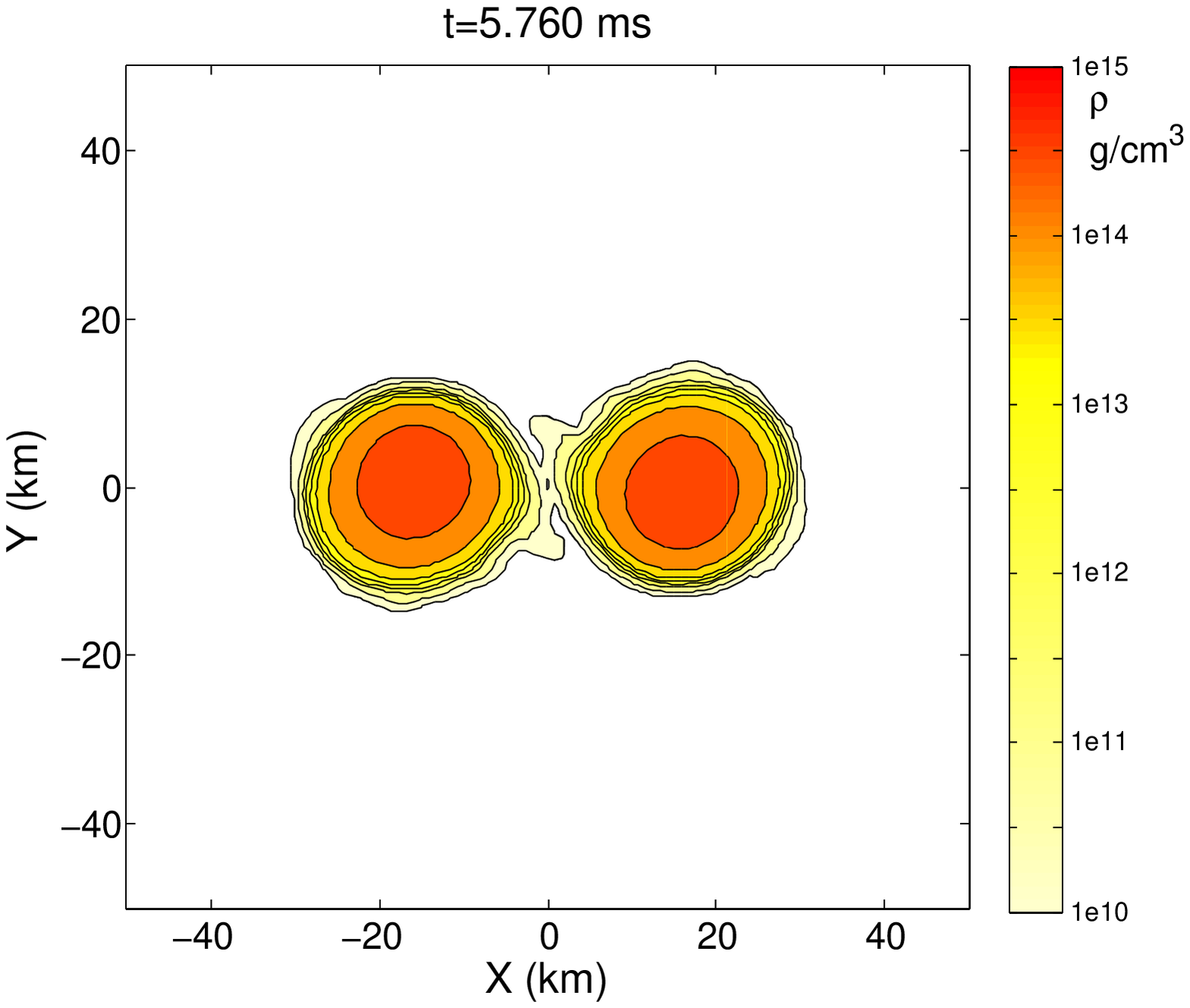}}
\hspace{.5 cm} 
{\label{fig:density-xy-em:4}
\includegraphics[width=.45\textwidth]{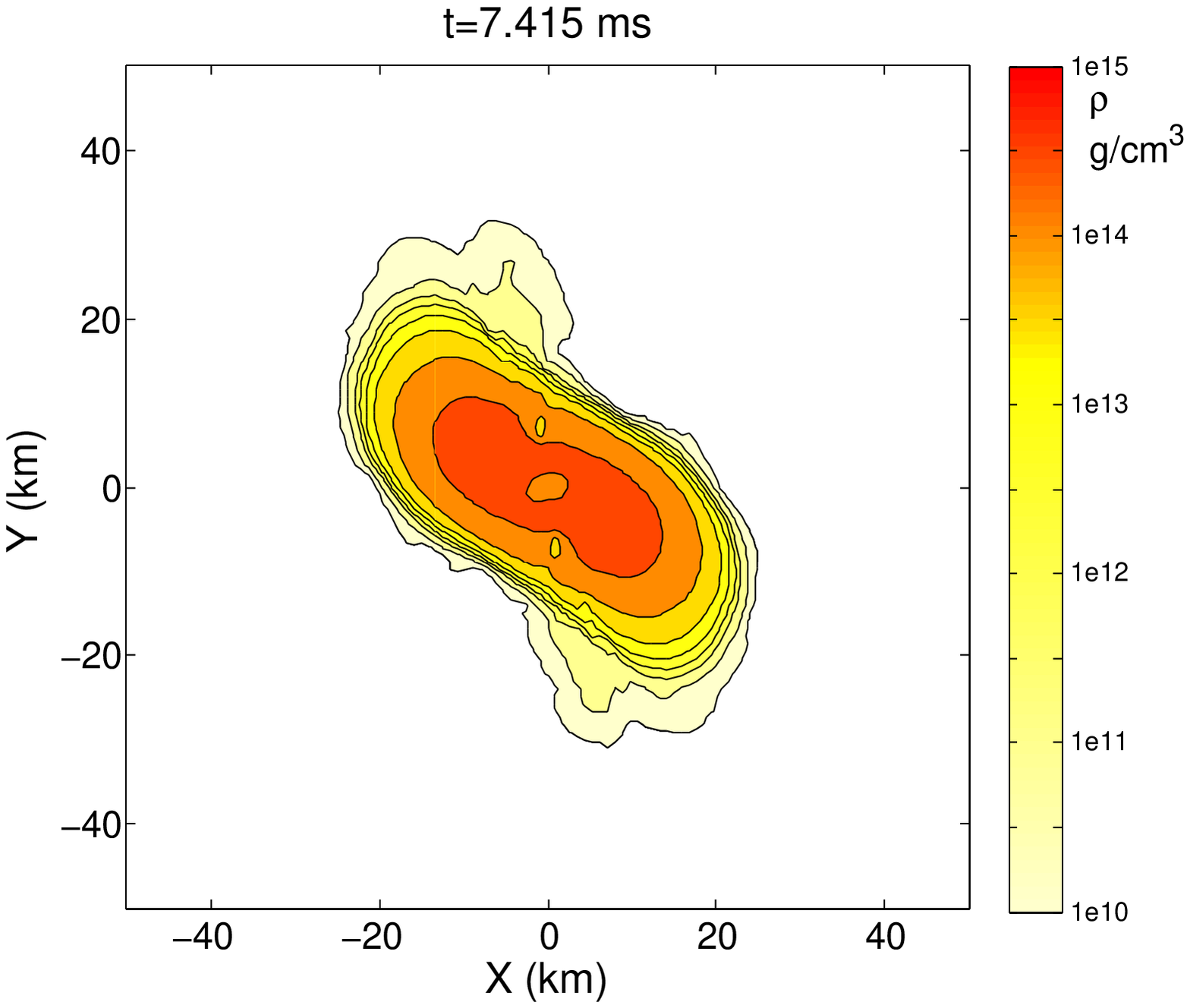}}
\vspace{.2 cm}
{\label{fig:density-xy-em:5}
\includegraphics[width=.45\textwidth]{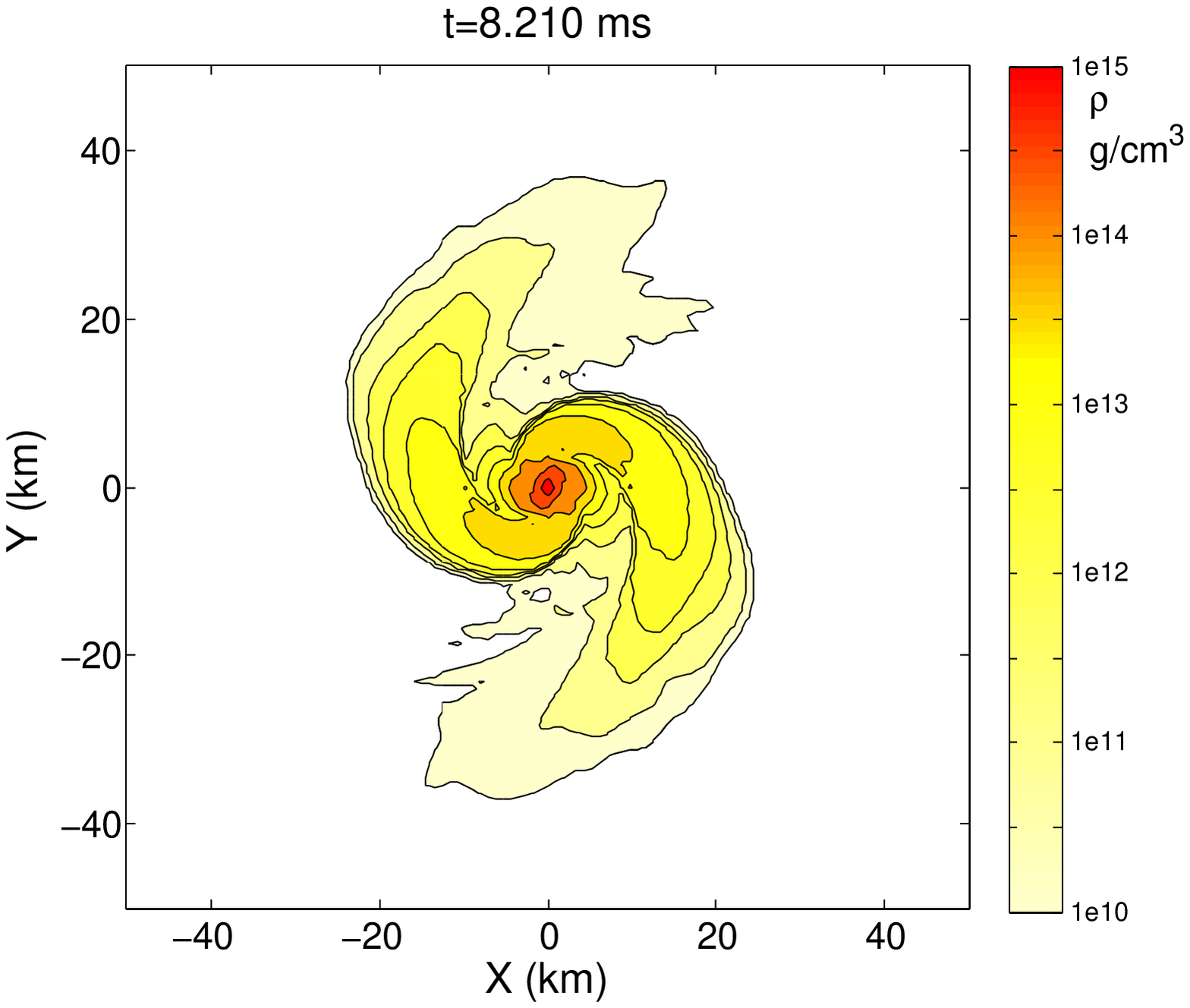}}
\hspace{.5 cm}
{\label{fig:density-xy-em:6}
\includegraphics[width=.45\textwidth]{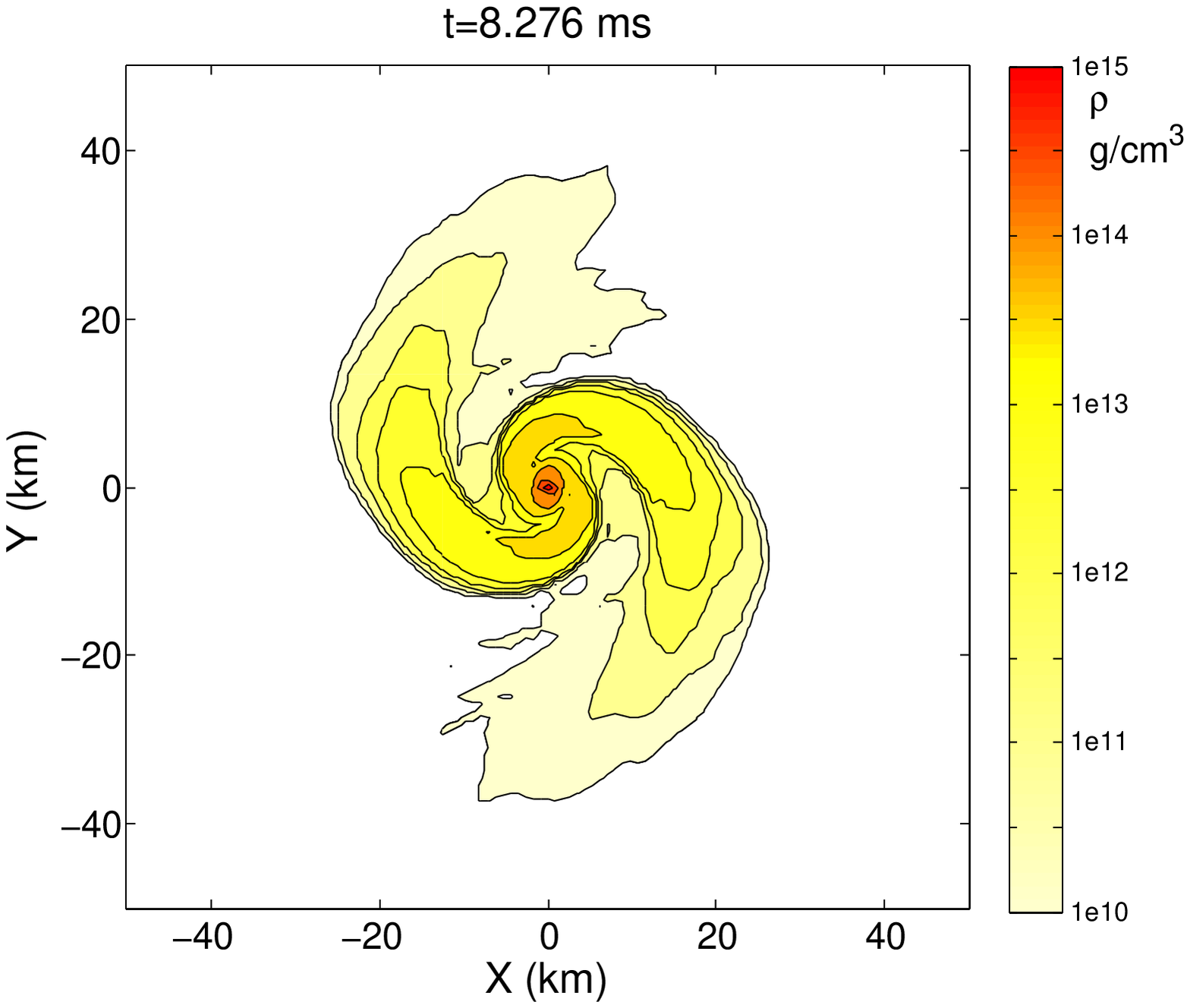}}
\end{center}
\caption{Isodensity contours for the \texttt{M3.6q1.00} model on the
  \ensuremath{(x,y)} plane. The times when the frames have been taken
  are shown on top of the plots while the color-code for the rest-mass
  density is indicated to the right of each plot. Additionally,
  isodensity contours are shown for the values of \ensuremath{\rho =
    10^{10}, 10^{11}, 10^{12}, 10^{12.5}, 10^{13}, 10^{13.5}, 10^{14},
    10^{14.5}, \unit[10^{15}]{g/cm^3}}. The third frame (at time t =
  \unit[5.760]{ms}) shows the onset of the merger, the last two frames
  (at times \ensuremath{t = \unit[8.210]{ms}}, \ensuremath{t =
    \unit[8.276]{ms}}) show the behaviour of the system during the
  collapse to a BH.}
\label{fig:density-xy-em}
\end{figure*}

\begin{figure*}[ht] 
\begin{center}
{\label{fig:density-xy-um:1} \includegraphics[width=.45\textwidth]{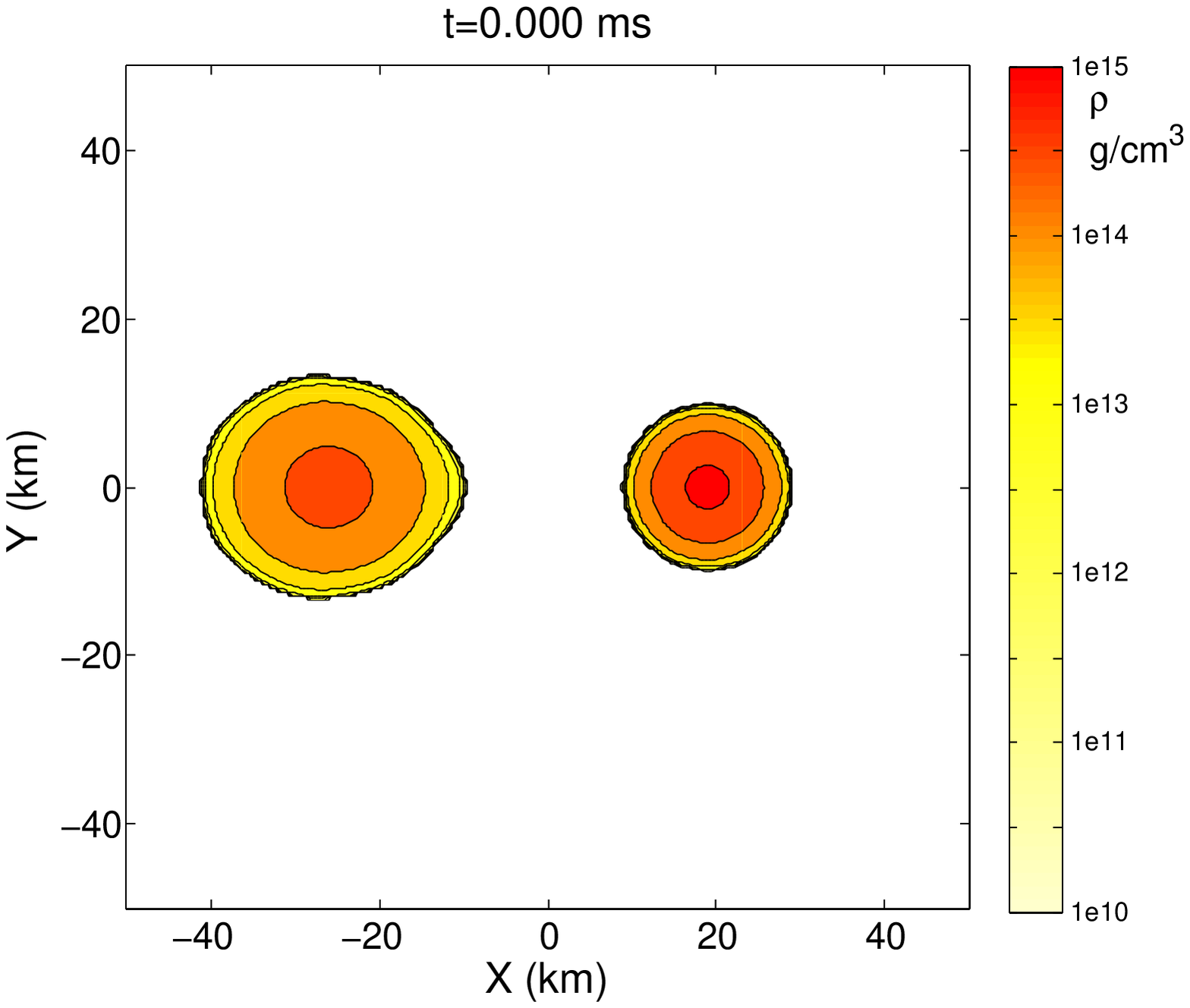}}
\hspace{.5 cm} 
{\label{fig:density-xy-um:2} \includegraphics[width=.45\textwidth]{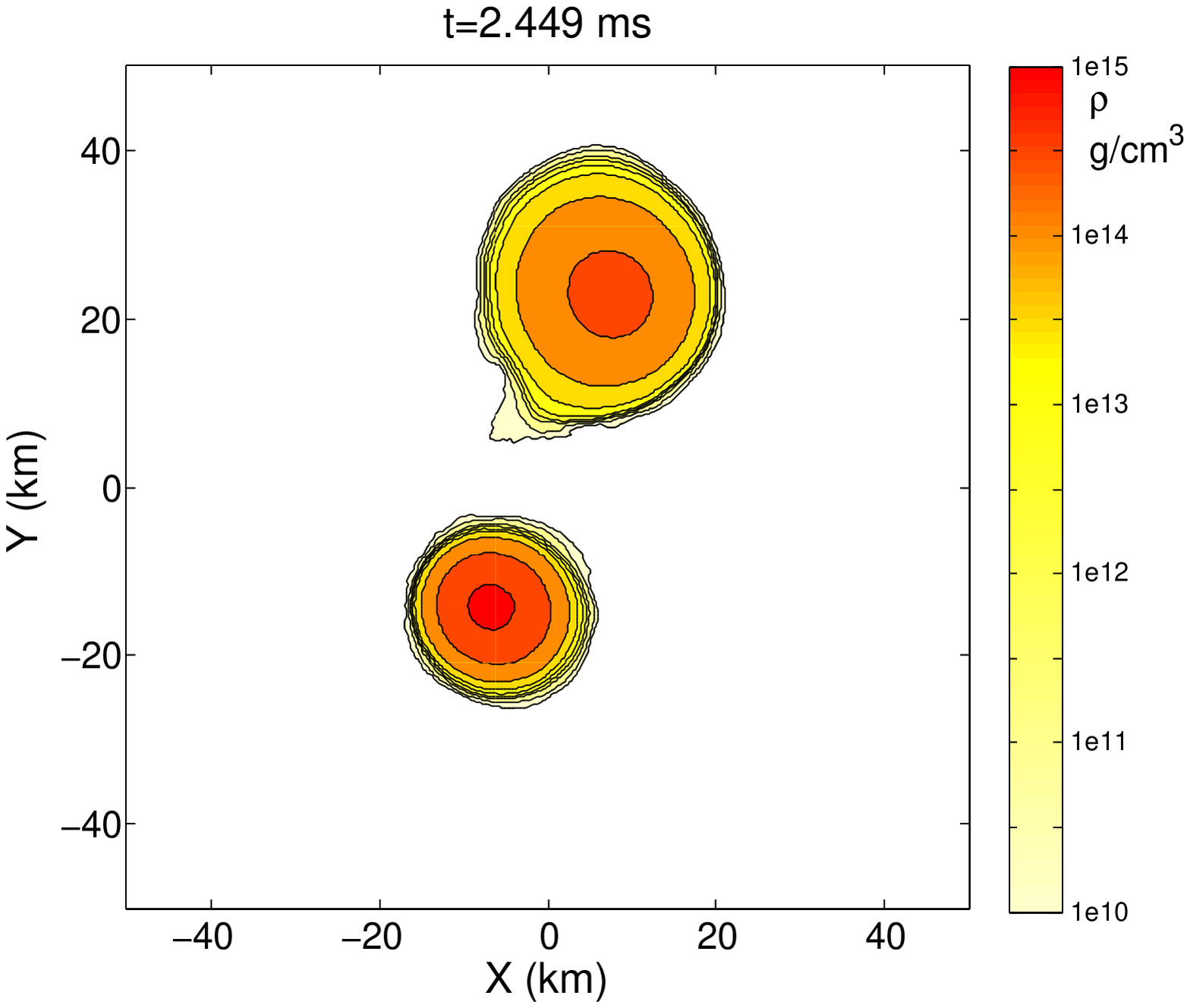}}
\vspace{.2 cm}
{\label{fig:density-xy-um:3} \includegraphics[width=.45\textwidth]{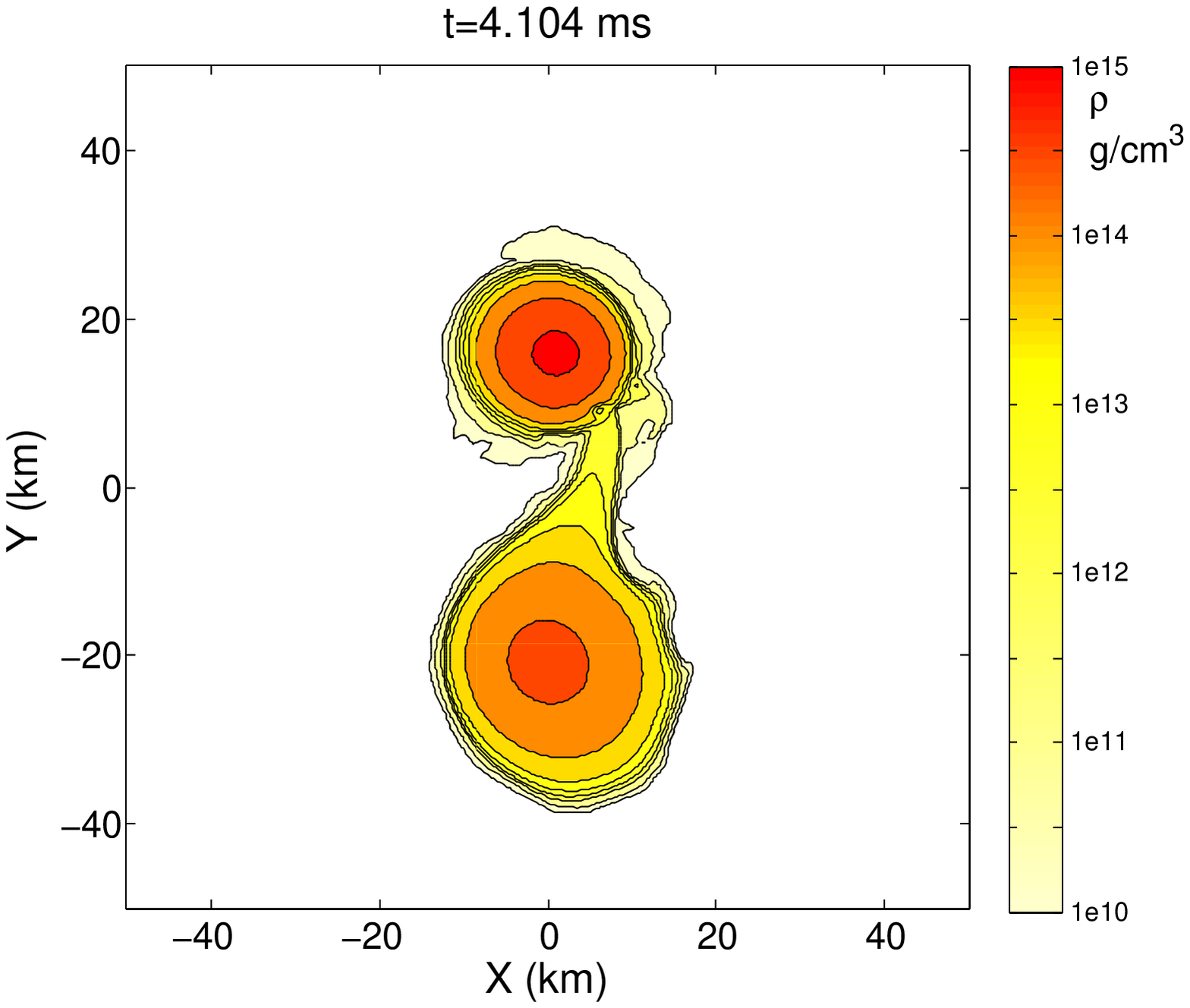}}
\hspace{.5 cm} 
{\label{fig:density-xy-um:4} \includegraphics[width=.45\textwidth]{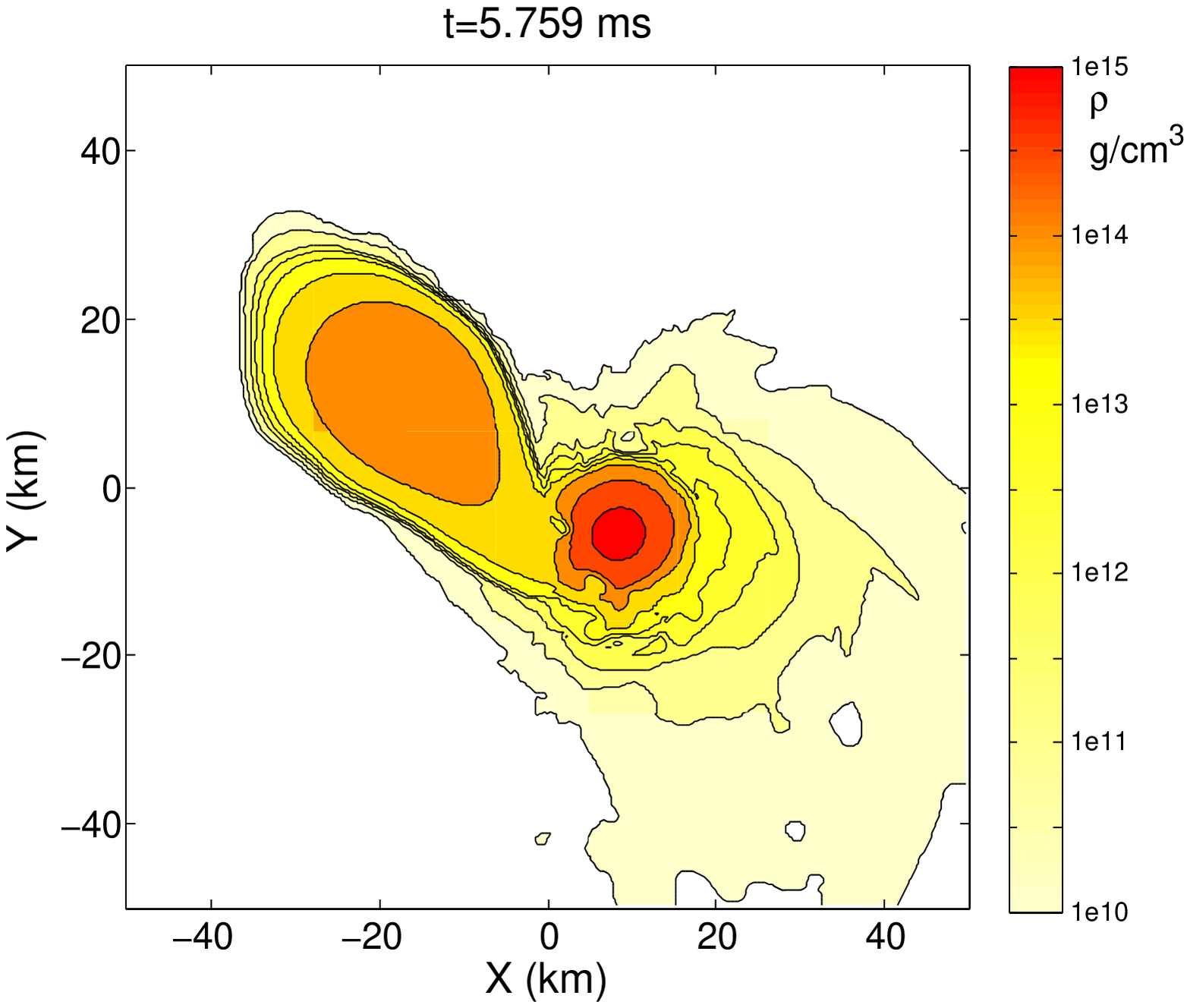}}
\vspace{.2 cm}
{\label{fig:density-xy-um:5} \includegraphics[width=.45\textwidth]{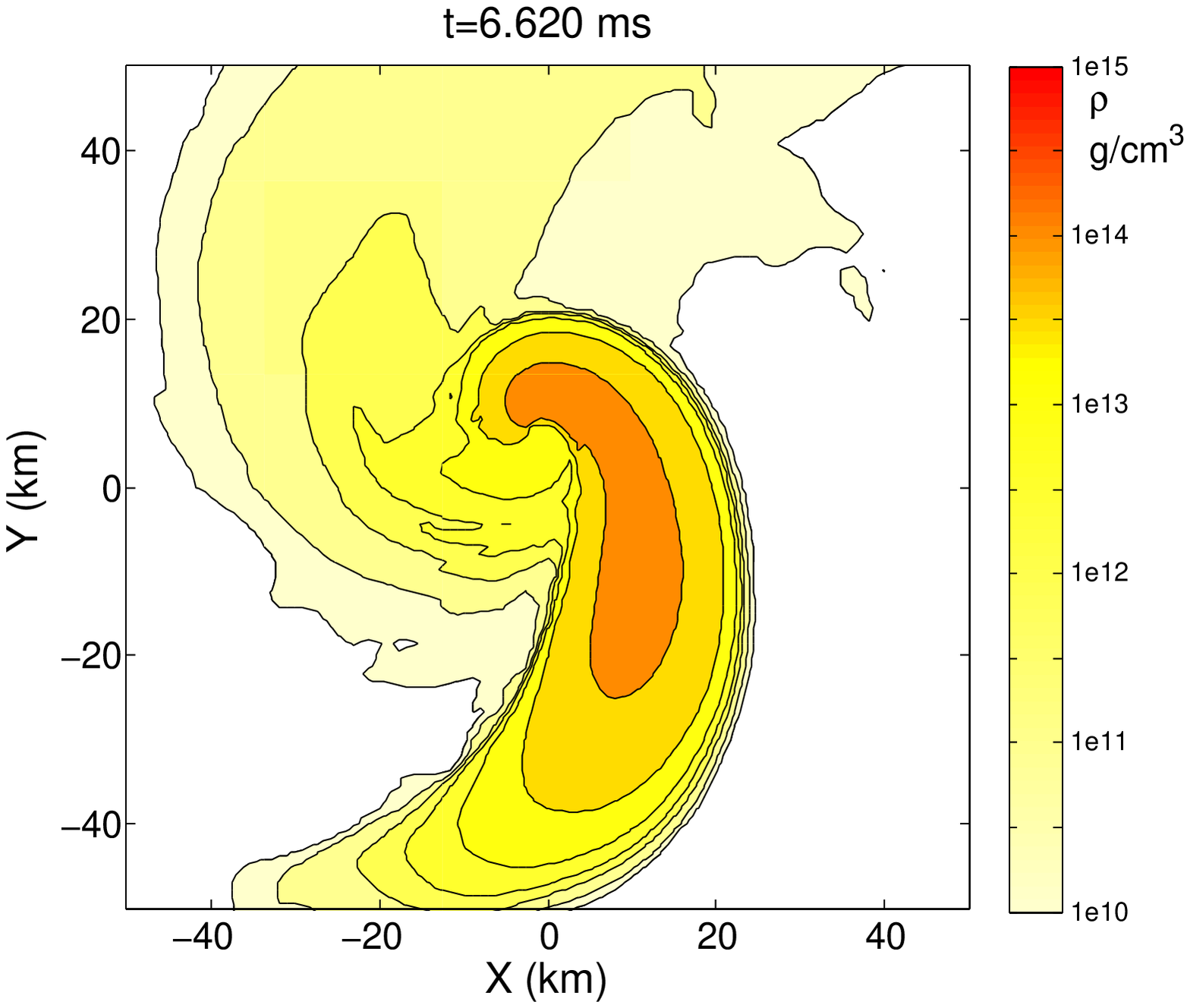}}
\hspace{.5 cm} 
{\label{fig:density-xy-um:6} \includegraphics[width=.45\textwidth]{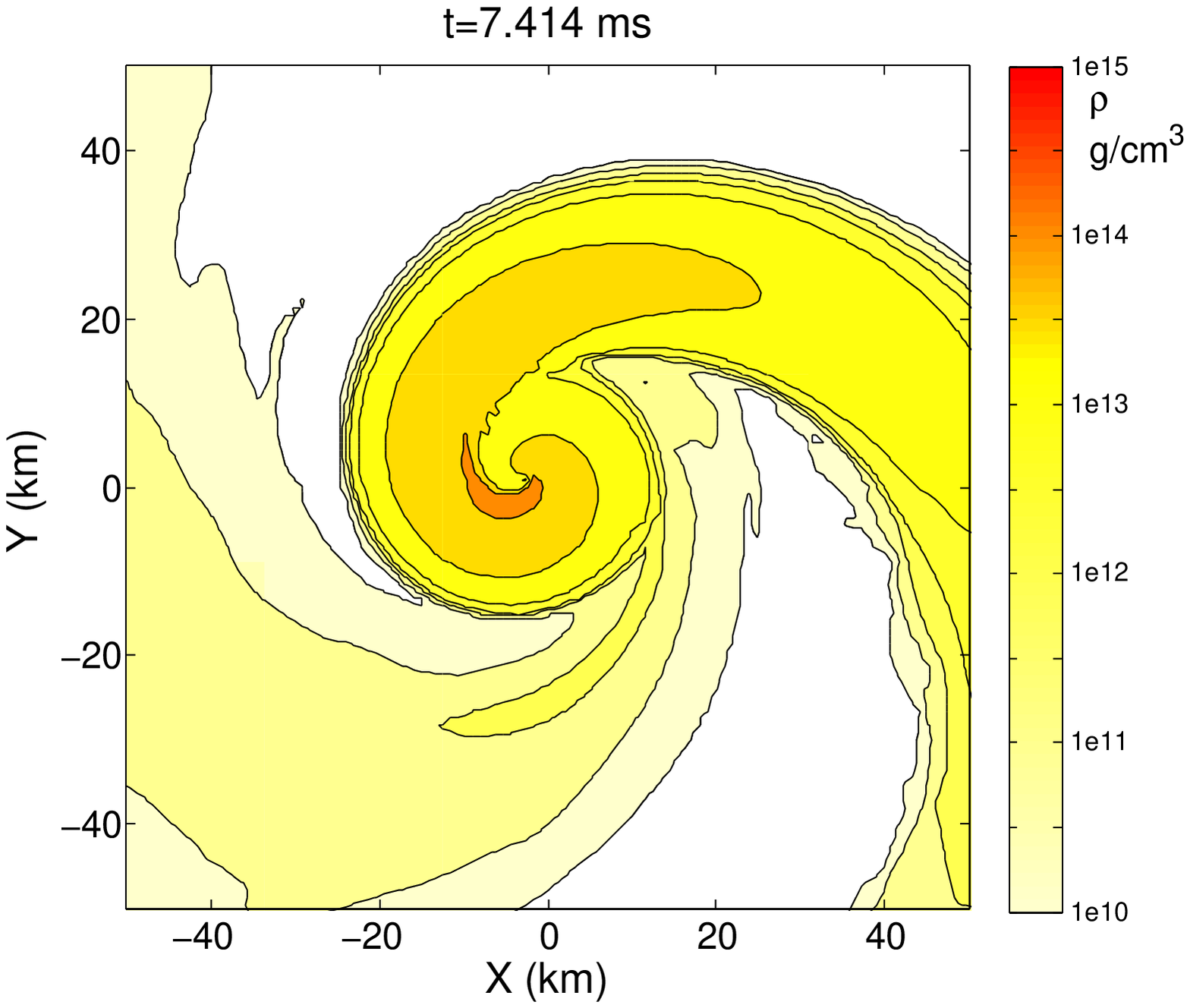}}
\end{center}
\caption{Isodensity contours for the \texttt{M3.4q0.70} model on the
  \ensuremath{(x,y)} plane. The times when the frames have been taken
  are shown on top of the plots while the color-code for the rest-mass
  density is indicated to the right of each plot. Additionally,
  isodensity contours are shown for the values of \ensuremath{\rho =
    10^{10}, 10^{11}, 10^{12}, 10^{12.5}, 10^{13}, 10^{13.5}, 10^{14},
    10^{14.5}, \unit[10^{15}]{g/cm^3}}. The third frame (at time t =
  \unit[4.104]{ms}) shows the onset of the merger, the last two frames
  (at times \ensuremath{t = \unit[6.620]{ms}}, \ensuremath{t =
    \unit[7.414]{ms}}) show the behaviour of the system during the
  collapse to a BH. Note that the computational domain is much larger
  than what is shown and extends to $\sim 360\,\km$ }
\label{fig:density-xy-um}
\end{figure*}

\section{Dynamics of the coalescence and merger}
\label{sec:CoalescenceDynamics}

\subsection{General dynamics}

In a previous work~\cite{Baiotti08}, we have investigated the dynamics
of the coalescence and merger of equal-mass binary NSs for models with
total baryonic mass \ensuremath{M_{{\mathrm{tot}}} = 2.912\,\Msun} and
\ensuremath{M_{{\mathrm{tot}}} = 3.250\,\Msun}. It was found that for
any of the two EOSs considered, binaries with (initial) total baryonic
mass below a certain limit do not collapse promptly to a BH but rather
yield an oscillating HMNS, which undergoes delayed collapse to a
BH. Independently of the mass ratio, all of the binaries under
consideration here have masses higher than those considered
in~\cite{Baiotti08} and all collapse promptly never leading to a HMNS
even if the EOS used here is a non-isentropic one (see discussion
in~\cite{Baiotti08} on the different qualitative behaviour between an
isentropic and a non-isentropic EOS). This absence of a HMNS,
however, is the direct consequence of the chosen initial data rather
than a feature of unequal-mass mergers and it has been here exploited
simply to reduce the computational costs by shortening the time of the 
collapse to a BH.

Figure~\ref{fig:density-xy-em} shows a selection of representative
isodensity contours on the equatorial plane for the equal-mass binary
\texttt{M3.6q1.00}. At the initial time, the stars are in their
quasi-equilibrium configuration at a coordinate separation of
\unit[45]{km}. The binary progressively speeds up while
inspiralling. After slightly more than two orbits have been completed
(namely after about \unit[5-6]{ms}), the stars merge and, about
\unit[2-3]{ms} later, an apparent horizon (which we search with the
code of~\cite{Thornburg2003:AH-finding_nourl}) is found. The
ideal-fluid EOS employed in the simulations allows for shock-heating
and an increase of the specific internal energy \ensuremath{\epsilon},
as shown in \cite{Baiotti08}; this, in turn, causes some matter to be
ejected from the rotating central object and to propagate into the
surrounding artificial atmosphere. The evolution of model \texttt{M3.6q1.00}
shows that matter is ejected in small amounts during the inspiral
phase and in larger amounts during the merger phase, when the shocks
are much stronger. Therefore, while small spiral arms can certainly be
observed in the outer regions during the merger phase (see the last
two snapshots of figure~\ref{fig:density-xy-em}), they do not have
sufficient angular momentum to reach distances as large as in the
unequal-mass models (see discussion below). Instead, the spiral arms
wind around the rapidly rotating central object formed by the two NS
cores. Quantitative results regarding the BH spin and the mass and
angular momentum of the remaining disk will be discussed in subsequent
sections.

To contrast the evolution of an unequal-mass binary, figure
\ref{fig:density-xy-um} shows the same selection of isodensity
contours on the equatorial plane as represented in figure
\ref{fig:density-xy-em}, only now for the \texttt{M3.4q0.70} model,
which has the smallest mass ratio considered in this work. The
asymmetry of the binary system is already apparent at the initial
time. The heavier star is much more compact than its extended less
massive companion, which is deformed already at the initial distance
by tidal forces. In addition, the center of mass does not coincide
with the point halfway between the centers of the stars, but it is
shifted toward the more massive star. During the inspiral phase, the
heavier and more compact star is only slightly affected by its
companion, whereas the latter is decompressed rapidly while being
accreted onto the heavier star. This is visible in the three
intermediate panels of figure~\ref{fig:density-xy-um}.  The tidal
disruption of the lower-mass NS when it still retains a large fraction
of its angular momentum results in an extended tidal tail, which,
unlike what happens in the equal-mass case, transfers angular momentum
outwards in a much more efficient way. This leads to the formation of
large spiral arms extending well beyond the domain shown in
figure~\ref{fig:density-xy-um} and ultimately to a more rapid ejection
of matter. Gravitationally bound matter travelling along the spiral
arms away from the central object will form a more massive accretion
torus around the central BH than that formed in the case of an
equal-mass, symmetric binary system. It should be noted that, although
the rest-mass density of the matter in these spiral arms is much
smaller than the central one, it has nevertheless densities $\rho
\gtrsim 10^{10}\,{\rm g/cm}^{-3}$ and thus well in a
general-relativistic regime.

\subsection{Properties of the black hole}
\label{sec:BlackHoleRecoil}

As mentioned above, because of the large initial mass of the system
and irrespective of the mass ratio, the merged object rapidly
collapses to a BH. Its mass and angular momentum have been computed
making use of the dynamical-horizon
formalism~\cite{Ashtekar:2004cn,Dreyer02a}, which provides a simple
and accurate measure of the BH properties also when this is subject to
the inflow of mass and angular momentum~\cite{Baiotti04}. In the case
of the equal-mass binary, because the disk resulting from the merger
has comparatively small mass, the BH settles rapidly to an
approximately stationary configuration, and the mass and spin of the
BH measured at formation, \ie $M=2.56\,\Msun$ and $a\equiv
J/M^2=0.745$, respectively, do not vary significantly throughout the
subsequent evolution of the system.  On the other hand, when
considering unequal-mass binaries, the mass and the spin of the BH
show, on the timescale of the simulations, a variation in time 
of $\sim 5\%$ and $\sim 2\%$, respectively, because of
the continued and intense accretion of both mass and angular
momentum. Table~\ref{tab:FinalProducts} shows the corresponding
parameters for all models at the final time of the evolution, which
is not the same for the different binaries considered.

We note that finding and tracking the apparent horizon in the case of
binaries with small mass ratio is far from being simple since the
asymmetry in the merger dynamics leads to a noticeable motion of the
``center-of-mass'' of the system. Hence, the location of the trial
surface for the apparent horizon cannot be simply associated to a
pre-existing black hole (as in the case of BH
binaries~\cite{Pollney:2007ss}) or to a pre-determined coordinate
location (as in the case of the collapse of a rotating
star~\cite{Baiotti04}). The end-result of this complication is that
the apparent horizon could not be tracked successfully in all the
models under consideration. This was the case of models
\texttt{M3.4q0.80} and \texttt{M3.4q0.70}, for which it was not
possible to measure the mass and spin of the corresponding
BH. Furthermore, in the case of the binary \texttt{M3.5q0.75} the
measurements were not made with the dynamical-horizon formalism but
rather by using the ratio of the polar to equatorial circumference of
the apparent horizon as discussed in detail
in~\cite{Baiotti04}. Cross-checking the two measures (\ie
apparent-horizon distortion and dynamical-horizon formalism) in the
cases where both are possible shows that they are equally
reliable (see also the extended discussion in~\cite{Baiotti04}).
Overall, the data available suggests the existence of a local maximum
of $a$ for $q\sim 0.9$, but more data is clearly necessary to confirm
this.

Interestingly, when inspecting carefully the apparent horizon in the
low-$q$ model \texttt{M3.5q0.75} it is possible to appreciate that its
appearance precedes the time when the two stellar cores merge and is
in contrast with what happens with models with high $q$. By
comparison, we believe the same happens also for the binary
\texttt{M3.4q0.70}, although in this case we were not able to detect an
apparent horizon. Of course these considerations have little physical
importance as the interior of the apparent horizon is causally
disconnected with what is astrophysically observable; nevertheless
this result provides another interesting example of the rich
phenomenology relative to the appearance and dynamics of trapped surfaces (see,
for instance, the discussion in section $4$
of~\cite{Szilagyi:2006qy}).

\begin{table}[ht]
\caption{\label{tab:FinalProducts} Columns $2-5$ report the properties
  of the final BH, \ie mass, angular momentum, spin parameter, and
  kick velocity, while columns $6-7$ report the measured torus masses
  $M_{\rm tor}$ and those inferred from relation
  (\ref{eqn:empirical-relation}), ${\widetilde M}_{\rm tor}$. Also
  shown in columns 8 and 9, respectively, are the numerical error
  $\epsilon_{\rm tor}\equiv |1 - (M_{\rm tor})_{_{\rm HR}}/(M_{\rm
    tor})_{_{\rm MR}}|$ as computed by comparing different resolutions
  (medium, \ie $\ensuremath{\Delta_6 = 0.19}$, and high, \ie
  $\ensuremath{\Delta_6 = 0.15}$) for each model and the relative
  error $\epsilon_{\rm fit} \equiv {|\widetilde{M}_{\rm tor} - M_{\rm
      tor}|/M_{\rm tor}}$ of the phenomenological expression for the
  mass of the torus with respect to the numerical data.  Clearly, the
  binaries with high mass ratio are not well described by relation
  (\ref{eqn:empirical-relation}) even though their numerical error is
  not very large.}}  \centering \small{\begin{tabular}[t]{lcccccccc}
\br
Model & $M$ & $J$ & $a\equiv J/M^2$ & $v_{\rm kick}$    &$M_{\rm tor}$   & $|{\widetilde{M}_{\rm tor}}|$ & $\epsilon_{\rm  tor}$ & $\epsilon_{\rm fit}$  \\
      &     &     &                 & $({\rm km/s})$ & $(\Msun)$    &                     & ($\Msun$)                  &\\
\mr
\texttt{M3.6q1.00} & $2.56$        & $4.90$        & $0.745$         & $  0.28$ & $0.0010$ & $0.021$ & $ 28\%$ & $> 100\%$ \\
\texttt{M3.7q0.94} & $2.64$        & $5.18$        & $0.743$         & $121.95$ & $0.0100$ & $0.048$ & $ 12\%$ & $> 100\%$ \\
\texttt{M3.4q0.91} & $2.99$        & $7.29$        & $0.815$         & $ 59.33$ & $0.0994$ & $0.103$ & $0.8\%$ & $89.6 \%$ \\
\texttt{M3.4q0.80} & $-$           & $-$           & $-$             & $ 56.22$ & $0.2088$ & $0.193$ & $1.5\%$ & $7.4 \%$  \\
\texttt{M3.5q0.75} & $3.00^{\dagger}$& $7.13^{\dagger}$& $0.792^{\dagger}$  & $ 18.05$ & $0.0802$ & $0.173$ & $2.5\%$ & $8.1 \%$  \\
\texttt{M3.4q0.70} & $-$           & $-$           & $-$             & $ 15.82$ & $0.2116$ & $0.202$ & $2.4\%$ & $4.6 \%$  \\ 
\br
\end{tabular}
\begin{flushleft}
${}^{\dagger}$ We could not compute the dynamical horizon for this
  model, so the reported values are calculated from the apparent
  horizon, with the method employed in Sections VA and VB1
  of ref.~\cite{Baiotti04}.
\end{flushleft}
\end{table}

Finally, also reported in table~\ref{tab:FinalProducts} is the recoil
velocity imparted to the BH at the end of the inspiral and computed
using the gravitational-wave emission as discussed
in~\cite{Koppitz-etal-2007aa,Pollney:2007ss} for binary BHs. We
recall, in fact, that together with energy and angular momentum,
gravitational radiation also carries away linear momentum. If the
binary system has a degree of asymmetry (either in the mass or in the
spin) then the trajectories of the two bodies will be (slightly)
different (\eg with the smaller body moving more rapidly and, hence,
being more efficient in beaming its emission) and the momentum loss in
one direction will not be balanced by an equal loss in the
diametrically opposite direction. This effect is well-known in binary
BHs, where the recoils from quasi-circular inspirals can be as larger
as $\sim 4000\,\km$ (see~\cite{Rezzolla:2008sd} for a recent review),
but it has never been reported before for binary NSs. The recoil
velocities reported in table~\ref{tab:FinalProducts} are clearly much
smaller than those measured for binary BHs. However, they could still yield
astrophysically interesting results being comparable or larger than
the escape velocity from the core of a globular cluster that is
$v_{\rm esc}\sim 50\,\km$~\cite{Webbink:1985}. Furthermore, and
possibly surprisingly, the values reported here for irrotational
binaries which have very little initial spin, are not much smaller
than those computed for non-spinning binary BHs (see,
\eg,~\cite{Gonzalez:2008} for a recent update) and have a local maximum
for $q\sim 0.9$.

\section{Torus Formation and Properties}
\label{sec:TorusFormation}

\subsection{General Dynamics}
\label{sec:DensityProfiles}

\begin{figure*}[t] 
\begin{center}
\includegraphics[width=0.45\textwidth]{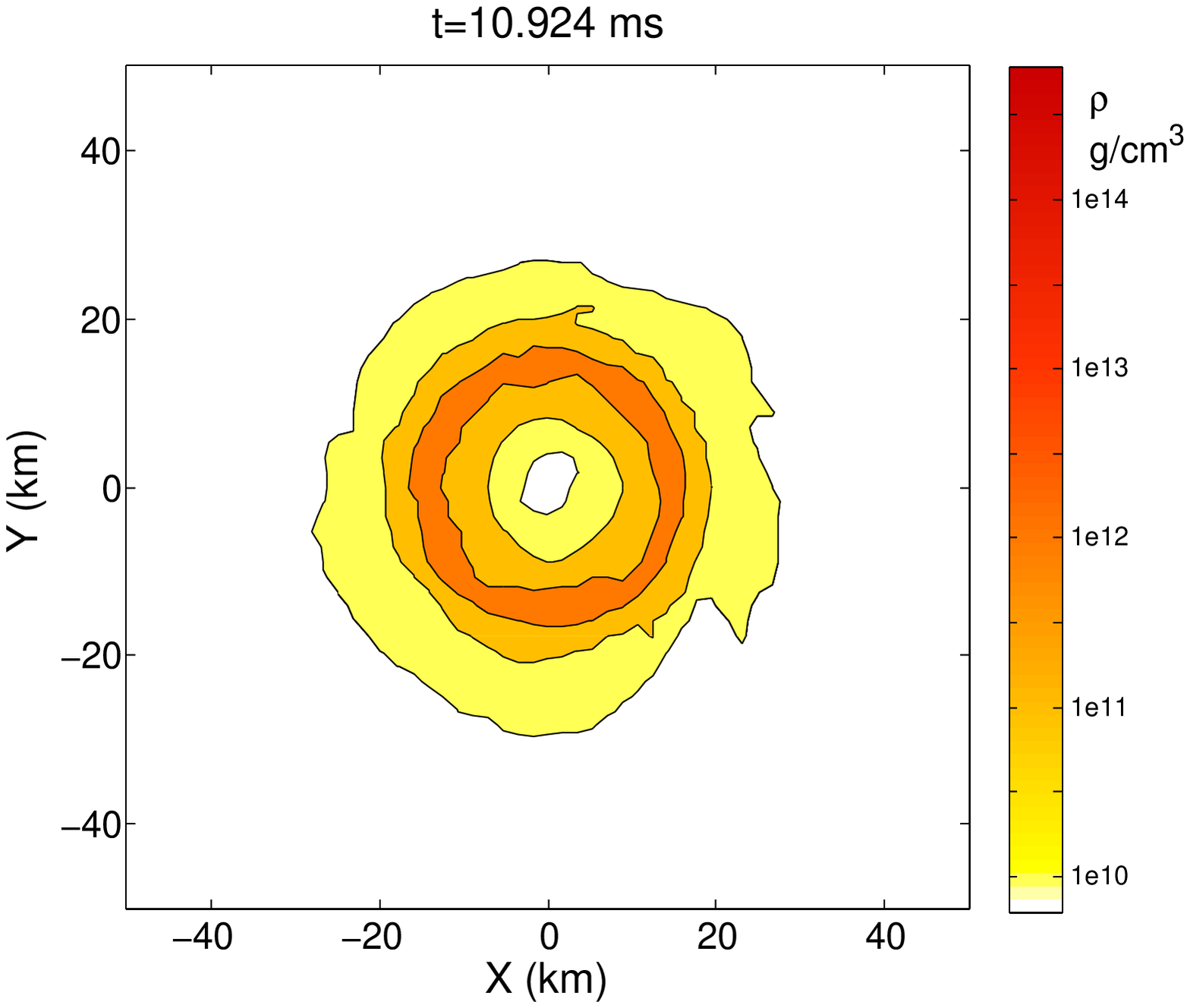}
\includegraphics[width=0.45\textwidth]{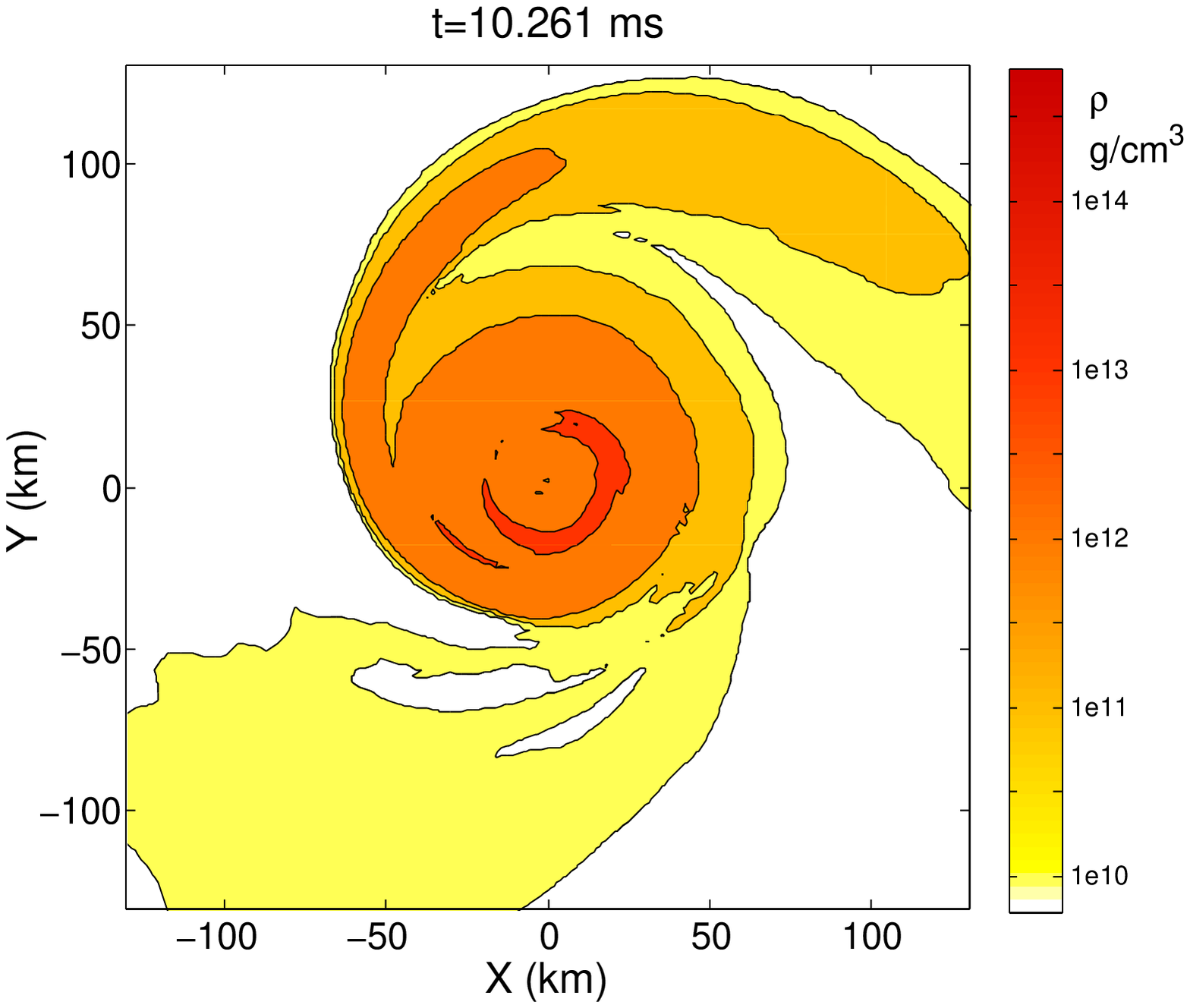}
\hfill
\includegraphics[width=0.45\textwidth]{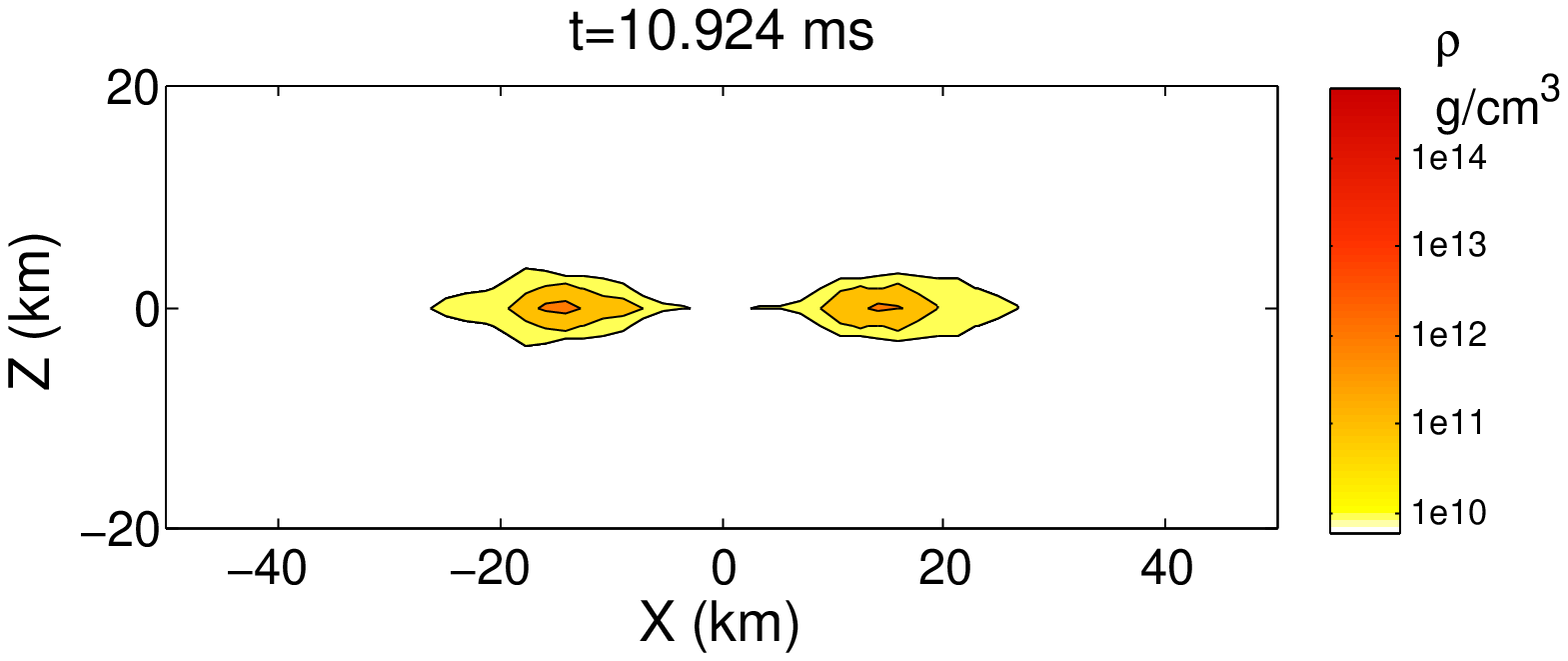}
\includegraphics[width=0.45\textwidth]{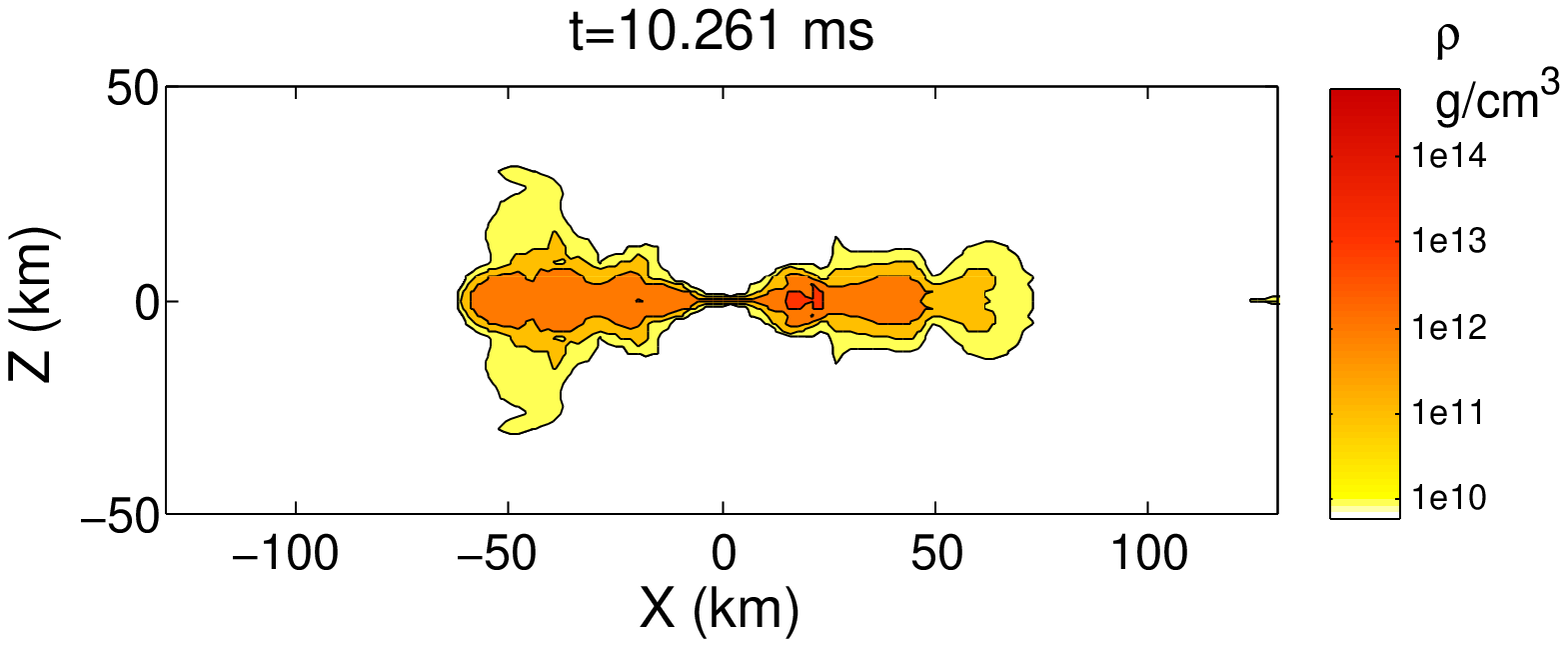}
\hfill
\end{center}
\caption{Isodensity contours for the binaries  \texttt{M3.6q1.00} (left panels) 
 and \texttt{M3.4q0.70} (right panels) showing the morphology of the tori at the 
 onset of the QSA on the \ensuremath{(x,y)} plane (upper rows) and on the 
 \ensuremath{(x,z)} plane (lower rows). Note that the disks in the two panels have 
 very different lengthscales, with the one for \texttt{M3.4q0.70}
  being about $3$ times larger than that for \texttt{M3.6q1.00}. The
  colormap used here is different from the one in
  figures~\ref{fig:density-xy-em} and~\ref{fig:density-xy-um}. Additionally, isodensity contours are
 shown for the values of \ensuremath{\rho = 10^{10}, 10^{11}, 10^{12},
 \unit[10^{13}]{g/cm^3}}.}
\label{fig:density-tori_1}
\end{figure*}

\begin{figure*}[t] 
\begin{center}
\vskip 0.5cm
\includegraphics[width=0.45\textwidth]{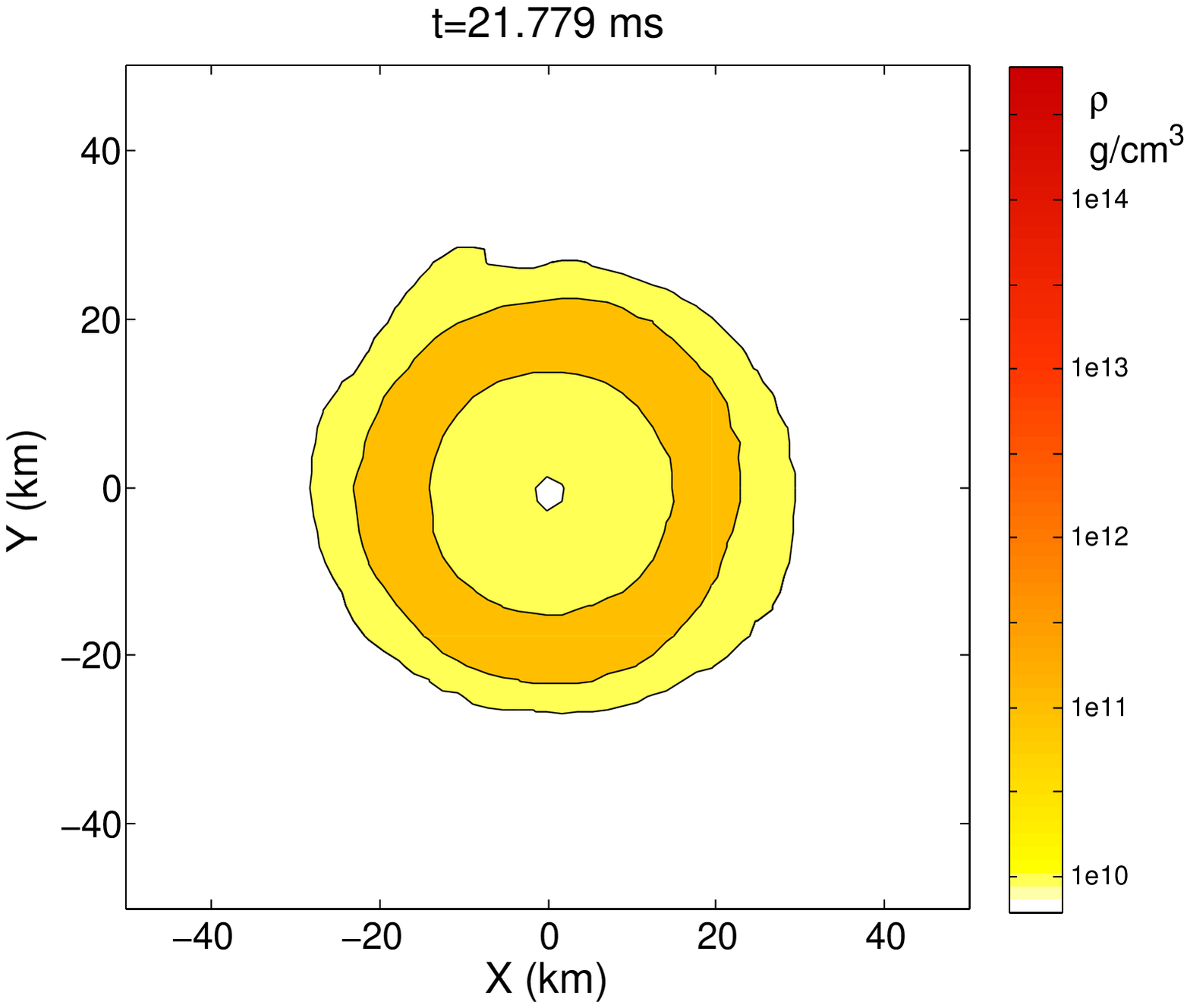}
\includegraphics[width=0.45\textwidth]{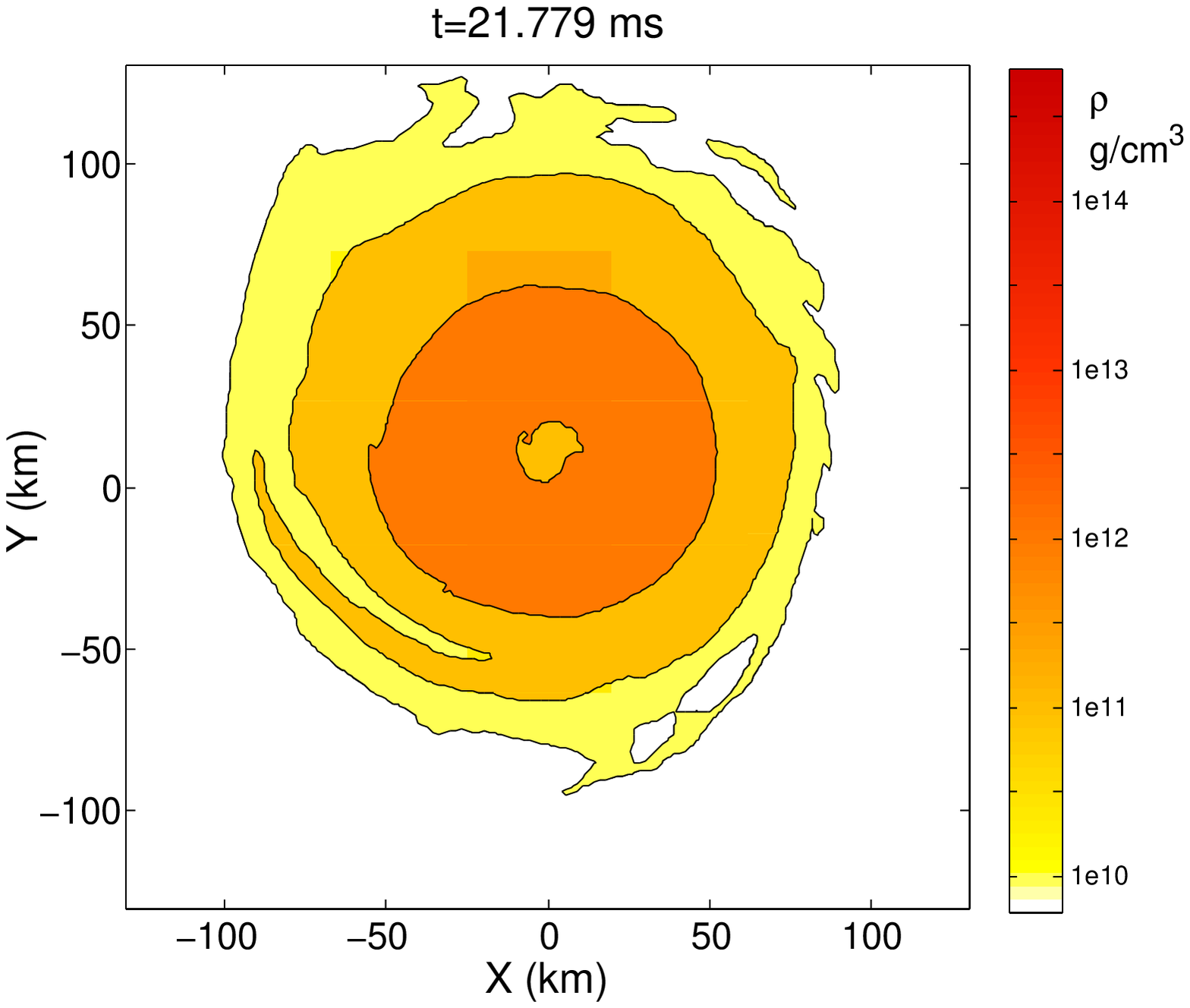}
\hfill
\includegraphics[width=0.45\textwidth]{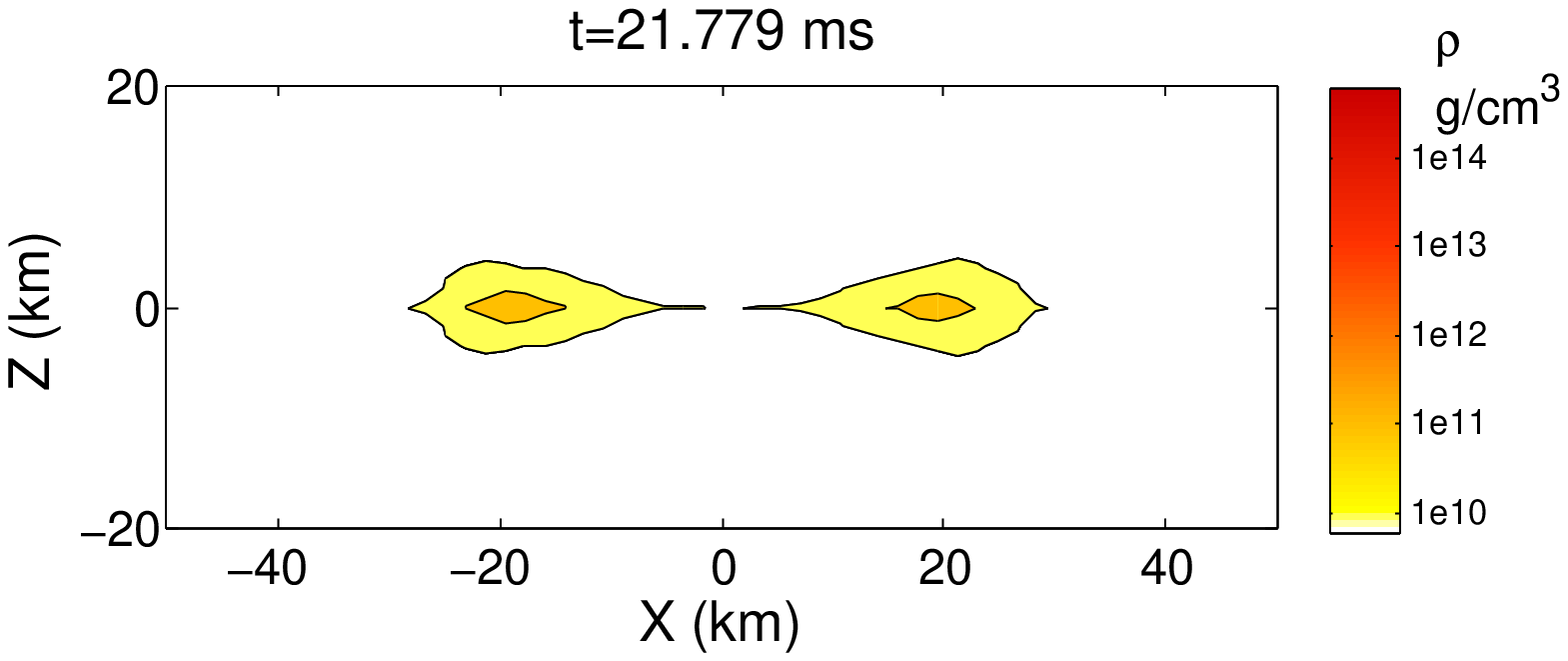}
\includegraphics[width=0.45\textwidth]{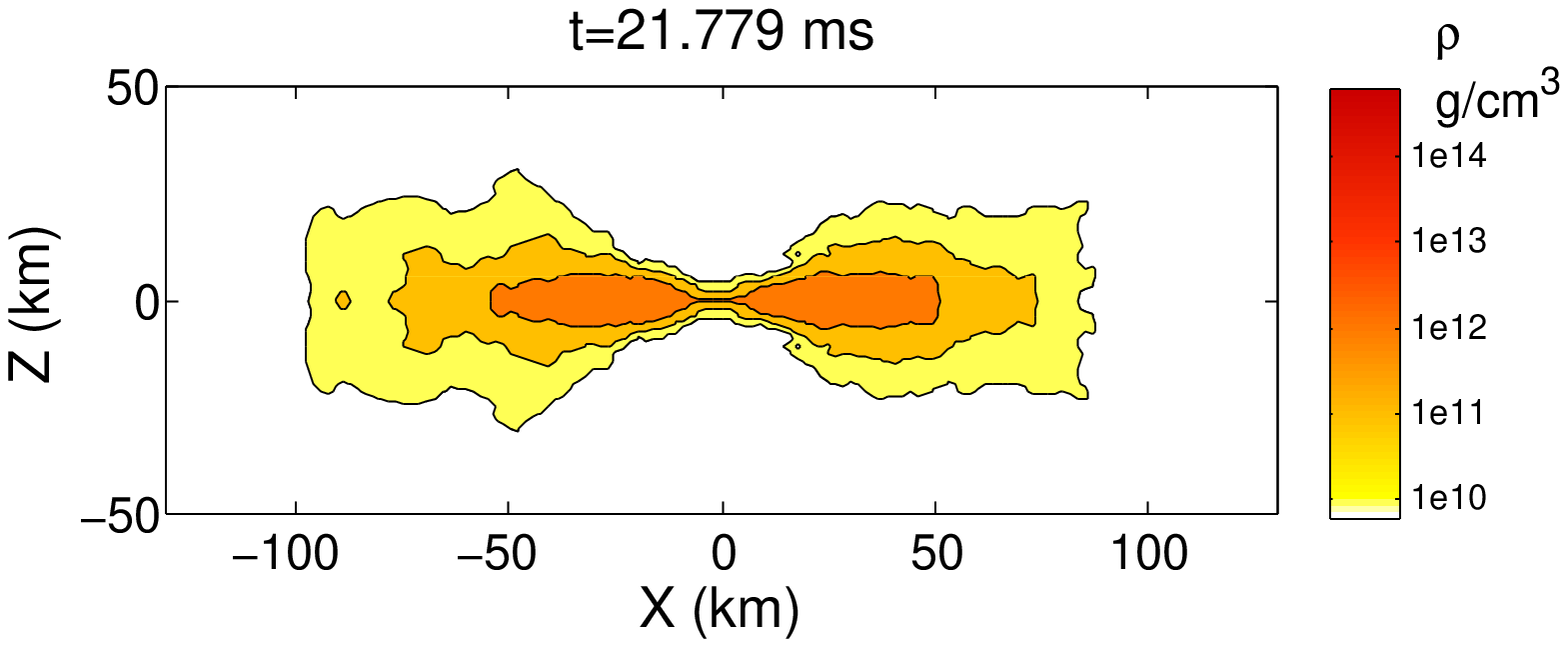}
\hfill
\end{center}
\caption{The same as figure~\ref{fig:density-tori_1} but showing the
  tori at the end of the simulation.}
\label{fig:density-tori_2}
\end{figure*}

In figures~\ref{fig:density-tori_1} and~\ref{fig:density-tori_2} we
show color-coded contours of the rest-mass density for models
\texttt{M3.6q1.00} (left panels) and \texttt{M3.4q0.70} (right
panels), either in the \ensuremath{(x,y)} plane (upper rows) and in
the \ensuremath{(x,z)} plane (lower rows). The snapshots in
figure~\ref{fig:density-tori_1}, in particular, correspond to the time
$t\sim 10\,\mss$ when the systems enter the regime of quasi-stationary
accretion (QSA, see below for definition), shortly after the formation
of the BH, while those in figure~\ref{fig:density-tori_2} refer to the
final time of the evolution, $t\sim 21\,\mss$. These figures allow for
a closer view of the morphological features of the disks, in
particular, their spatial dimensions and thickness, and are a natural
continuation of the dynamics already shown in figures
\ref{fig:density-xy-em} and \ref{fig:density-xy-um}, although they use
a different colormap that has been tuned to yield a better
contrast in the density profiles.

The large morphological differences between these two extreme models
are clearly visible in these figures.  The equal-mass model produces a
highly symmetric, geometrically thin disk, similar to the ones already
observed for other equal-mass initial data in \cite{Baiotti08}. The
unequal-mass model, on the other hand, at the time of the onset of the 
regime of QSA is characterized by the presence of a large spiral arm, 
which has not yet been accumulated onto the central disk
surrounding the formed BH. The asymmetry in the distribution of matter
at this stage is also apparent from the color map of the rest-mass
density. Only at the end of the simulation the disk of the
unequal-mass binary acquires a more axisymmetric shape. The diameters
of the disks and their heights perpendicular to the horizontal plane
differ in a significant way between the two models. More specifically,
at the end of the evolution and using the $\rho=10^{10}\,{\rm
  g/cm}^{3}$ isodensity contour as the reference value below which
material is not considered part of the disk, our simulations yield
disk diameters of $\sim 50$ km for model \texttt{M3.6q1.00} and $\sim
150$ km for model \texttt{M3.4q0.70}. The corresponding vertical scale
is $\sim 5$ km and $\sim 35$ km,
respectively.\footnote{Of course it should be noted that the spatial
  dimensions reported here depend on the cut-off chosen for the
  rest-mass density. Using cut-offs smaller than $\rho=10^{10}\,{\rm
    g/cm}^{3}$ would lead to considerably larger estimates for the
  sizes of the tori.} Taking into account all the models of our
sample, we find that both scales increase as the mass ratio
decreases. Even more worth noticing is the fact that, in the cases
considered, while the tori differ in size by about a factor $\sim 3$,
they differ by a factor $\sim 200$ in mass, while having comparable
mean rest-mass densities (see further discussion in
sections~\ref{sec:MassEvolution} and~\ref{sec:TorusMass}).

\subsection{Rest-mass Evolution}
\label{sec:MassEvolution}

\begin{figure*}[t] 
\begin{center}
\includegraphics[width=.65\textwidth]{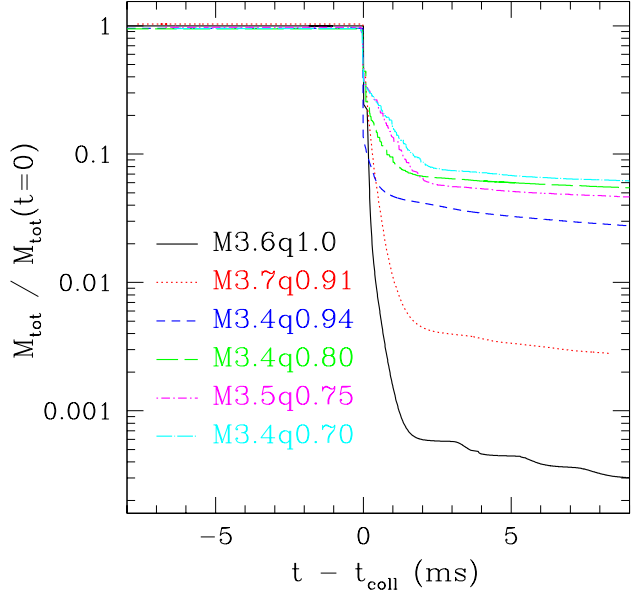}
\end{center}
\caption{Evolution of the total rest masses
  \ensuremath{M_{\mathrm{tot}}} normalized to their initial values for 
  all the models considered. The order of magnitude of the
  mass fraction in the accretion torus can be read off the logarithmic
  mass scale on the vertical axis. The curves referring to different
  models have been shifted in time to coincide at $t_{\rm coll}$,
  which represents the time when the very rapid decrease of the total
  rest mass takes place. Note that this time is not physically
  relevant (the apparent horizon is usually found earlier than $t_{\rm
    coll}$) and simply corresponds to when the very large amount of
  rest-mass accumulated in a very few cells is numerically
  dissipated.}
\label{fig:m-total}
\end{figure*}

In order to establish how the asymmetry in the mass of the two NSs in
the binary leads to tori with different masses we show in
figure~\ref{fig:m-total} the evolution of the total rest mass, defined
as
\begin{eqnarray} \label{eqn:total-rest-mass}
M_{\mathrm{tot}} &=& \int\limits_V \rho\,W\sqrt{\gamma}\,d^3x \;\;=\;\; \int\limits_V D\sqrt{\gamma}\,d^3x\,,
\end{eqnarray}
normalized to its initial value and for the different models. In
this equation $W\equiv \alpha u^t$ is the Lorentz factor, $\alpha$
being the lapse function, and $\gamma$ is the determinant of the
spatial metric. All the curves in figure~\ref{fig:m-total} have been
shifted in time to coincide at $t_{\rm coll}$, which represents the
(collapse) time at which a rapid decrease of the total rest-mass takes
place following the formation of a BH. Note that, in practice, the
collapse time is different for all models, ranging from around
\unit[6]{ms} for model \texttt{M3.4q0.70} to around \unit[11]{ms} for
model \texttt{M3.4q0.80}.

Figure~\ref{fig:m-total} shows that all models conserve the baryonic
mass almost perfectly (\ie with losses of $\lesssim 10^{-6}$) up until
the formation of the apparent horizon, after which most of the rest
mass disappears in the singularity. One obvious result which can be
deduced from figure~\ref{fig:m-total} is that the mass of the
resulting accretion disk is larger for smaller values of
$q$. However, this trend is not entirely monotone in the figure as it
is also influenced by the initial total baryonic mass of the
binary. The particular values of the tori masses computed for all
models are reported in table~\ref{tab:FinalProducts}. While the
equal-mass model produces a disk of barely $10^{-3}\Msun$, models
\texttt{M3.4q0.80} and \texttt{M3.4q0.70} produce significantly more
massive tori with masses of about $0.2\Msun$. A more detailed
discussion of the mass in the tori will be made in
section~\ref{sec:TorusMass}.

As the apparent horizon is formed, a substantial part of the rest-mass
is still outside it, although it will accrete rapidly onto the
BH. This makes the mere definition of what is the torus and its mass
rather arbitrary and we decide therefore to define the torus mass
$M_{\rm tor}$ as the total rest mass outside the apparent
horizon when the disk enters a regime of quasi-steady accretion (QSA),
a regime which is found in all models investigated. More specifically,
we compute the accretion rate as
\begin{eqnarray} \label{eqn:accretion-rate}
    \dot M_{\mathrm{tot}} &=& \frac{d}{dt}\int\limits_V \rho\,W\sqrt{\gamma}\,d^3x
\end{eqnarray}
and define the onset of the QSA as the point in time when the
condition ${\dot M_{\mathrm{tot}}}/M_{\mathrm{tot}} < 10^{-6}(G c^{-3}
\Msun)^{-2}$ is satisfied for the first time. In other words, we
define the onset of the QSA as the time when the accretion has stabilized and
the matter is moving  on essentially circular orbits. This
definition is again somewhat arbitrary, but has the advantage of
allowing for a systematic comparison of the differences in the
properties of the accretion tori produced by the several models
considered in this work.

\begin{figure*}[t] 
{\label{fig:accretion-rate:1} \includegraphics[width=0.5\textwidth]{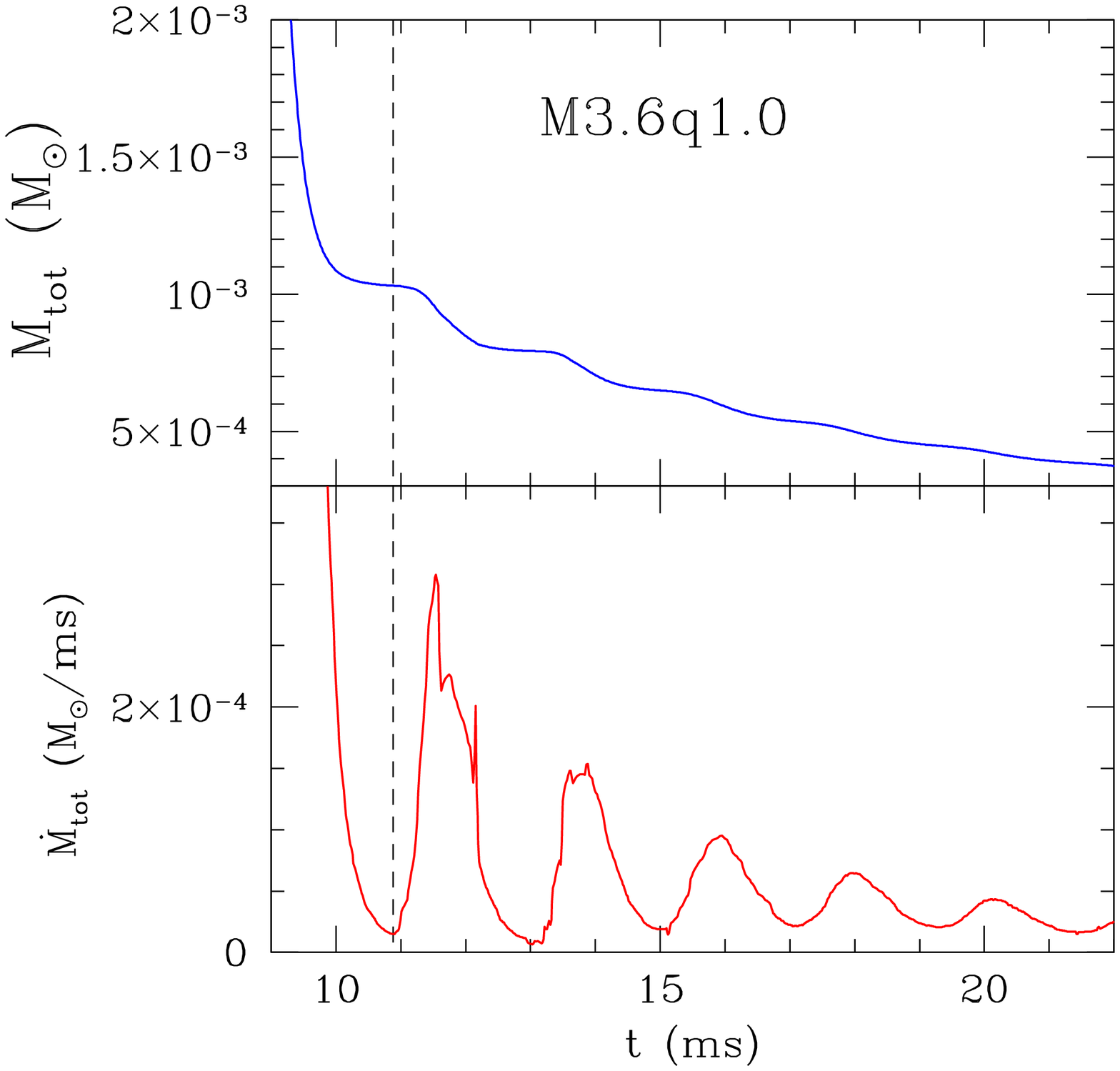}}
{\label{fig:accretion-rate:6}
  \includegraphics[width=0.5\textwidth]{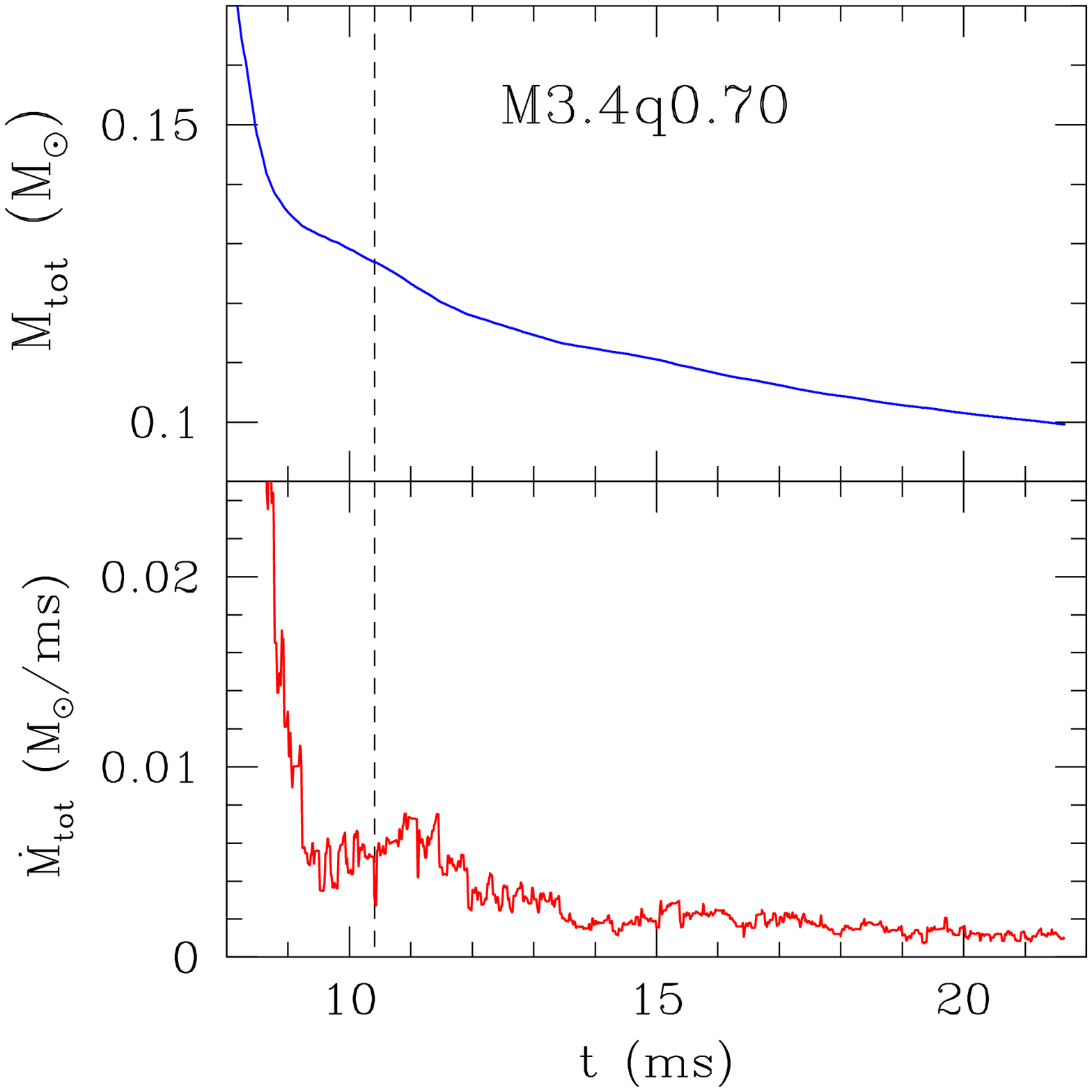}}
\caption{Evolution of the total baryonic mass $M_{\mathrm{tot}}$
  (upper rows) and of the accretion rate ${\dot M}_{\mathrm{tot}}$
  (lower rows) in the regime of QSA for the representative models
  \texttt{M3.6q1.00} (left panel) and \texttt{M3.4q0.70} (right
  panel). Indicated with a vertical dashed line is the onset of the
  QSA. Note that both $M_{\mathrm{tot}}$ and ${\dot M}_{\mathrm{tot}}$
  differ by almost two orders of magnitude when comparing equal and
  unequal-mass binaries.}
\label{fig:accretion-rate}
\end{figure*}

Figure~\ref{fig:accretion-rate} shows the evolution of the total rest
mass $M_{\mathrm{tot}}$ (upper rows) and the mass accretion rate
${\dot M_{\mathrm{tot}}}$ (lower rows) in the regime of QSA
for the two extreme models of our sample, \texttt{M3.6q1.0} (left
panel) and \texttt{M3.4q0.70} (right panel). Also indicated with a
vertical dashed line is the onset of the QSA and it should be noted
that both $M_{\mathrm{tot}}$ and ${\dot M}_{\mathrm{tot}}$ differ by
almost two orders of magnitude when comparing equal and unequal-mass
binaries. An aspect of the evolution of the accretion rate which is
quite evident in figure~\ref{fig:accretion-rate} is the sharp
difference between equal- and unequal-mass binaries. The equal-mass
case, in fact, shows an accretion rate (and indeed the whole evolution
of the torus) that is subject to quasi-periodic oscillations as the
torus moves in and out at about the radial epicyclic frequency. The
mass flux of the unequal-mass model, on the other hand, is rather
constant in time and this reflects a very different distribution of
angular momentum in the tori. Both of these aspects will be further
discussed in the following sections.

\subsection{Density evolution}
\label{sec:TorusDynamics}

Once the BH is formed, an effective gravitational potential well
builds up in which the torus undergoes radial oscillations. In the
case of an equal-mass binary, the well is essentially axisymmetric and
the dynamics of oscillating relativistic tori in equilibrium and in
axisymmetry have been analyzed extensively in a series of
papers~\cite{Zanotti03b,Rezzolla_qpo_03b,Zanotti05,Montero07} in the
test-fluid approximation (where the self-gravity of the disk is
neglected), with and without magnetic fields, and for the cases of
Schwarzschild and Kerr BHs. These papers have shown that, upon the
introduction of perturbations in the tori, a long-term oscillatory
behavior is found, lasting for tens of orbital periods. These
oscillations correspond to axisymmetric $p$-mode oscillations whose
lowest-order eigenfrequencies appear in the harmonic sequence
$2$:$3$. This harmonic sequence is present with a variance of $\sim
10\%$ for tori with a constant distribution of specific angular
momentum and of $\sim 20\%$ for tori with a power-law distribution of
specific angular momentum. More recently, those studies have been
extended by~\cite{Montero2008}, where systems formed by a BH (in the
puncture framework) surrounded by (marginally stable) self-gravitating
disks have been evolved in axisymmetry. Even in this case, the ratio
of the fundamental oscillatory mode and the first overtone also shows
approximately the $2$:$3$ harmonic relation found in earlier
works~\cite{Rezzolla_qpo_03b,Zanotti05,Montero07}.

The dynamics of the BH--torus system produced by the merger of model
\texttt{M3.6q1.00} is considerably more complicated than that
considered in the test-fluid studies, for which initial configurations
in stable equilibrium could be found. However, despite the fact that these
systems are studied {\it ab-initio} as the end-products of highly
dynamical events, it is remarkable that so much of the phenomenology
studied and reported
in~\cite{Zanotti03b,Rezzolla_qpo_03b,Zanotti05,Montero07} continues to
apply also here. Unfortunately, although the simulation extends to
$\sim 26\,\mss$, the timeseries is much too short to provide a firm
evidence of the presence of the $2$:$3$ harmonic relation, although
the spectral analysis of the data indicates that excess power is
present at such frequencies.

\begin{figure*}[t] 
\begin{center}
{\label{fig:rho-max:1} \includegraphics[width=.45\textwidth]{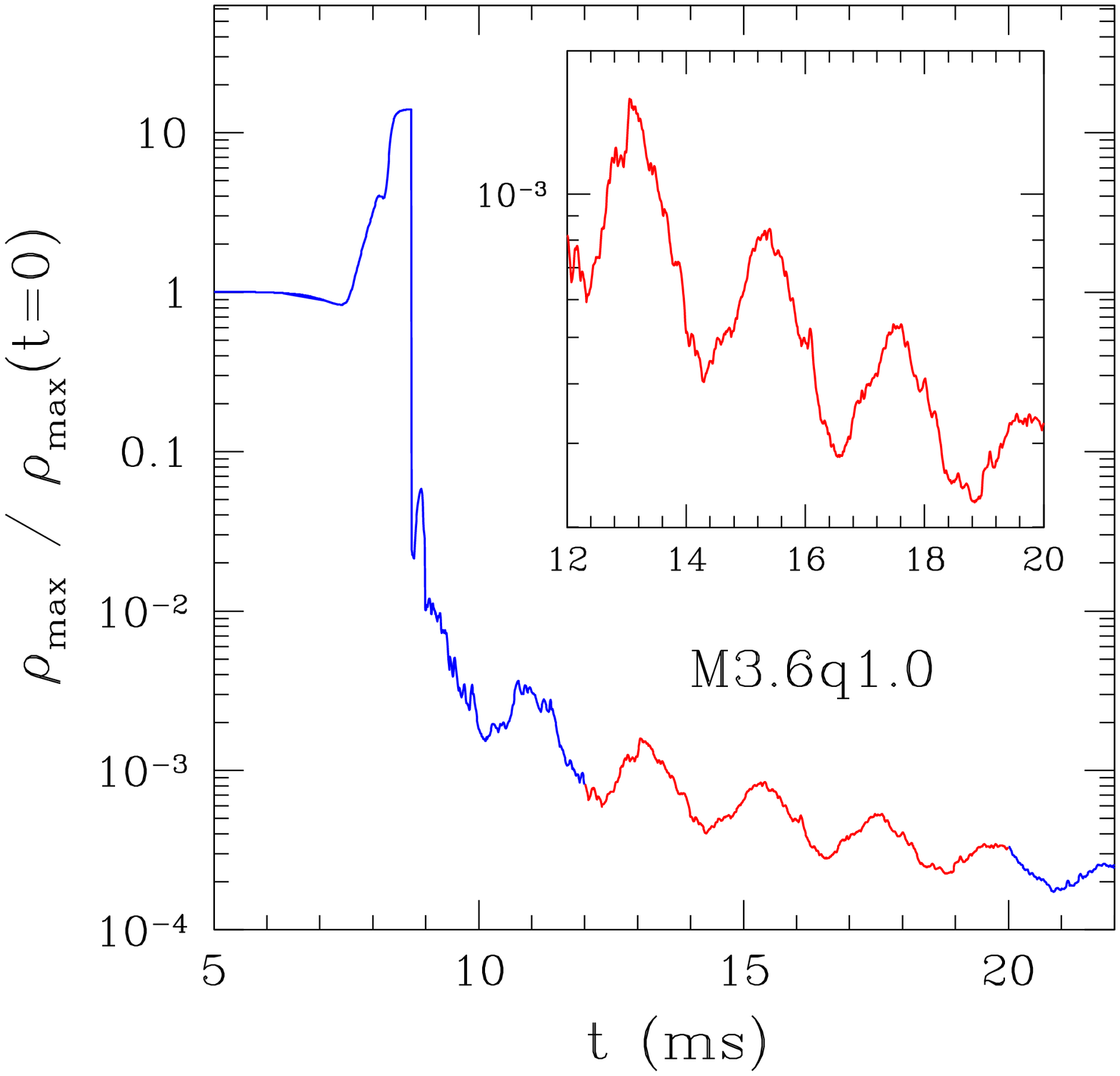}}
\hspace{.5 cm} 
{\label{fig:rho-max:6} \includegraphics[width=.45\textwidth]{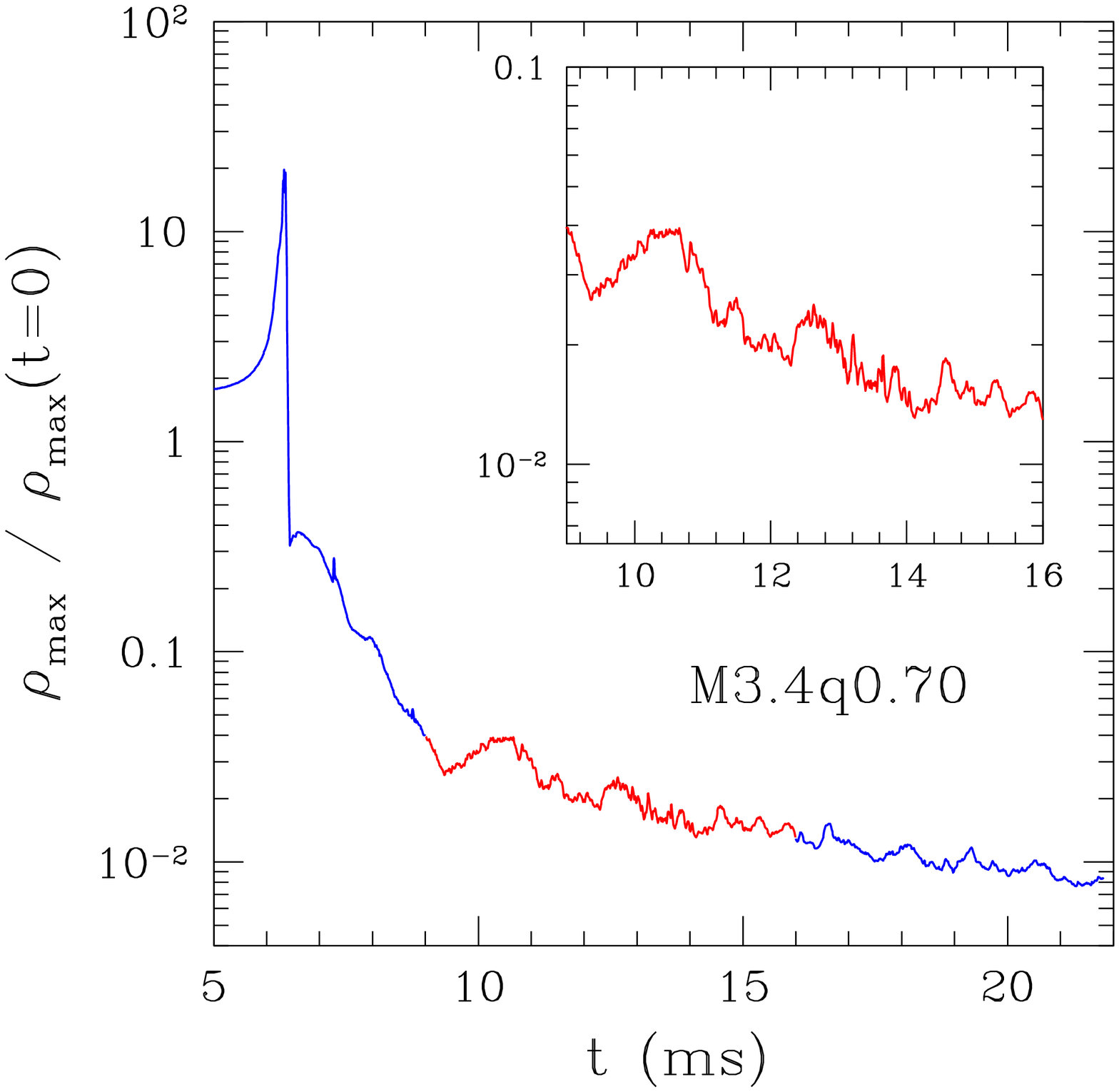}}
\end{center}
\caption{Evolution of the maximum of the rest-mass density
  \ensuremath{\rho_\mathrm{max}}, normalized to its initial value for
  the representative models \texttt{M3.6q1.00} (left panel) and
  \texttt{M3.4q0.70} (right panel). The rapid drops take place well
  after an apparent horizon has been formed and are caused by the
  numerical methods which are no longer able to resolve the very large
  gradients in the very central grid cells. The two insets provide a
  magnified view of the evolution of the density in the torus and help
  to contrast the periodic accretion produced in the case of
  equal-mass binaries and the QSA for the unequal-mass binaries.}
\label{fig:rho-max}
\end{figure*}

To provide additional evidence that the harmonic behaviour is not just
in the accretion rate, figure~\ref{fig:rho-max} shows the evolution of
the maximum of the rest-mass density \ensuremath{\rho_\mathrm{max}},
normalized to the corresponding initial value, for the two extreme
models of our sample, \texttt{M3.6q1.00} and \texttt{M3.4q0.70}.  The
equal-mass model \texttt{M3.6q1.00} (represented in the left panel of
figure \ref{fig:rho-max}) shows the most regular and pronounced
oscillatory behavior, as was already evident in the time evolution of
the total rest mass and accretion rate in
figure~\ref{fig:accretion-rate}.  The two insets in this figure
magnify these features in the QSA regime and, in the case of
\texttt{M3.6q1.00}, they highlight the presence of both maxima and
minima corresponding to configurations when the torus
reaches the point of closest approach to the BH (periapsis or
pericenter) and of farthest excursion (apoapsis or apocenter),
respectively.  A similar trend can be hinted also for the
\texttt{M3.4q0.70} model on the right panel of Fig~\ref{fig:rho-max},
although the quality of the oscillations is smaller in this case, most
likely because in this case the enhanced tidal disruption during the
merger phase leads to a more complex dynamics. Interestingly, such
oscillations seem to become more regular during the final stages of
the evolution, \ie for $t \gtrsim 17\,\mss$, as the torus reaches a
more axisymmetric configuration.

\begin{figure*}[ht] 
\begin{center}
{\label{fig:rest-mass-dens-xt:1}\includegraphics[width=0.45\textwidth]{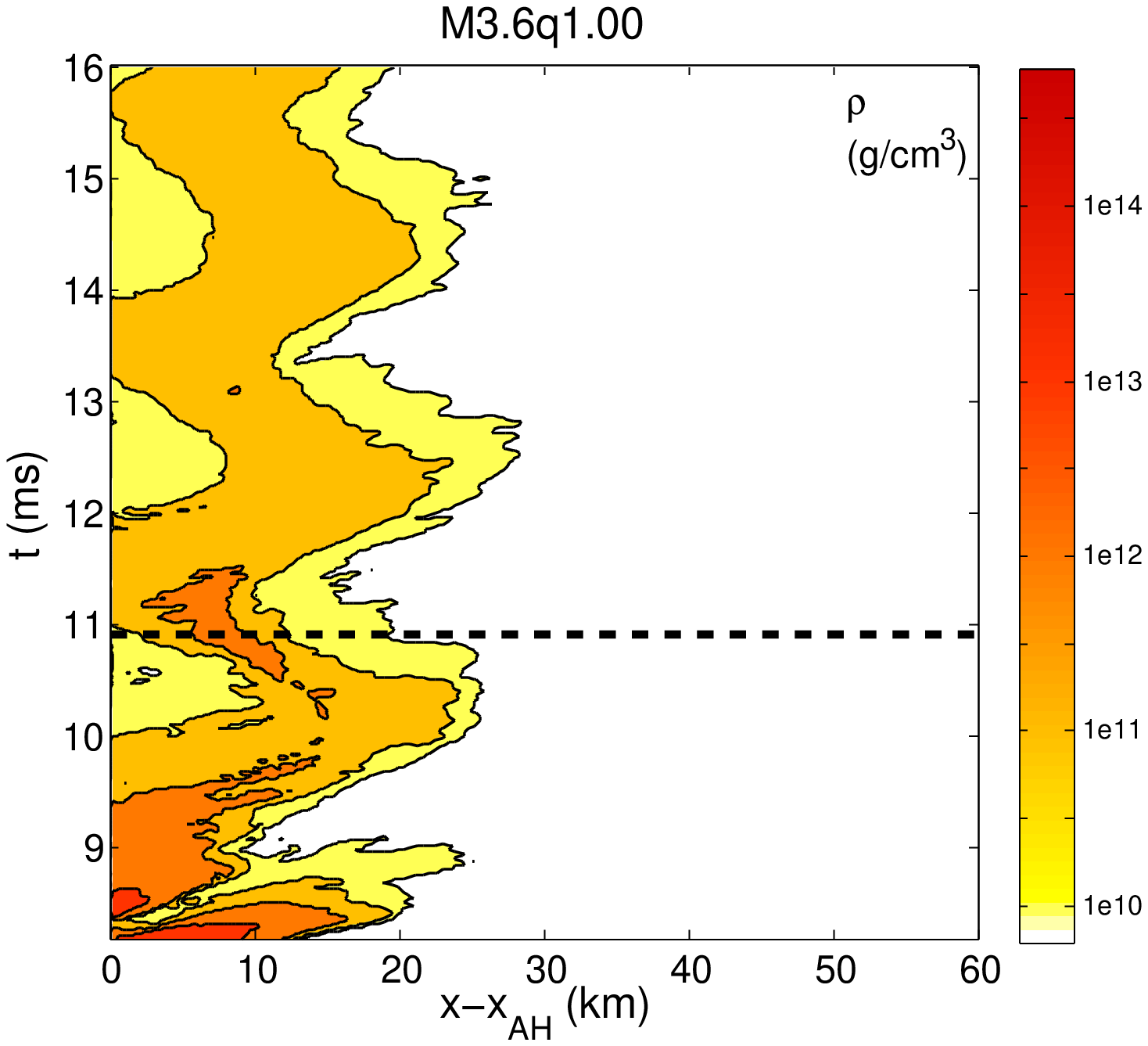}}
{\label{fig:rest-mass-dens-xt:2}\includegraphics[width=0.45\textwidth]{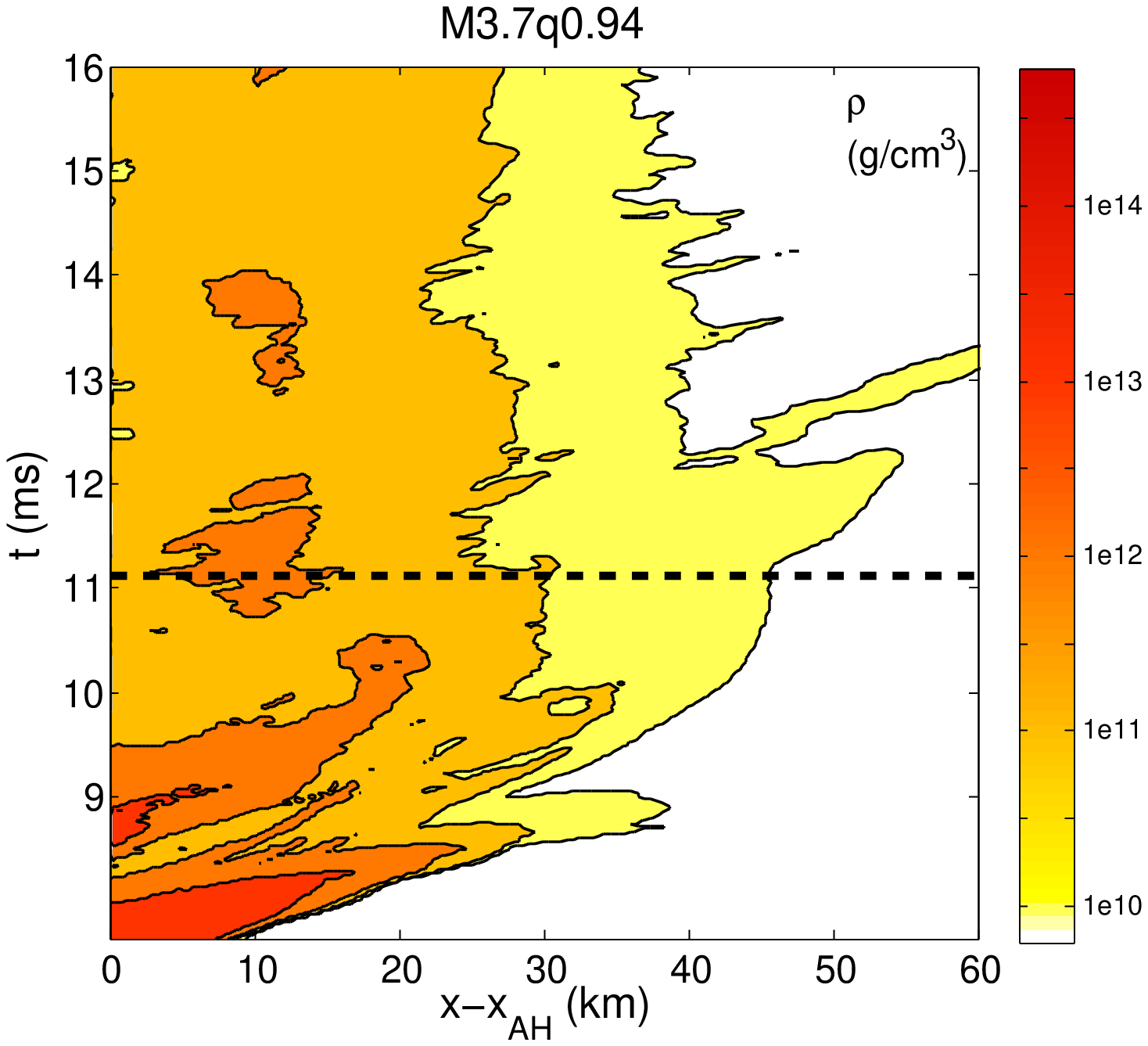}}
{\label{fig:rest-mass-dens-xt:3}\includegraphics[width=0.45\textwidth]{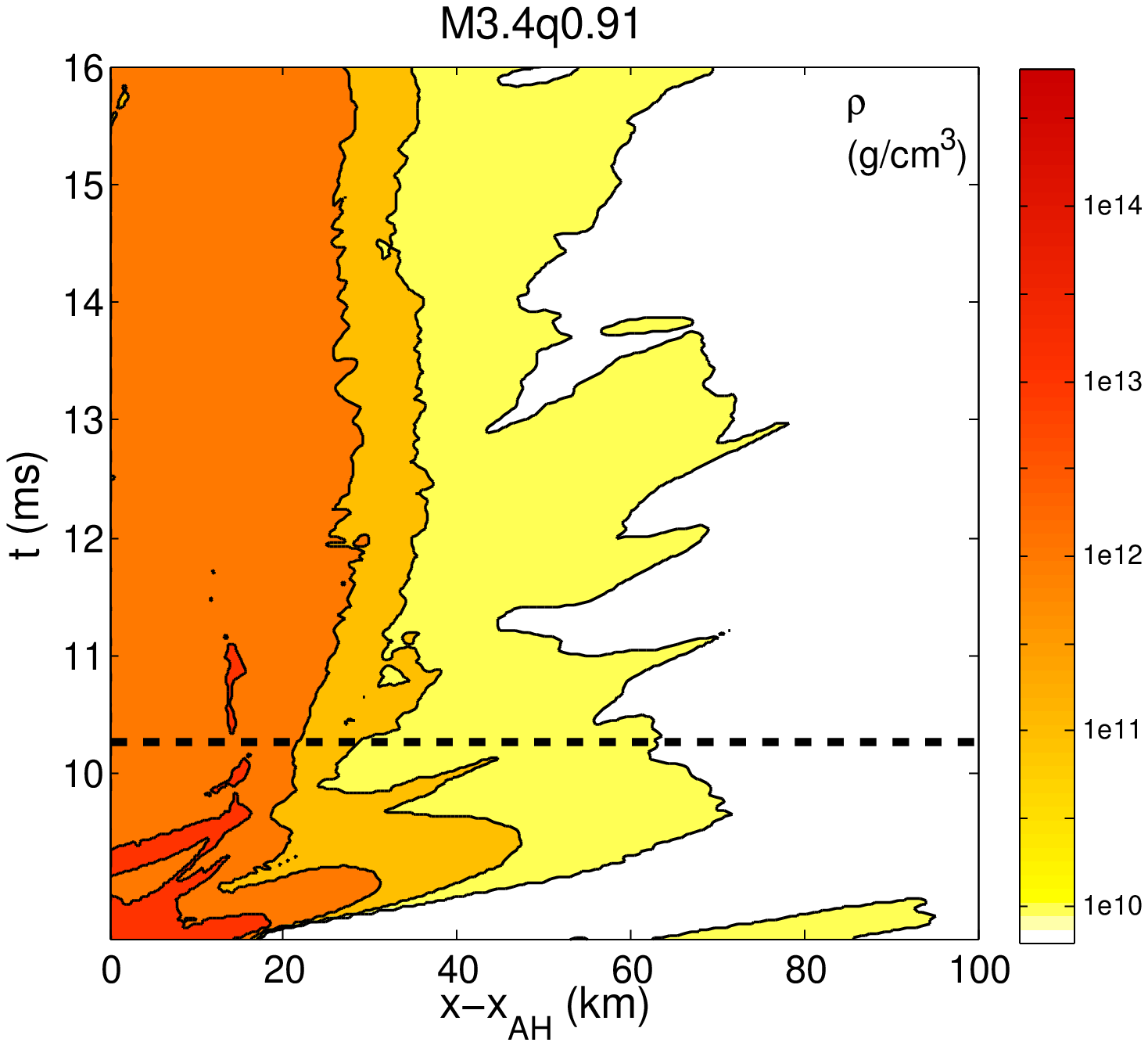}}
{\label{fig:rest-mass-dens-xt:4}\includegraphics[width=0.45\textwidth]{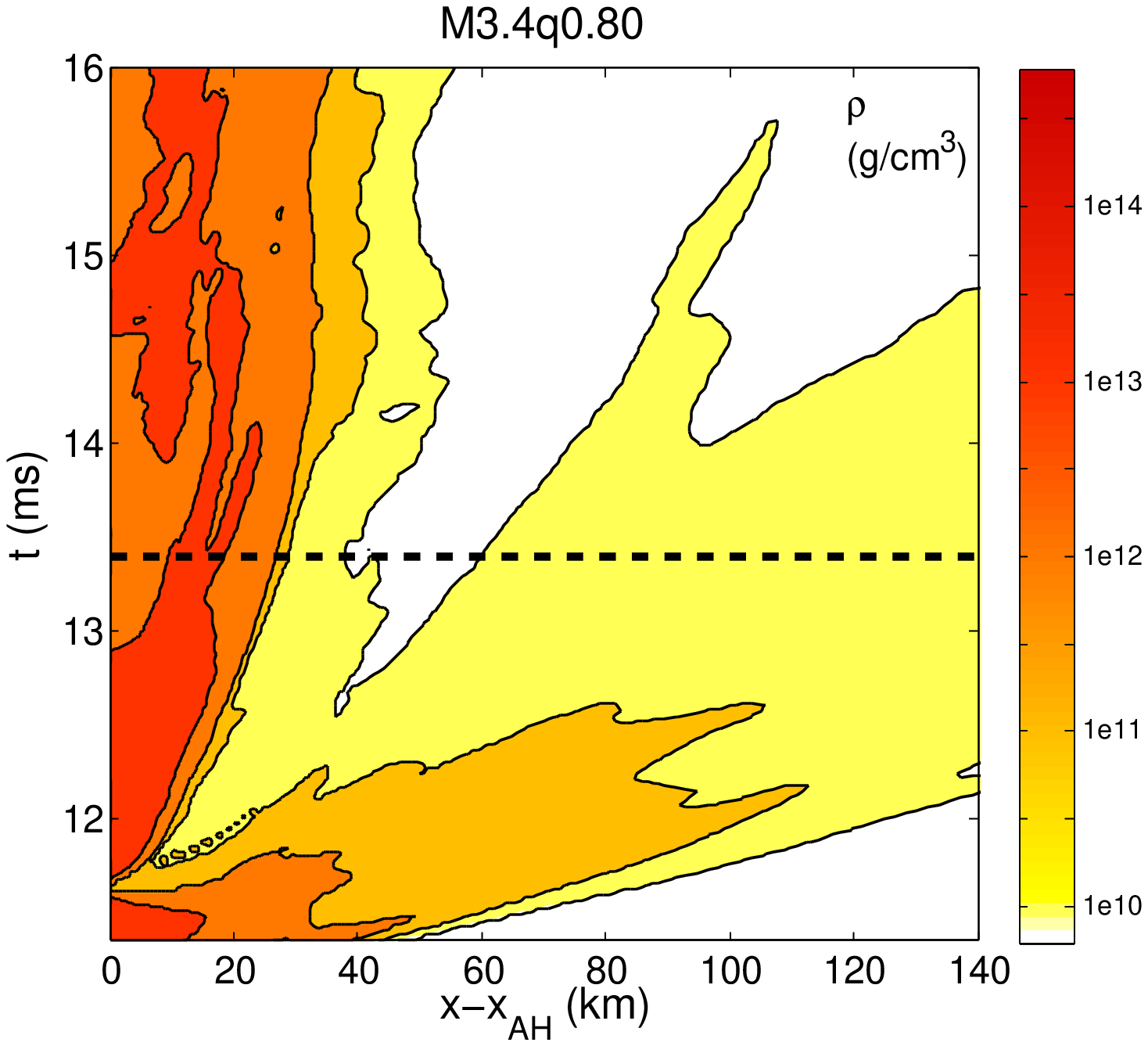}}
{\label{fig:rest-mass-dens-xt:5}\includegraphics[width=0.45\textwidth]{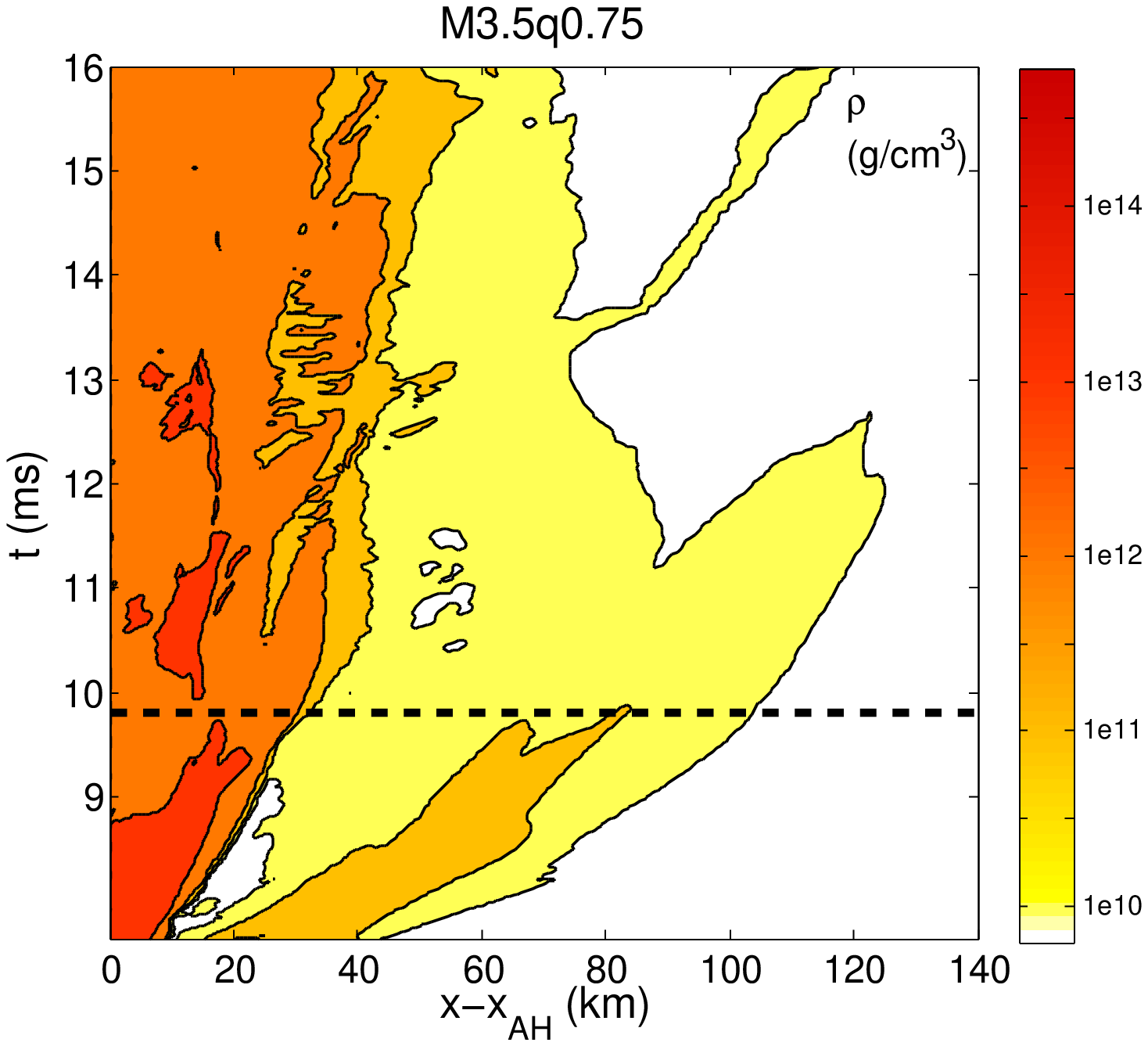}}
{\label{fig:rest-mass-dens-xt:6}\includegraphics[width=0.45\textwidth]{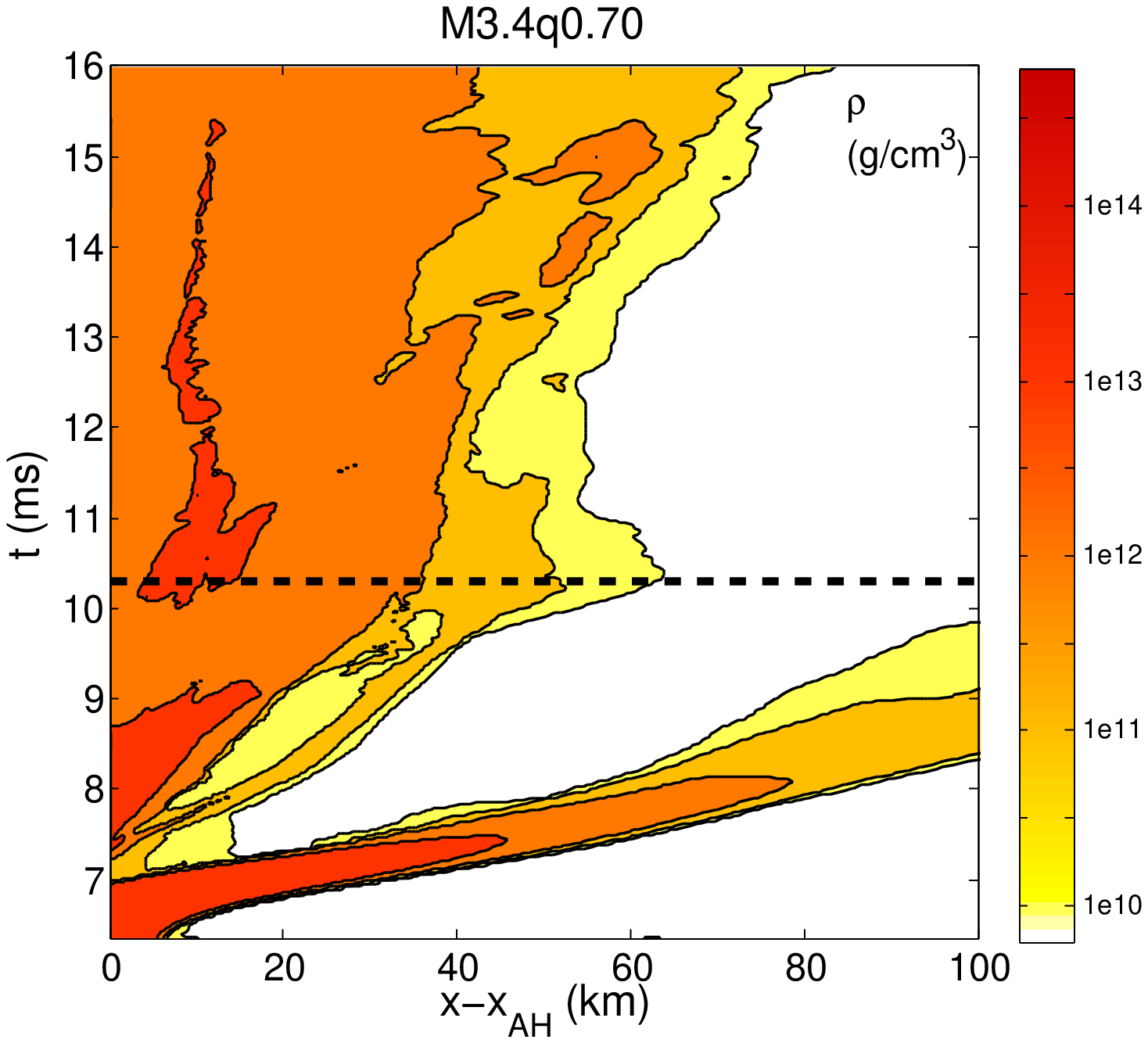}}
\end{center}
\caption{Evolution of the rest-mass density \ensuremath{\rho} along
  the positive \ensuremath{x} axis in a frame comoving with the BH. The
  panels show the color-coded rest-mass density embedded in a
  spacetime diagram with the \ensuremath{(x-x_{_{\rm AH}})} coordinate
  on the horizontal axis, being $x_{_{\rm AH}}$ the position of the
  apparent horizon, and the coordinate time \ensuremath{t} on the
  vertical axis. The color code is indicated to the right of each
  plot. Additionally, isodensity contours are shown for the values of
  \ensuremath{\rho = 10^{10}, 10^{11}, 10^{12},
    \unit[10^{13}]{g/cm^3}}. For models \texttt{M3.4q0.80} and
  \texttt{M3.4q0.70}, where the horizon could not be tracked,
  $x_{_{\rm AH}}$ represents a guess for the border of the
  horizon. For each model, the dotted horizontal line marks the onset
  of the regime of QSA.}
\label{fig:rest-mass-dens-xt}
\end{figure*}

A novel technique to analyze the evolution of the tori and to gain
some insight on their dynamics is that offered by spacetime diagrams
for observers comoving with the black hole. This is shown in
figure~\ref{fig:rest-mass-dens-xt}, which reports the evolution of
the color-coded rest-mass density embedded in a spacetime diagram with
the \ensuremath{(x-x_{_{\rm AH}})} coordinate on the horizontal axis,
where $x_{_{\rm AH}}$ is the position of the apparent horizon, and the
coordinate time \ensuremath{t} on the vertical axis. The color code is
indicated to the right of each plot and isodensity contours are shown
for the values of \ensuremath{\rho = 10^{10}, 10^{11}, 10^{12},
  \unit[10^{13}]{g/cm^3}}. Note that while these values are the same
for all the panels, the spatial dimensions vary considerably. For each
model, the dotted horizontal line marks the onset of the regime of
QSA.

By comparing the spacetime diagrams for all models it is evident that
only the equal-mass model \texttt{M3.6q1.00} shows a global
oscillatory movement with respect to the location of the BH
horizon. The movement is indeed global as all the isodensity contours
plotted oscillate simultaneously and the maximum and minimum radial
extensions reached by the disk (as signalled by the location of the
$10^{10}\, {\rm g/cm}^3$ contour) are $\sim 25\,\km$ and $\sim
15\,\km$, respectively. It is these oscillations that produce the periodic
increase in the maximum rest-mass density reported in
figure~\ref{fig:rho-max} and it is easy to appreciate that in this
case the average density in the disk is less than about $10^{12}\,
{\rm g/cm}^3$.

Scrolling through the different panels in
figure~\ref{fig:rest-mass-dens-xt} it is possible to appreciate that
the dynamics of the torus is strongly influenced by the mass
ratio. More specifically, models \texttt{M3.4q0.80},
\texttt{M3.5q0.75}, \texttt{M3.4q0.70}, show very rapid expansions
corresponding to the ejection of the large spiral waves discussed in
the previous sections. As we will comment later on, most of this
matter is still bound but it nevertheless reaches distances which are
several hundreds of $\km$ away from the BH, leading to tori that have
spatial dimensions as large as $\sim 80\,\km$. Furthermore, noticeably
higher average rest-mass densities are reached in the three low-$q$
models. As a result, while tidal disruption sweeps away a large
fraction of the external layers of the less massive star in the
binary, the corresponding tori are still able to retain the inner and
denser regions; this is particularly the case for models
\texttt{M3.4q0.80} and \texttt{M3.4q0.70}, where the tori reach
maximum densities as high as $\sim 10^{14}\, {\rm g/cm}^3$. 

We finally note that, because the matter in the spiral arms is bound, it
will eventually fall back onto the tori, where it may lead to enhanced
accretion and consequently to a new and \textit{delayed} gamma-ray
emission as the one recently observed
in~\cite{Giuliani2010}. Determining the dynamics of the fall-back
material is therefore of great importance astrophysically and the
focus of our attention in future work.

\subsection{Dynamical Instabilities}
\label{sec:TorusStability}

As mentioned in the Introduction, current models of GRBs assume that
the central engine is a system consisting of a BH and a thick disk,
either formed at the late stages of the coalescence of two NSs or
after the gravitational collapse of a massive star. The energy supply
comes from the energy released by the accretion of disk material on to
the BH and from the rotational energy of the BH itself, which can be
extracted, for instance, via the Blandford-Znajek
mechanism~\cite{Blandford1977}. This vast amount of energy (of the order of
$10^{53}$--$10^{54}$ erg, depending on the mass of the disk and on the
BH rotation and mass) is sufficient to power a GRB if the energy
released can be converted into $\gamma$-rays with an efficiency of
about a few percent. This scenario requires a stable enough system to
survive for a few seconds. In particular, the internal-shock
model~\cite{rees:94} implies that the duration of the energy release
by the source has a duration comparable with the observed duration of
the GRB. Any instability which might disrupt the system on shorter
timescales, such as the so-called runaway
instability~\cite{nishida:96a}, could pose a severe problem for the
accepted GRB models. The runaway instability was first pointed out 
in~\cite{Abramowicz83} and operates as follows: If the torus is
initially filling its Roche lobe, transfer of mass onto the BH is
possible through the cusp located at the $\mathrm{L}_{1}$ Lagrange
point. As a result of accretion, the mass of the BH increases, thus
leading to a change in the gravitational field of the system and
ultimately to a change in the position of the cusp. This can move either
inwards (towards the BH) or outwards (away from the BH), and when the latter
happens it leads to an increase in the mass transfer and hence to the
runaway accretion of the torus on a timescale of a few milliseconds.

The runaway instability has been investigated under different
assumptions and approximations (see~\cite{Font02a,Montero09} and
references therein). Early simplified studies based on stationary
models showed that, on one hand, the self-gravity of the disk
favours the instability, and, on the other hand, there are also
parameters which may help to stabilize the disk, such as the rotation
of the BH and the radial distribution of specific angular
momentum. The first time-dependent, general relativistic
hydrodynamical axisymmetric simulations of the runaway instability of
tori around BHs were performed
by~\cite{Font02a,Font02b,Zanotti03,Daigne04}, who treated the dynamics
of the gravitational field in an approximate way and neglected the
self-gravity of the torus. Overall,~\cite{Font02a,Zanotti03,Daigne04}
found that tori with constant distribution of specific angular
momentum were unstable while non-constant (power-law) angular momentum
disks were stable. More recently, in~\cite{Montero09} the first
simulations in full general relativity of marginally-stable
self-gravitating tori in axisymmetry were performed with the purpose
of evaluating the influence of the torus self-gravity on the runaway
instability. The results of~\cite{Montero09} indicate that the tori
are indeed stable irrespective of the angular momentum
distribution. It is therefore interesting that the results presented
in figure~\ref{fig:rest-mass-dens-xt}, which are not restricted to
axisymmetry but are however constrained to much shorter timescales,
reach the same conclusion: Self-gravitating tori around BHs, as those
produced by the merger of binary NSs, are stable at least on the
dynamical timescales investigated here. Additional considerations on
the stability of the tori are presented in the following section.

\begin{figure*}[ht] 
\begin{center}
{\label{fig:spec-ang-mom-xt:1}\includegraphics[width=0.45\textwidth]{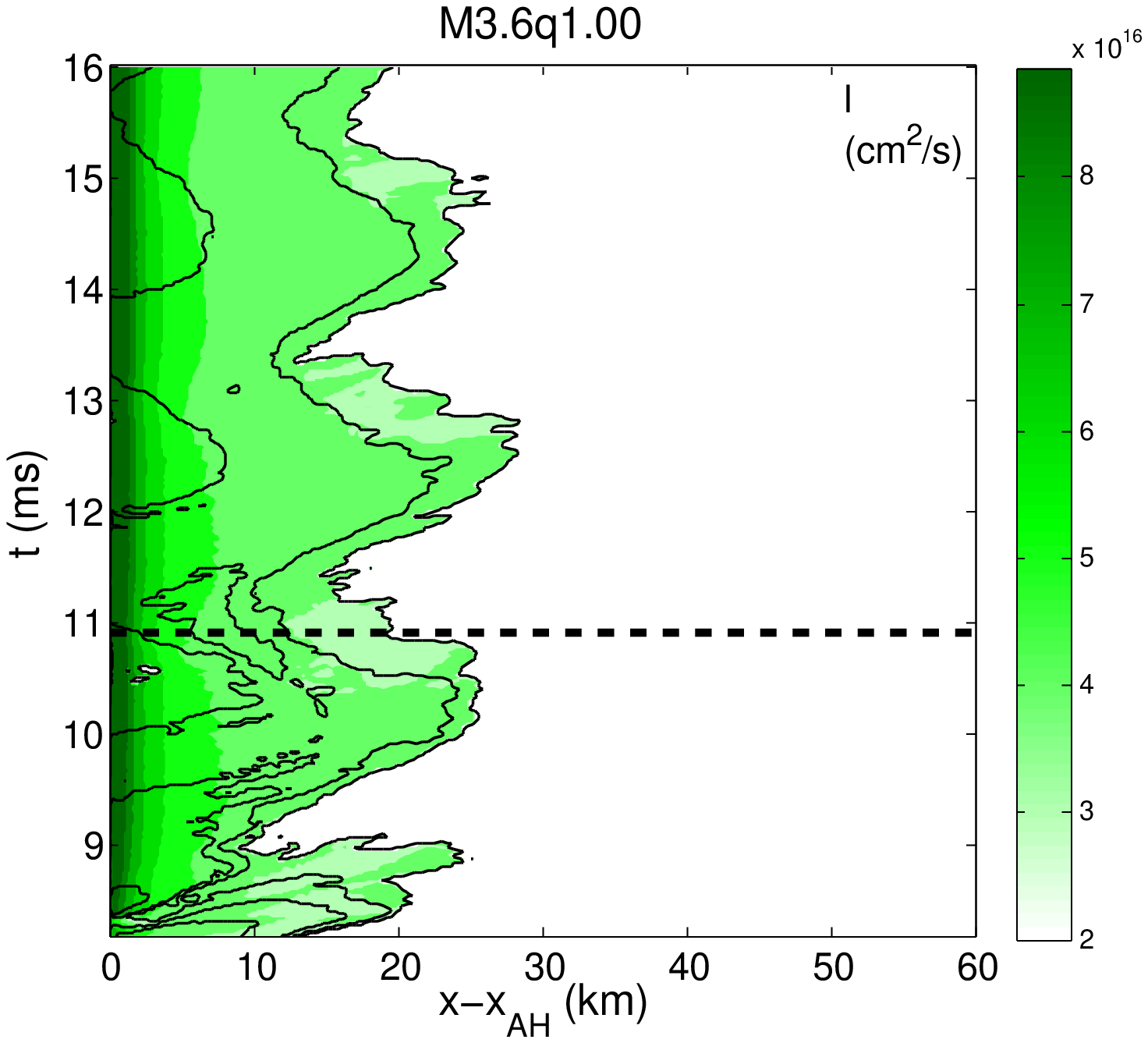}}
{\label{fig:spec-ang-mom-xt:2}\includegraphics[width=0.45\textwidth]{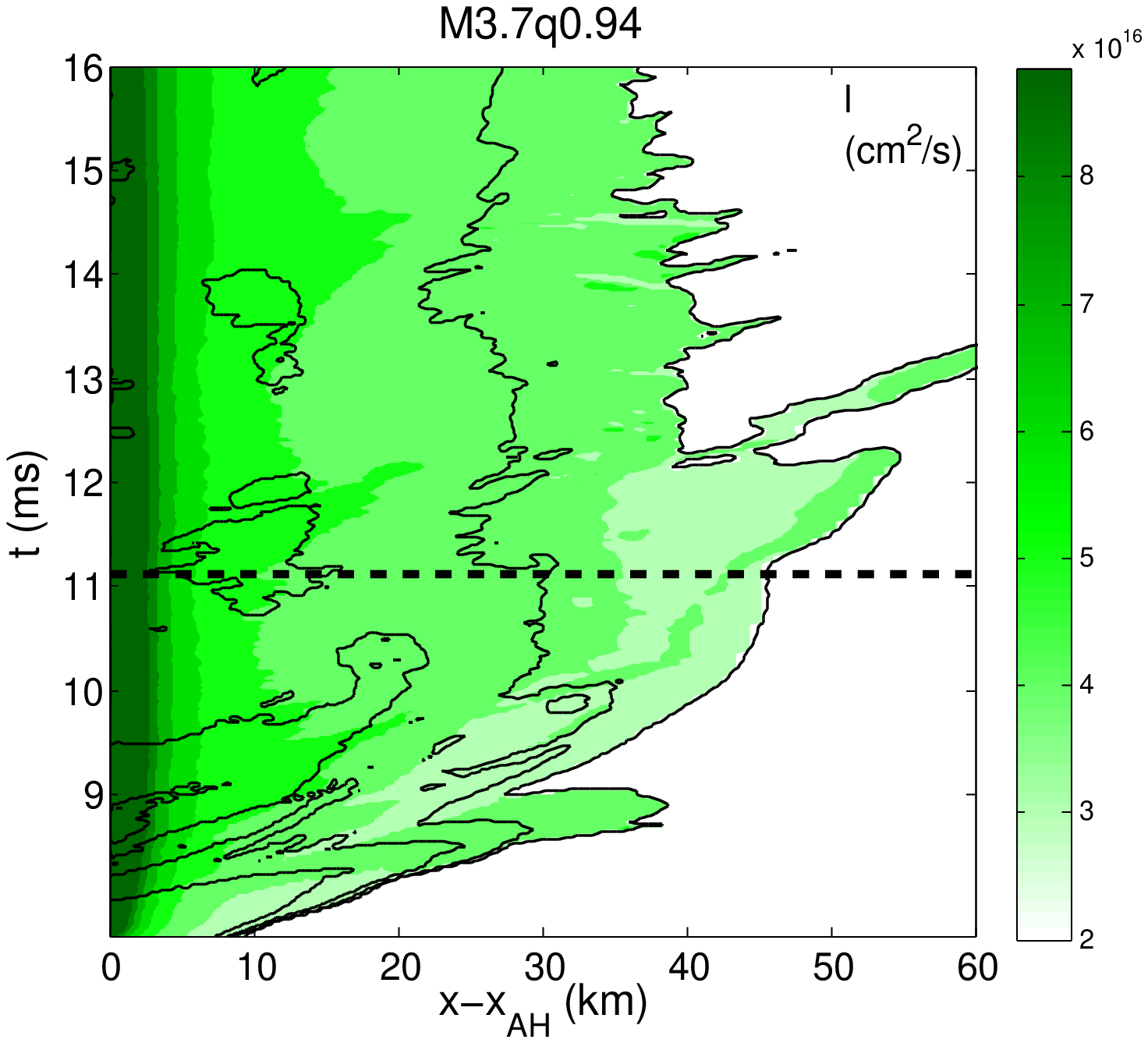}}
{\label{fig:spec-ang-mom-xt:3}\includegraphics[width=0.45\textwidth]{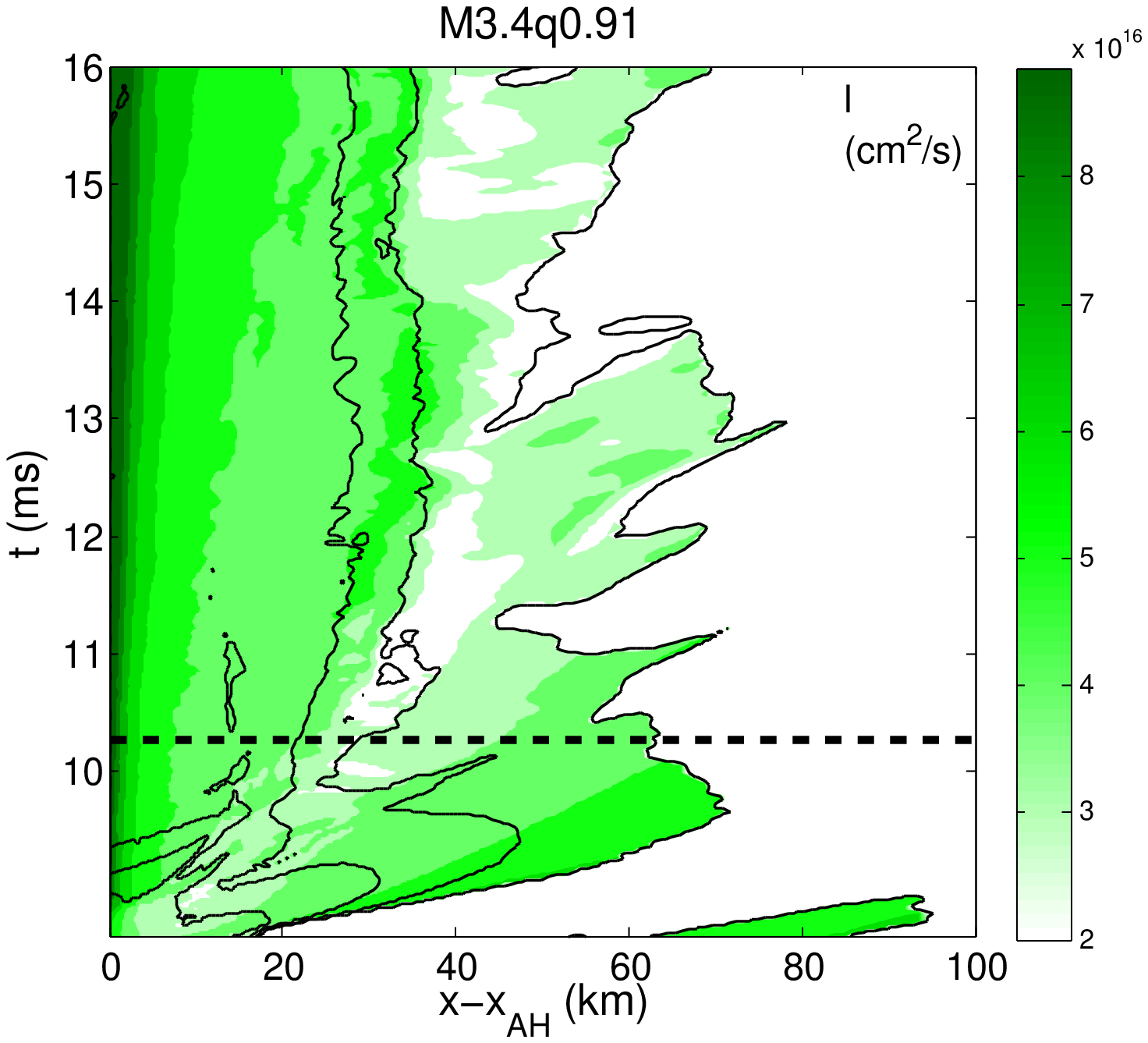}}
{\label{fig:spec-ang-mom-xt:4}\includegraphics[width=0.45\textwidth]{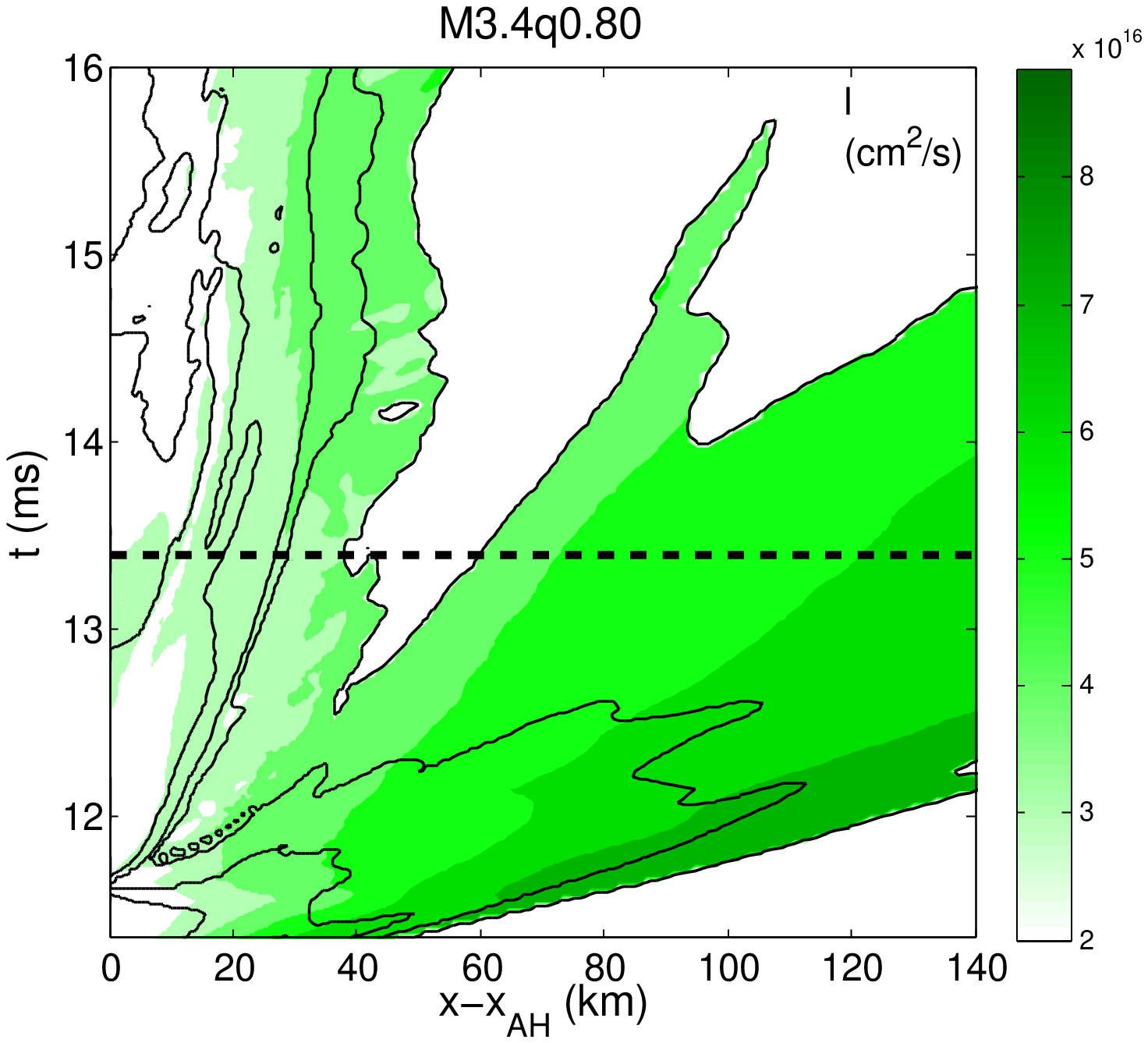}}
{\label{fig:spec-ang-mom-xt:5}\includegraphics[width=0.45\textwidth]{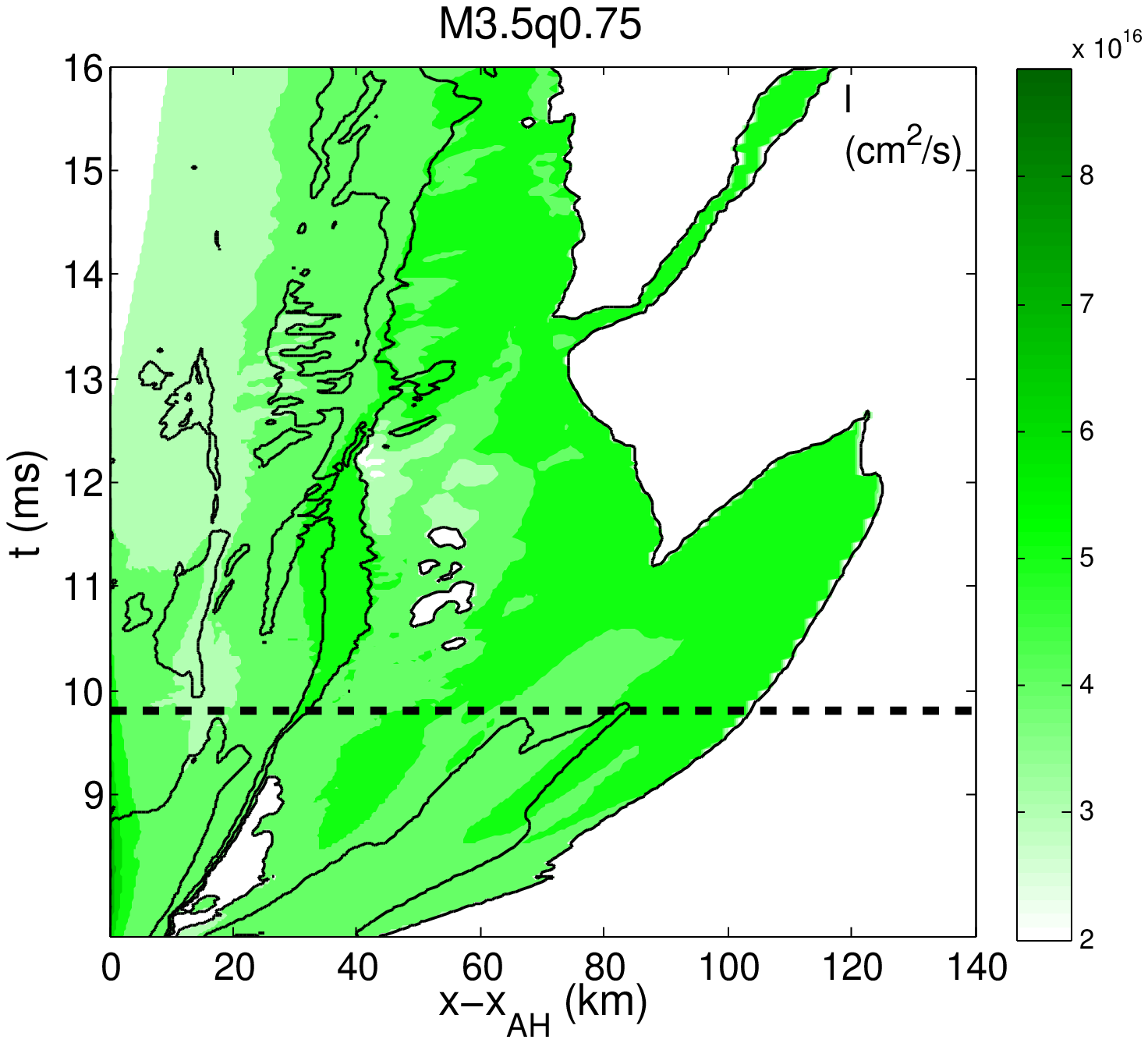}}
{\label{fig:spec-ang-mom-xt:6}\includegraphics[width=0.45\textwidth]{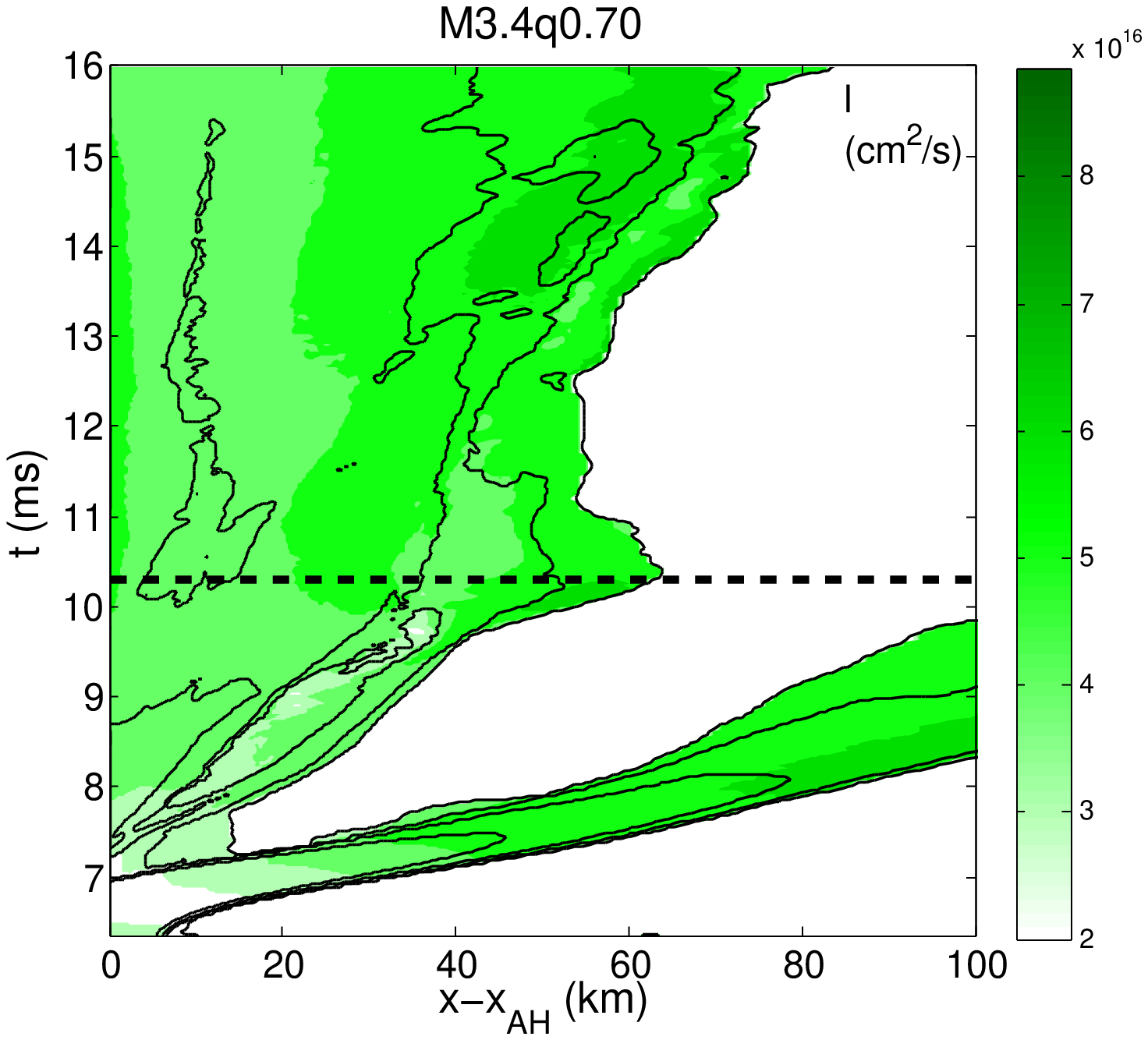}}
\end{center}
\caption{The same spacetime diagrams as in
  figure~\ref{fig:rest-mass-dens-xt} but for the evolution of the
  specific angular momentum $\ell=-u_{\phi}/u_t$. Note that the
  isocontours in this case refer to the rest-mass density and are the
  same as in figure~\ref{fig:rest-mass-dens-xt}.}
\label{fig:spec-ang-mom-xt}
\end{figure*}

\subsection{Specific Angular-Momentum Evolution}

Besides the rest-mass density, another quantity whose evolution is
useful to understand the dynamics of the tori is the the specific angular 
momentum. This quantity plays an important
role in defining the dynamics of point particles around black holes
and in defining the equilibrium of non-self gravitating tori around
black holes~\cite{Abramowicz78}. As mentioned above, we define the
specific angular momentum as $\ensuremath{\ell}\equiv -u_{\phi}/u_t$.
We also note that a similar but distinct definition of the specific angular
momentum was used in~\cite{Shibata06a}, namely $j=hu_{\phi}$. The two
definitions have the same Newtonian limit of $j_{\rm Newt}=\ell_{\rm
  Newt}=\Omega r^2$, $\Omega$ being the angular velocity. However, 
  it is important to stress that only the definition
used here yields the correct zero radial epicyclic frequency for tori
with constant specific angular momentum [see eqs.~(43) and (45)
  of~\cite{Rezzolla_qpo_03b}].

Figure \ref{fig:spec-ang-mom-xt} shows the evolution of the
specific angular momentum, where the different panels show the
color-coded specific angular momentum for an observer comoving with
the BH. The color code is indicated to the right of each plot and in
addition the same isodensity contours reported in
figure~\ref{fig:rest-mass-dens-xt} are shown here to aid to follow the
dynamics of the matter. The most striking feature to note when
scrolling through the different panels in
figure~\ref{fig:spec-ang-mom-xt} is that the radial distribution
changes radically but systematically when going from the equal-mass
binary \texttt{M3.6q1.00} over to the most extreme unequal-mass binary
considered \texttt{M3.4q0.70}. In particular, while the specific
angular momentum is decreasing outwards in models \texttt{M3.6q1.00},
\texttt{M3.7q0.94}, and \texttt{M3.4q0.91}, it is Keplerian and increasing
outwards as $\sim x^{1/2}$ for the remaining models (see also the
discussion in the following section). Furthermore, the spacetime plots
show that the matter located in the outer regions of the disks
acquires the largest values of the specific angular momentum. This is
particularly visible in the early evolution of model
\texttt{M3.4q0.80}, in which a large spiral arm develops extending
beyond the computational boundary, and also in the late evolution of
model \texttt{M3.4q0.70}, when the corresponding disk reaches the
largest radial extension (\cf~right panel of
figures~\ref{fig:density-tori_1}
and~\ref{fig:density-tori_2}). Broadly speaking, our simulations show
that, in agreement with the results of~\cite{Shibata06a}, the smaller
the value of $q$, the more the angular momentum is transported
outwards by a torque from the non-axisymmetric object that forms after
the merger.

\begin{figure*}
\begin{center}
\includegraphics[width=.65\textwidth]{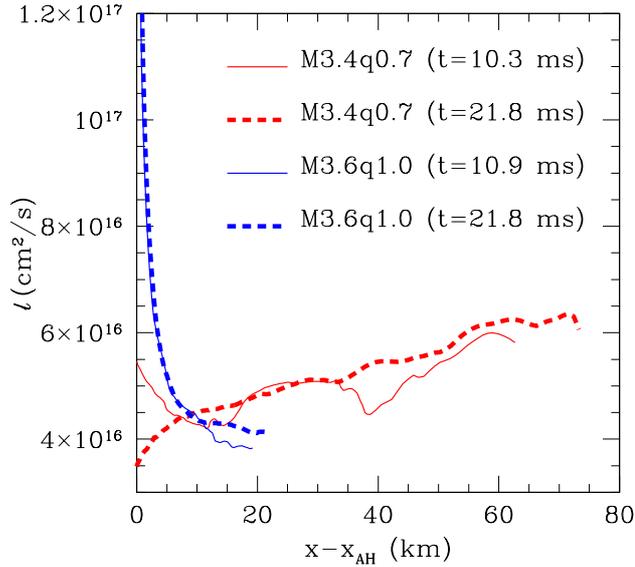}
\end{center}
\vskip -1.0cm
\caption{Profiles along the $x$-axis of the specific angular momentum
  of the tori produced by the binaries \texttt{M3.6q1.00} (blue lines
  extending to $\lesssim 20\,\km$) and \texttt{M3.4q0.70} (red lines
  extending up to $\gtrsim 70\,\km$). The profiles are computed in a
  frame comoving with the BH and for densities $\rho > 10^{10}\,{\rm
    g/cm}^3$. Different line types refer either to the onset of the
  QSA (\ie $t \sim 10\,\mss$, thin solid lines) or to the end of the
  simulation (\ie $t \sim 22\,\mss$, thick dashed lines). Note the
  markedly different behaviour and that the specific angular momentum
  for the unequal-mass case increases outwards.}
\label{fig:ell-profiles}
\end{figure*}

To better highlight the different behaviour of $\ell$ for different
mass ratios we show in figure~\ref{fig:ell-profiles} the profiles
along the $x$-axis for the tori produced by the binaries
\texttt{M3.6q1.00} (blue lines extending to $\lesssim 20\,\km$) and
\texttt{M3.4q0.70} (red lines extending up to $\gtrsim 70\,\km$). The
profiles are computed in a frame comoving with the BH and for
densities $\rho > 10^{10}\,{\rm g/cm}^3$. Different line types refer
either to the onset of the QSA (\ie $t \sim 10\,\mss$, thin solid
lines) or to the end of the simulation (\ie $t \sim 22\,\mss$, thick
dashed lines). Quite clearly, the specific angular momentum decreases
outward at all times for the equal-mass binary, while it increases
outward for the unequal-mass one (although it was initially decreasing
at the innermost parts). At this point it is worth remarking that
Rayleigh's criterion against axisymmetric perturbations of rotating
fluids requires that $d\ell/dx \geq 0$ for a dynamical
stability~\cite{Tassoul-1978:theory-of-rotating-stars}. While this
criterion is clearly satisfied by model \texttt{M3.4q0.70}, it is
equally-clearly violated by \texttt{M3.6q1.00}, which is nevertheless
stable. We believe this difference is due to the fact that Rayleigh's
criterion assumes that the motion is stationary and purely
azimuthal. While this is essentially the case for the unequal-mass
binary, which does not show clear evidence of epicyclic oscillations,
it does not hold true for the unequal-mass binary, which shows instead
large radial epicyclic oscillations. The nonlinear stability of
\texttt{M3.6q1.00}, but also of \texttt{M3.7q0.94} and
\texttt{M3.4q0.91}, seems therefore to indicate that Rayleigh's
criterion can and should be extended to account for fluids which are
subject to large radial excursions.

\begin{figure*}[ht] 
\begin{center}
{\label{fig:ang-vel-xt:1}\includegraphics[width=0.45\textwidth]{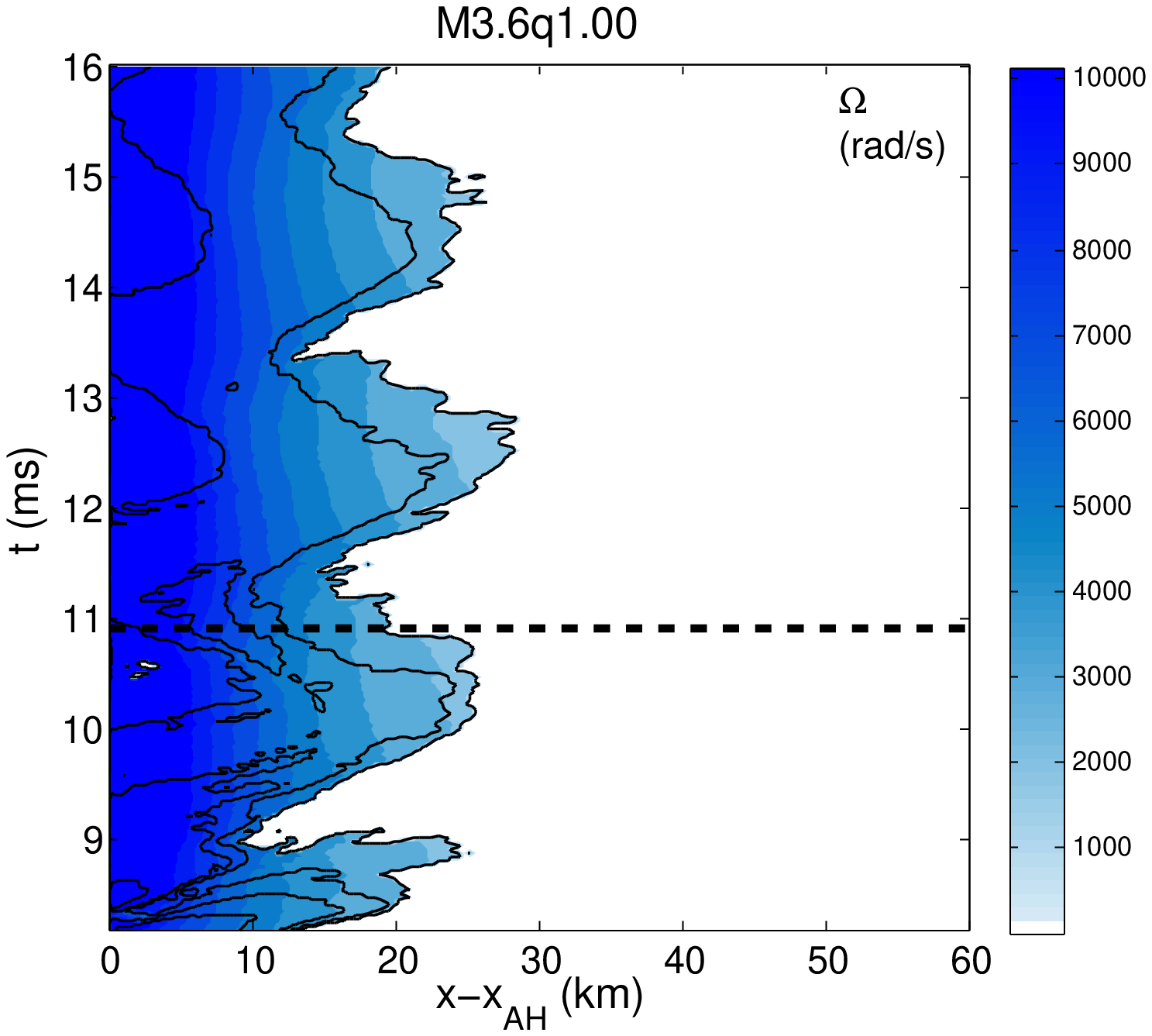}}
{\label{fig:ang-vel-xt:2}\includegraphics[width=0.45\textwidth]{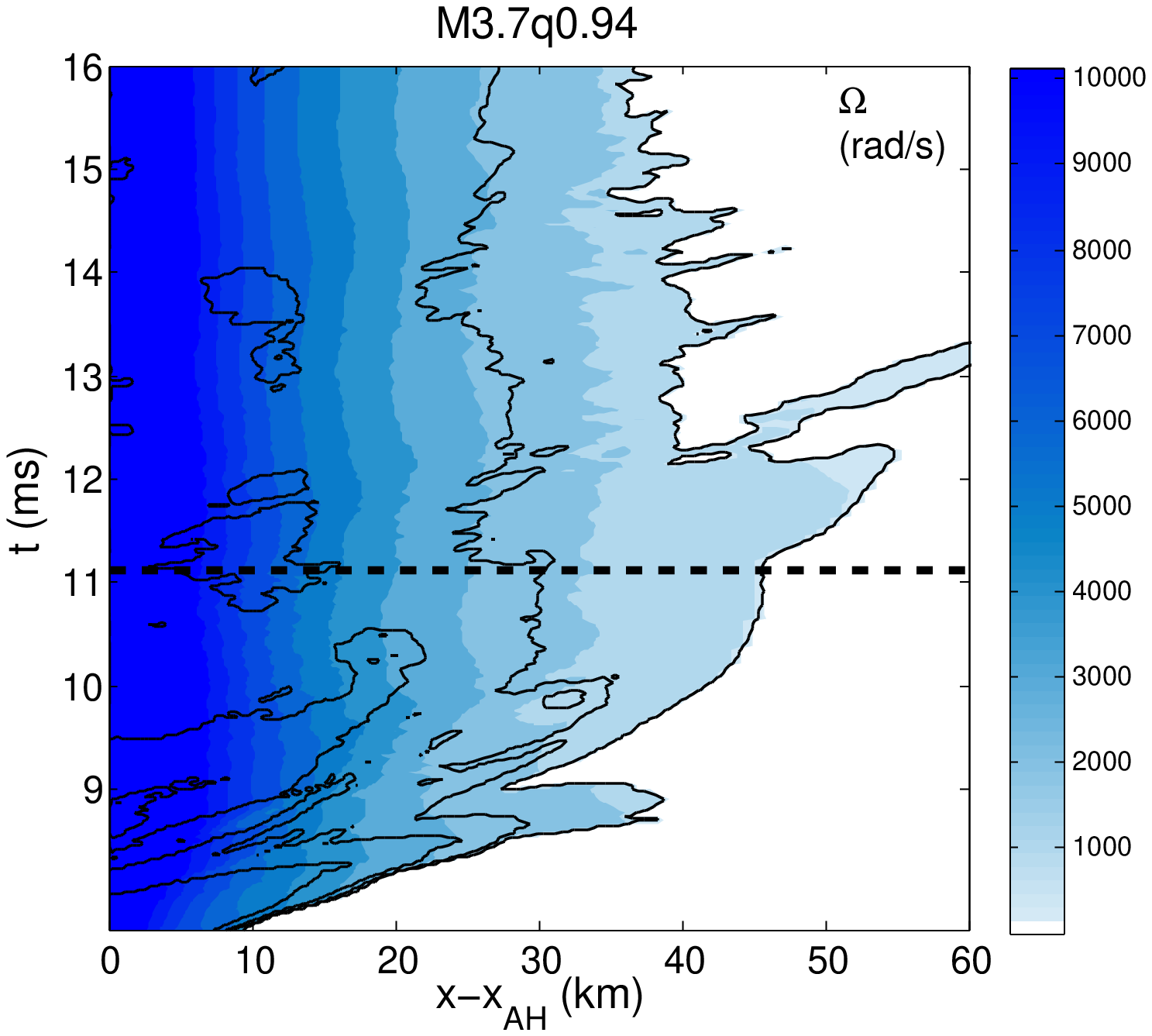}}
{\label{fig:ang-vel-xt:3}\includegraphics[width=0.45\textwidth]{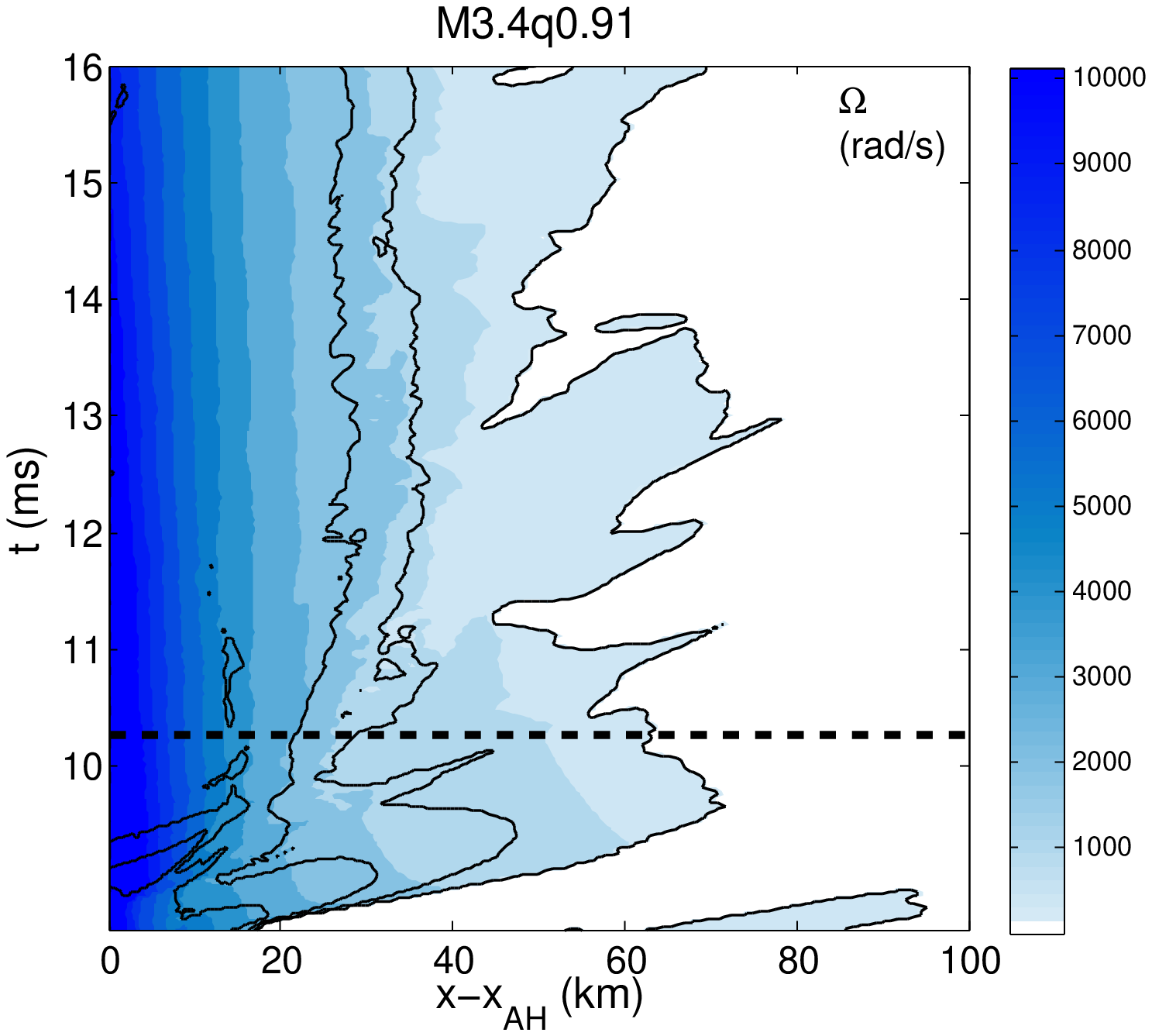}}
{\label{fig:ang-vel-xt:4}\includegraphics[width=0.45\textwidth]{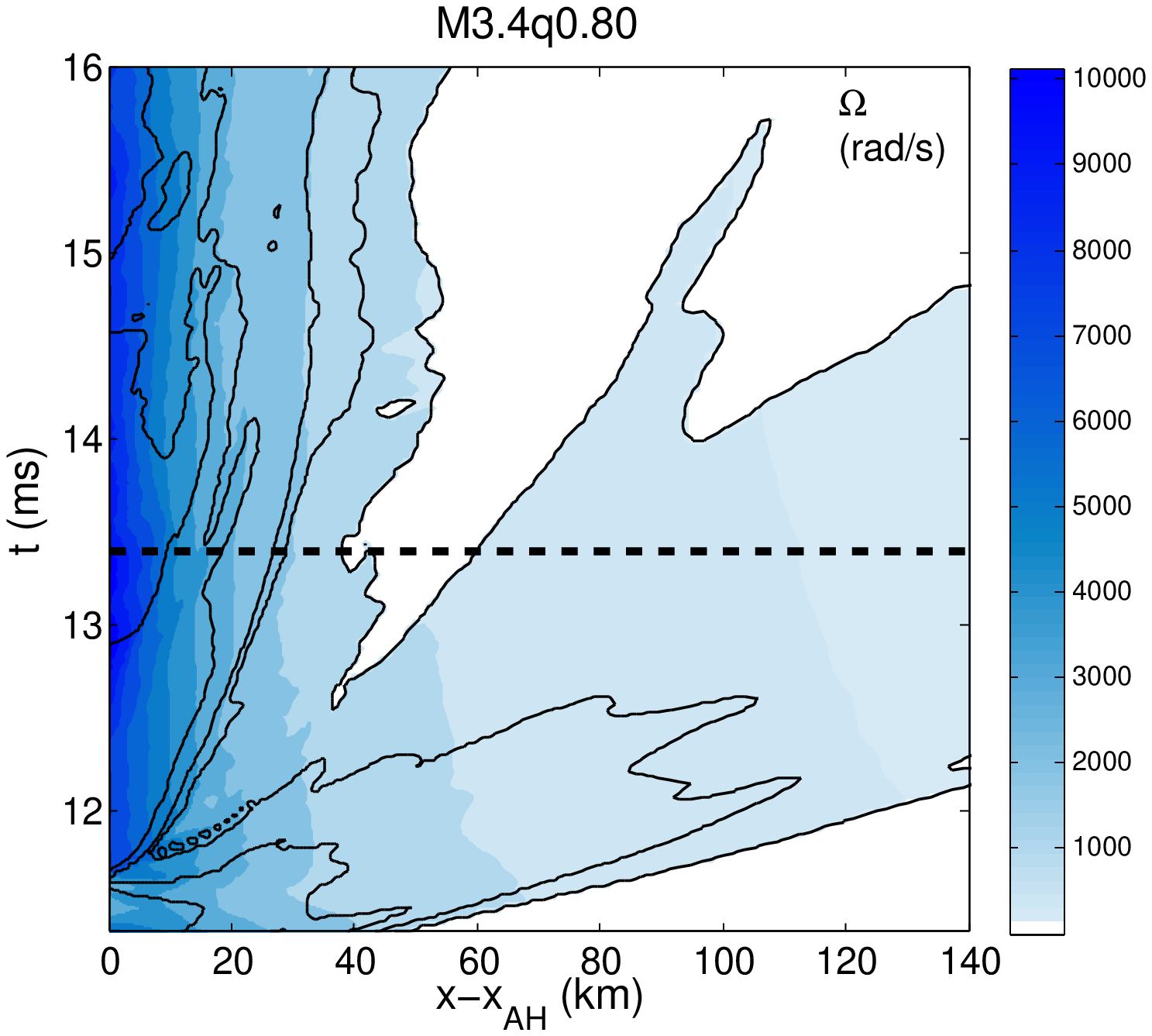}}
{\label{fig:ang-vel-xt:5}\includegraphics[width=0.45\textwidth]{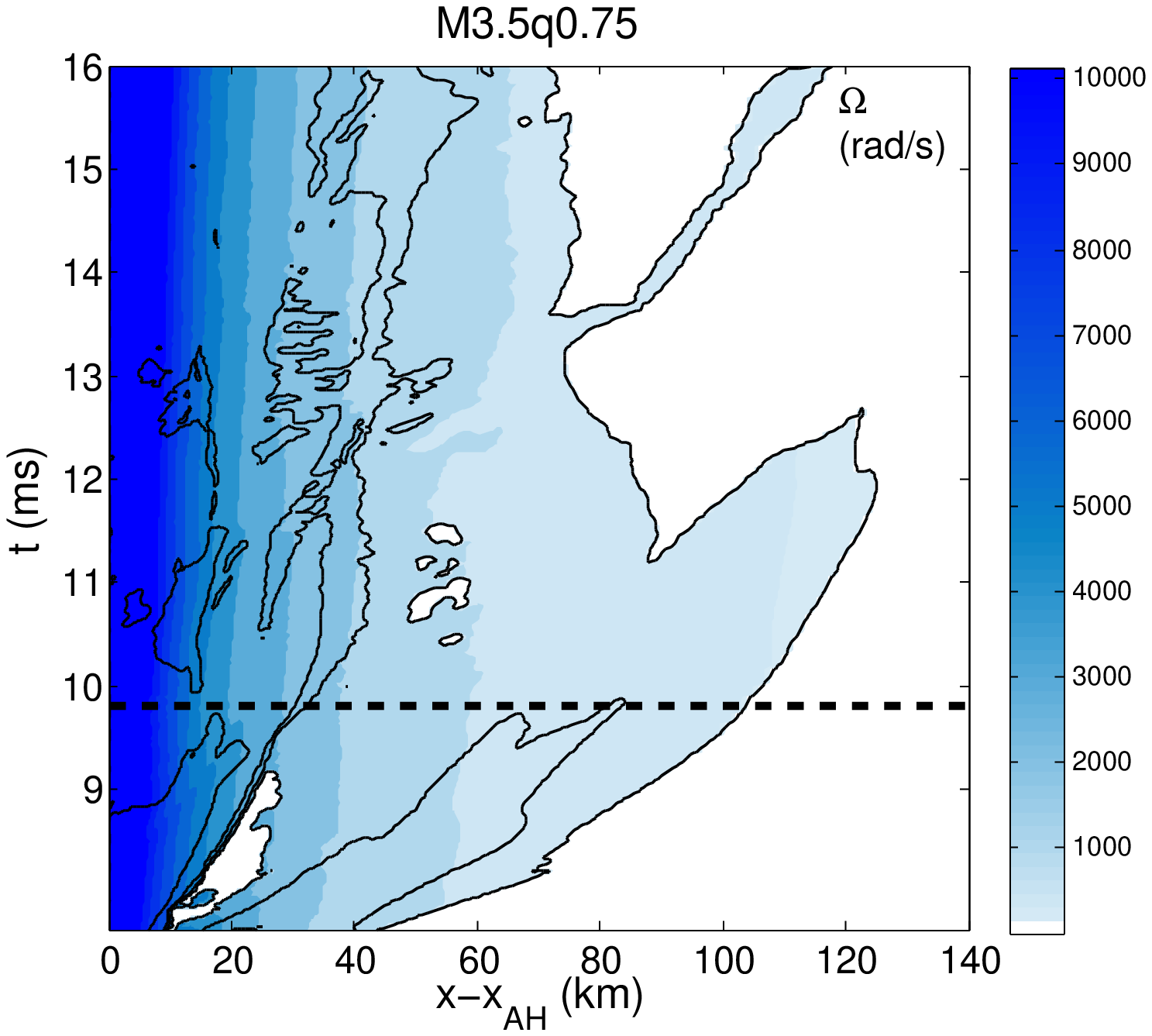}}
{\label{fig:ang-vel-xt:6}\includegraphics[width=0.45\textwidth]{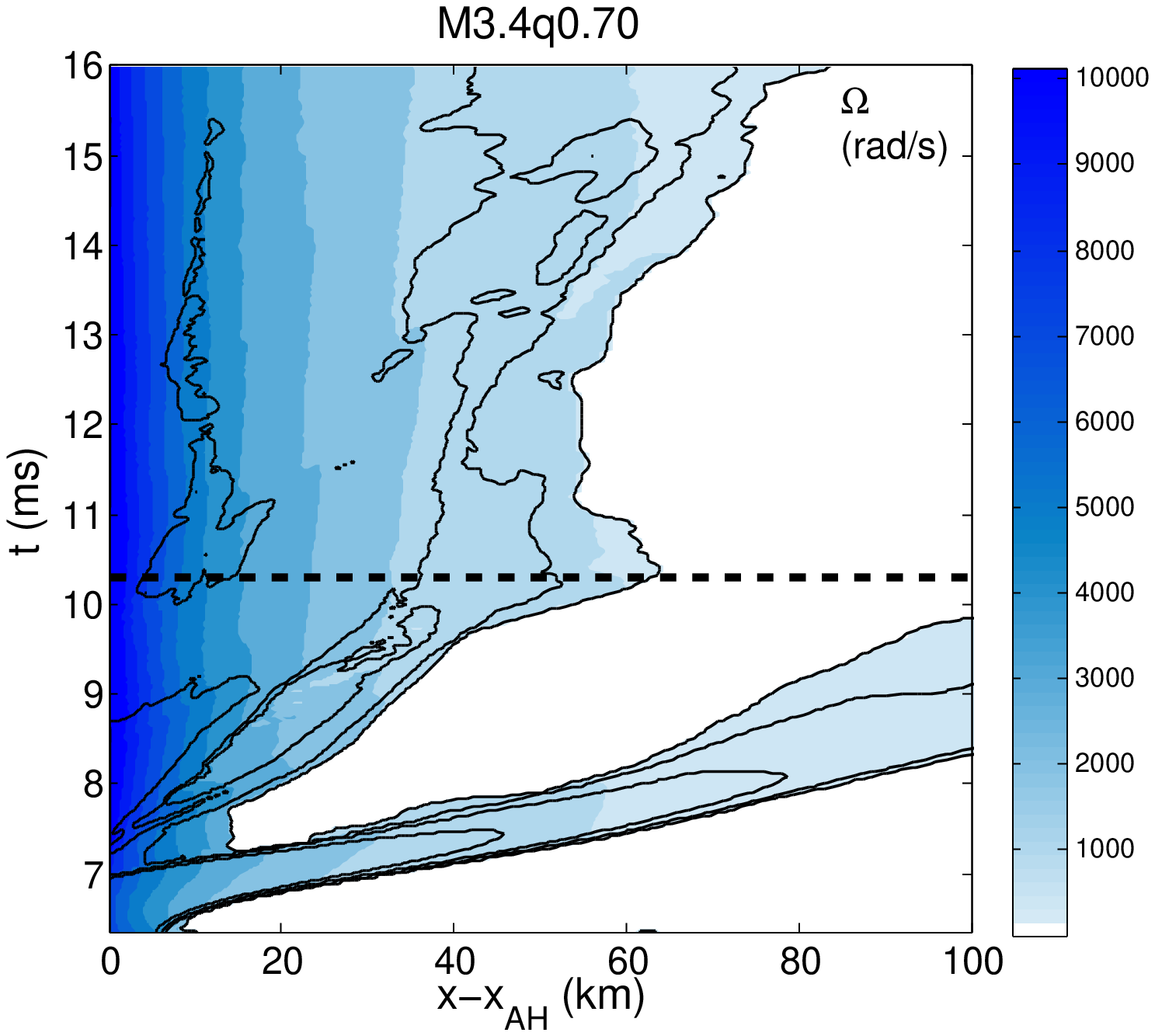}}
\end{center}
\caption{The same spacetime diagrams as in
  figure~\ref{fig:spec-ang-mom-xt} but for the evolution
  of the angular velocity $\Omega$. Note that the isocontours in this
  case refer to the rest-mass density and are the same as in
  figure~\ref{fig:rest-mass-dens-xt}.}
\label{fig:ang-vel-xt}
\end{figure*}

\begin{figure*} 
\begin{center}
\includegraphics[width=.65\textwidth]{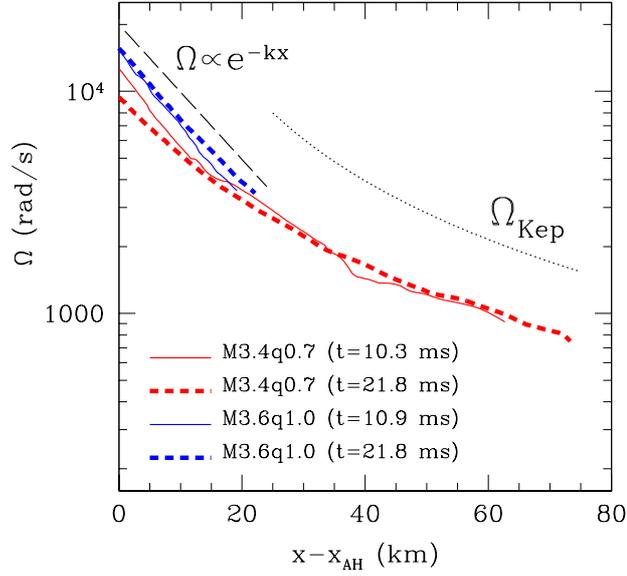}
\end{center}
\vskip -1.0cm
\caption{The same as in figure~\ref{fig:ell-profiles} but for the
  angular velocity. Shown as reference with a dotted line is the
  Keplerian angular velocity $\Omega_{\rm Kep}$, which matches very
  well the outer parts of the torus from the unequal-mass
  binary. Shown instead with a long-dashed line is an exponentially
  decaying profile, which instead reproduces well the profile for the
  equal-mass binary.}
\label{fig:omg-profiles}
\end{figure*}

\subsection{Angular-velocity Evolution}

In analogy with figures~\ref{fig:rest-mass-dens-xt} and
~\ref{fig:spec-ang-mom-xt}, figure~\ref{fig:ang-vel-xt} shows the
spacetime diagram for the evolution of the angular velocity
$\Omega\equiv u^{\phi}/u^t$ for all models of our sample. It is
straightforward to notice that for all models the angular velocity
decreases with the radial distance from the apparent horizon. While
this is qualitatively in agreement with the results
of~\cite{Shibata06a}, it is worth noting that the radial fall-off is
very different as the mass ratio is varied among the different
binaries. This is shown in figure~\ref{fig:omg-profiles}, which reports
the profiles of $\Omega$ along the $x$-axis for the tori produced by
the binaries \texttt{M3.6q1.00} (blue lines extending to $\lesssim
20\,\km$) and \texttt{M3.4q0.70} (red lines extending up to $\gtrsim
70\,\km$). As before, the profiles are computed in a frame comoving
with the BH and for densities $\rho > 10^{10}\,{\rm g/cm}^3$ and
different line types refer either to the onset of the QSA (\ie $t \sim
10\,\mss$, thin solid lines) or to the end of the simulation (\ie $t
\sim 22\,\mss$, thick dashed lines). It is then clear that while the
equal-mass binary has an exponentially decaying profile (\cf
long-dashed line), \ie $\Omega \propto \exp[-k(x-x_{_{\rm AH}})] \sim
\exp[-0.07(x-x_{_{\rm AH}})]$, which does not change significantly
with time, the unequal-mass binary reaches at the end of the
simulation a profile which is, especially in the outer parts,
essentially Keplerian, \ie with $\Omega_{\rm Kep} \sim x^{-3/2}$ (\cf
dotted line). This feature, which is also shared by the other low-$q$
binaries, explains the scaling of the specific angular momentum as
$\ell \sim x^{1/2}$ and provides firm evidence that the tori produced
in this case will be dynamically stable.

\subsection{Matter Ejection}

\begin{figure*}[ht] 
\begin{center}
{\label{fig:ut-xt:1} \includegraphics[width=0.45\textwidth]{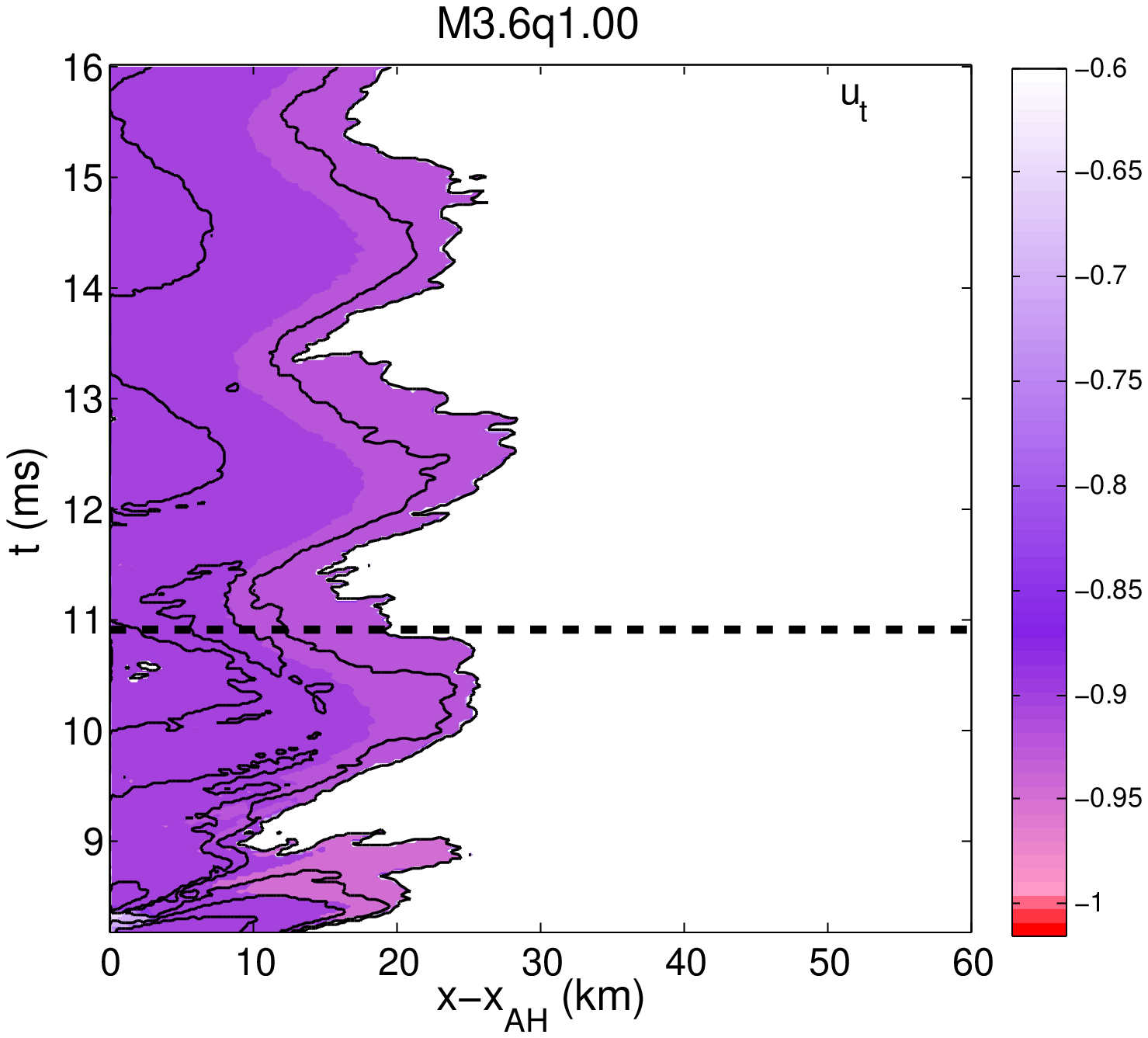}}
{\label{fig:ut-xt:2} \includegraphics[width=0.45\textwidth]{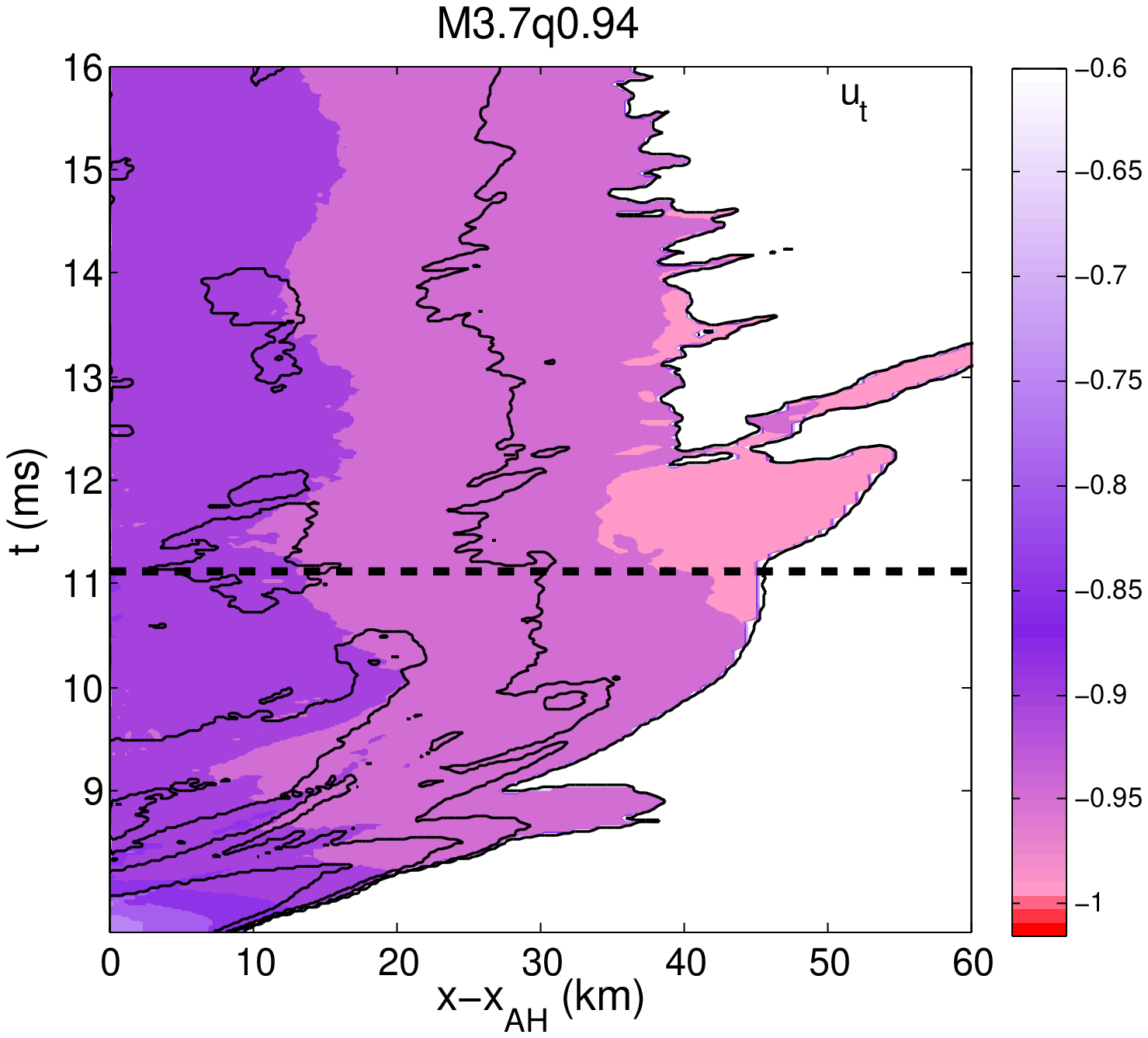}}
{\label{fig:ut-xt:3} \includegraphics[width=0.45\textwidth]{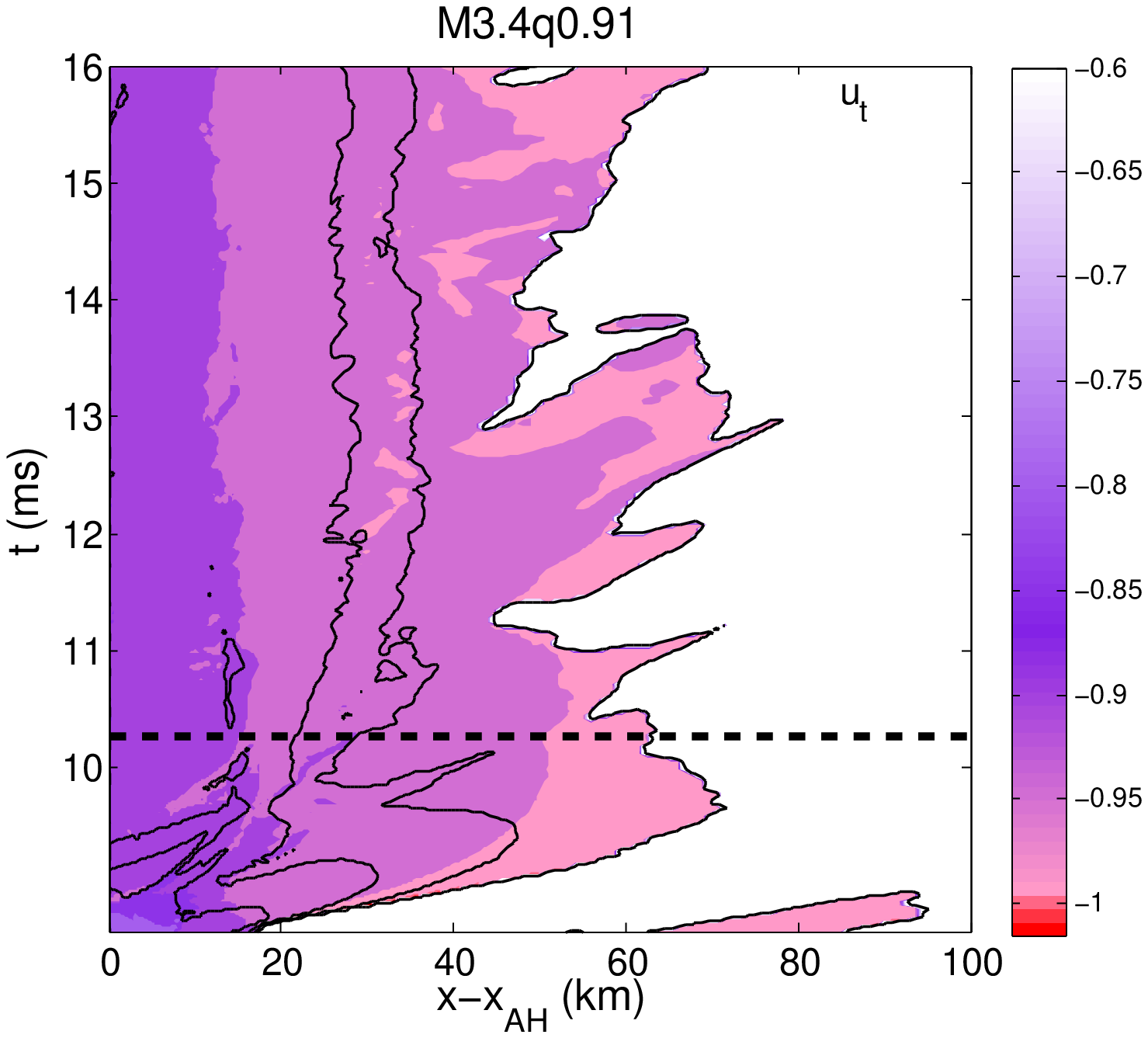}}
{\label{fig:ut-xt:4} \includegraphics[width=0.45\textwidth]{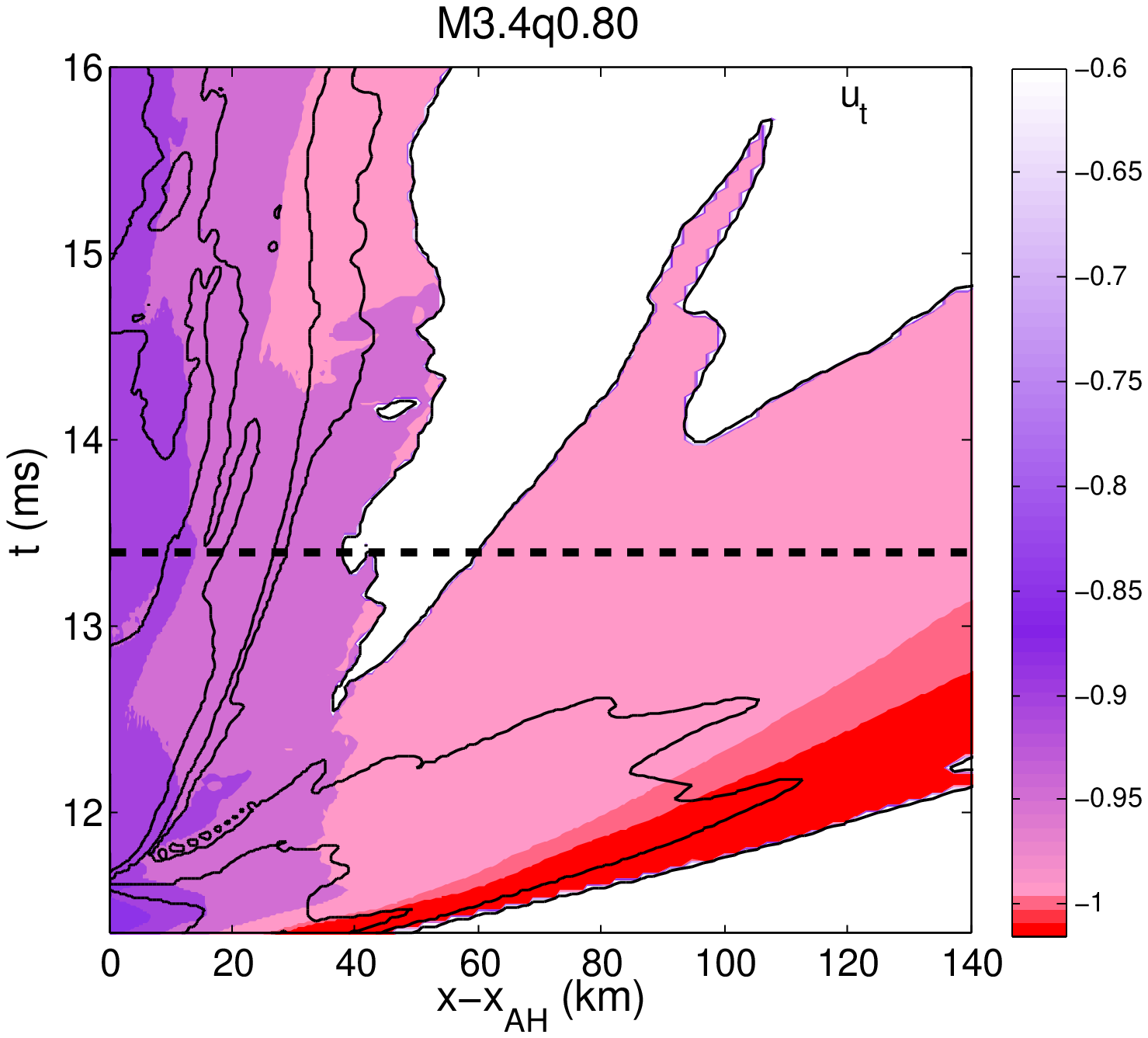}}
{\label{fig:ut-xt:5} \includegraphics[width=0.45\textwidth]{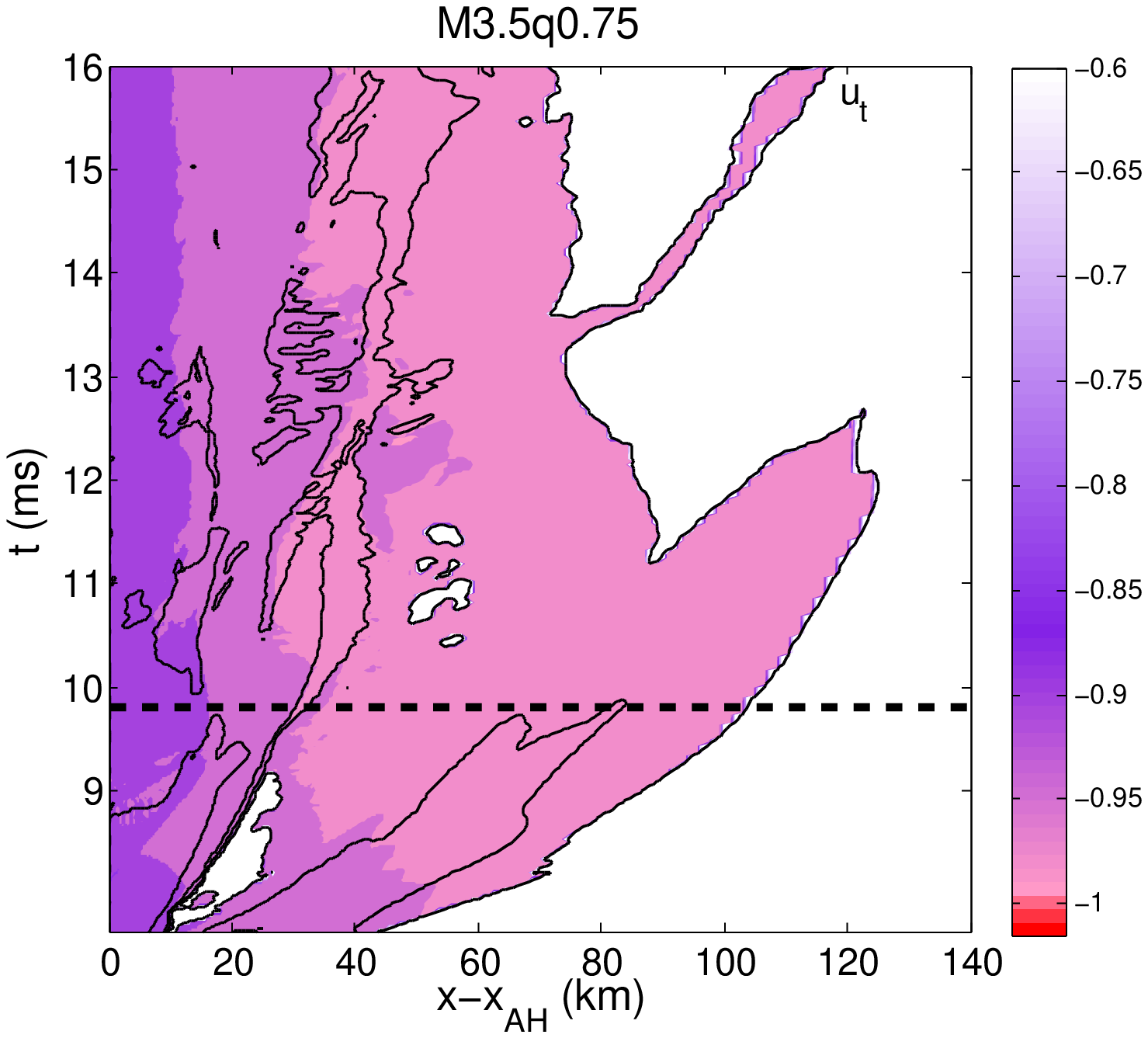}}
{\label{fig:ut-xt:6} \includegraphics[width=0.45\textwidth]{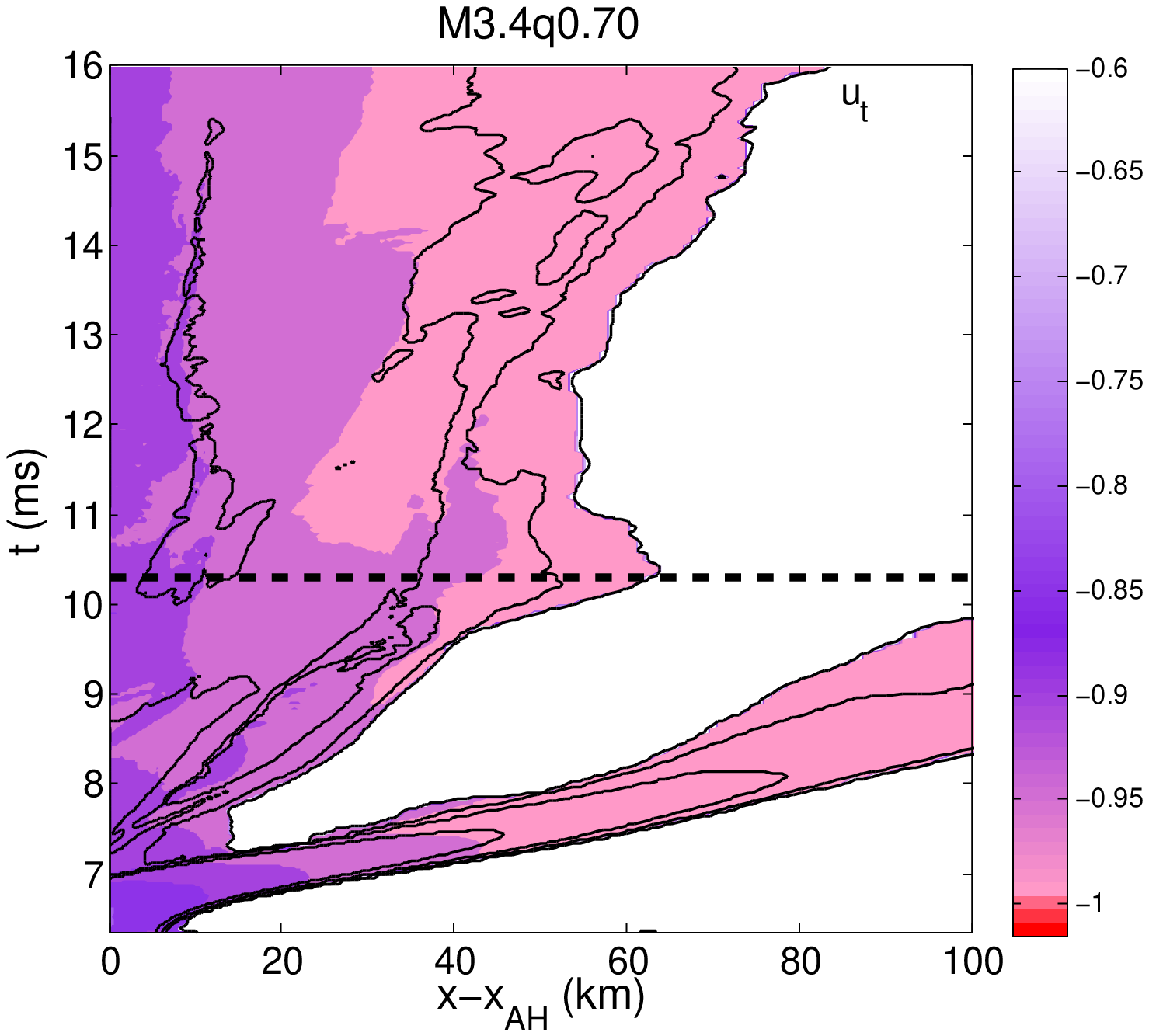}}
\end{center}
\caption{The same spacetime diagrams as in
  figure~\ref{fig:spec-ang-mom-xt} but for the evolution of local
  fluid energy \ensuremath{u_t}. Note that the isocontours in this
  case refer to the rest-mass density and are the same as in
  figure~\ref{fig:rest-mass-dens-xt}.}
\label{fig:ut-xt}
\end{figure*}

As a final but nevertheless important aspect of the formation and
evolution of the tori, we consider whether or not a part of the
rest-mass of the system is ejected during the merger and the
subsequent evolution.  To determine whether a fluid particle is bound
or unbound we use the covariant time component of the 4-velocity
\ensuremath{u_t} and recall that, in an axisymmetric and stationary
spacetime, the value of \ensuremath{u_t} for a particle moving along a
geodesic is conserved. If the particle is unbound, it moves outwards
and \ensuremath{-u_t = W > 1} at infinity, where \ensuremath{\alpha
  \equiv 1, \beta_i \equiv 0}. The local condition \ensuremath{u_t >
  -1} thus provides a necessary although not sufficient, condition for
a fluid element to be bound; stated differently, if a particle reaches
infinity it is because it has $u_t < - 1$. Furthermore, this condition
is exact only in an axisymmetric and stationary spacetime, and our
spacetimes attain these properties only in the final stages of the
evolution. Nevertheless, this is a useful condition for a first
estimate of the amount of matter ejected and a in-depth discussion on
the assumptions implicit in this criterion and on how it applies if
one accounts for external forces are presented
in~\cite{Ansorg:1998}. (Note that the alternative criterion for bound
flows, namely $hu_t>-1$, would yield similar results since in the
relevant regions $h\sim 1$.)

Figure~\ref{fig:ut-xt} shows the evolution of \ensuremath{u_t}
embedded in a spacetime diagram much like the ones presented before
for the rest-mass density, the specific angular momentum and the
angular velocity. For all models under consideration, the criterion
\ensuremath{u_t > -1} is well fulfilled, namely all the matter in the
tori is bounded, except for model \texttt{M3.4q0.80} which clearly
shows in the early stages of its evolution, that a certain amount of
unbound matter is ejected before reaching the regime of QSA. Only for
the outermost, very low-density regions of the tori (which are not
shown in the spacetime diagrams of figure~\ref{fig:ut-xt}) values of
\ensuremath{u_t \leq -1} are encountered in the other models and are
probably the manifestation of an outflowing wind caused by the very
large temperatures of those regions. As a final remark we note that
although the total amount of matter ejected in this way is rather
small and only of the order of $\sim 10^{-4}\,\Msun$, it can
nevertheless act as the site for the production of the neutron-rich
heavy elements that are formed by rapid neutron capture (\ie the
r-process) (see~\cite{Freiburghaus:1999} and references
therein). Performing such calculations and thus determining to what
extent binary NS mergers contribute to the whole observed r-process
material in the Galaxy requires a fully developed reaction network and
is outside the scope of this study, but will be the focus of our
future research.

\subsection{A phenomenological expression for the mass in the torus}
\label{sec:TorusMass}

As mentioned above, determining the amount of rest-mass in the torus
may be one of the most important aspects of this research for the
impact it has on the modelling of the emission in SGRBs.
Table~\ref{tab:FinalProducts} also reports the mass of the torus and
since the latter slightly decreases in time, we have arbitrarily
chosen the time of $t~\sim 17$ ms as a reference (it is about the
latest time for which we have data from all the simulations). Because
of the importance of the information and because of the scarcity of
the numerical data available, it would be valuable to derive a
phenomenological expression for the mass in the torus which can be
constructed on basic expectations and that can be constrained by using
the numerical data.

Following this spirit, we first search for a phenomenological
expression for the torus mass which will depend only on the mass ratio
and on the total mass of the binary, \ie $\widetilde{M}_{\rm tor}=
\widetilde{M}_{\rm tor}(q,M_{\mathrm{tot}})$. Next, we exclude the
trivial case in which the total mass is larger than the maximum mass
of the binary system $M_{\mathrm{max}}$ (based on the maximum allowed
mass for isolated stars with the given EOS); in practice we impose
that $\widetilde{M}_{\rm tor} (q, M_{\mathrm{tot}} \geq
M_{\mathrm{max}})= 0$ for any value of $q$. Finally we impose the
expectation that the mass of the torus should depend, at least at
lowest order, on the mass ratio (this was already noted
by~\cite{Shibata06a}) and yield the torus with the smallest possible mass
for an equal-mass binary. Collecting all these constraints,
our~\textit{ansatz} is
\begin{eqnarray} 
\label{eqn:empirical-relation}
\widetilde{M}_{\rm tor}(q,M_{\mathrm{tot}})
        &=&
        c_1 (1-q) (M_{\mathrm{max}}-M_{\mathrm{tot}}) +
        c_2(M_{\mathrm{max}}-M_{\mathrm{tot}}) \nonumber\\
        &=&
        \left[c_3(1+q)M_{*}-M_{\mathrm{tot}}\right] \left[c_1 (1-q) + c_2 \right]\,,
\end{eqnarray}
where in the second expression we have written the maximum mass of the
binary in terms of the maximum mass $M_*$ of an isolated nonrotating
star, \ie $M_{\mathrm{max}} = c_3(1+q)M_*$.

Note that as introduced in expression (\ref{eqn:empirical-relation}),
the coefficients $c_2$ and $c_3$ have a direct physical interpretation:
$c_2$ is proportional to the mass of the torus for equal-mass binaries, while
$c_3$ parametrises the excess of maximum mass that can be supported in the
binary because of the stabilizing effect produced by the nonzero spin
of the stars and of the tidal potential (\ie $c_3$ is expected to be
slightly larger than $1$). The three coefficients, $c_1, c_2$, and
$c_3$ can then be computed by comparing
expression~(\ref{eqn:empirical-relation}) with the numerical data
reported in table~\ref{tab:FinalProducts} as well as the one computed
in~\cite{Baiotti08} for equal-mass binaries. The fitting procedure
then yields $c_1 = 1.115 \pm 1.090, c_2 = 0.039 \pm 0.023, c_3 = 1.139
\pm 0.149$, with a reduced $\chi^2\simeq 2\times 10^{-3}$.
 
\begin{figure*}[ht]
\centering
\includegraphics[width=0.55\textwidth,angle=-90]{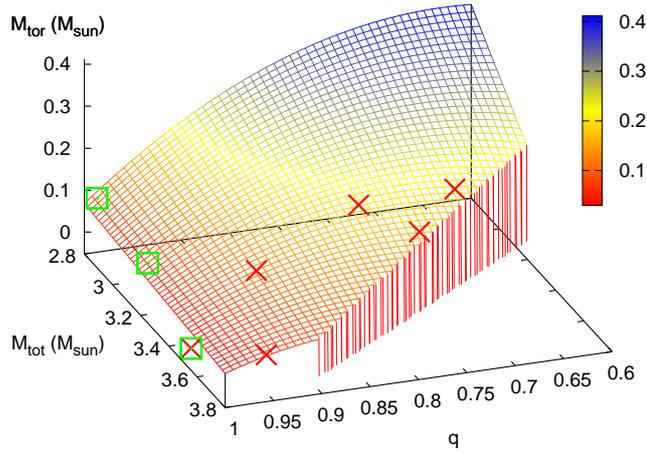}
\caption{Different symbols show the torus mass \ensuremath{M_{\rm
      tor}} measured either in the simulations reported here (red
  crosses) or in those reported in~\cite{Baiotti08} (green
  squares). Also shown in the parameter space
  \ensuremath{(q,M_{\mathrm{tot}})} considered here is the
  phenomenological modelling ${\widetilde M}_{\rm tor}$ suggested by
  expression~(\ref{eqn:empirical-relation}). Note that to highlight
  the functional behaviour of the phenomenological expression, the
  $x-$ and $y-$axes are shown as decreasing when moving to the right
  and to the left, respectively.}
\label{fig:parameter-space}
\end{figure*}

Figure \ref{fig:parameter-space} shows the torus mass
\ensuremath{M_{\rm tor}} either as measured in the simulations
reported here (red crosses) or in those presented in~\cite{Baiotti08}
(green squares) and against the phenomenological modelling
${\widetilde M}_{\rm tor}$ suggested by
expression~(\ref{eqn:empirical-relation}) in the region where
${M}_{\rm tot} \leq M_{\mathrm{max}}$. Note that to highlight the
functional behaviour of the phenomenological expression, the $x-$ and
$y-$axes are shown as decreasing when moving to the right and to the
left, respectively. Overall, the figure shows rather generically that:
1) The mass of the torus increases with the asymmetry in the mass
ratio; 2) Such an increase is not monotonic and for
sufficiently small mass ratios the tidal disruption leads to tori that
have a smaller mass for binaries with the same total mass; 3)
Tori with masses $\lesssim 0.21\,M_{\odot}$ have been measured and
even more massive ones, \ie with masses up to $\sim 0.35\,M_{\odot}$,
are possible for mass ratios $q\sim 0.75-0.85$. 

We note that somewhat similar considerations about the mass of the
torus were made also in~\cite{Shibata06a}, where a different
phenomenological expression for the mass of the torus was
proposed. When applied to the data computed here, the expression
suggested in~\cite{Shibata06a} does not reproduce well the data and
yields rather large errors. There are a number of reasons that could
justify these differences and that are related to the different
initial data chosen (ref.~\cite{Shibata06a} has only two initial total
masses which are smaller than those considered here), to the different
EOSs employed (ref.~\cite{Shibata06a} uses cold but realistic EOSs in
contrast to the ideal-fluid chosen here), and to the different
numerical techniques adopted (ref.~\cite{Shibata06a} uses a uniform
grid with rather coarse resolution in place of the mesh-refined
grid employed here). All these differences make the comparison between
the two calculations rather difficult, although they also motivate a
closer comparison using at least the same initial data and the same
EOSs, and which will be the subject of our future work. However,
common conclusions of both calculations are that: The mass of the
torus can be as large as $\sim 0.1\,M_{\odot}$ and larger; It
increases with the mass asymmetry in the binary; It is larger for
systems with smaller total mass. We believe these features are robust
and will be also present when different initial data and EOSs are
considered.

A final note of caution must be mentioned: Although
figure~\ref{fig:parameter-space} indicates a very good match between
the data and expression~(\ref{eqn:empirical-relation}), it also shows
that the latter is inaccurate for $q\simeq 1$, where the tori masses
are much smaller and the prediction leads to small but negative
values; luckily, the regime where~(\ref{eqn:empirical-relation}) is
less accurate is also the least interesting one from an astrophysical
point of view. Most importantly, however, it is clear that the attempt
to produce a phenomenological description for the mass of the torus
after having investigated only a small portion of the space of the
parameters (especially with respect to the total mass of the binary)
and after using as support only $8$ simulations is a very demanding
task and potentially a flawed one. However, because we believe that
expression (\ref{eqn:empirical-relation}) is a reasonable description
of the expected results, we foresee that it will reveal its robustness
as additional simulations are performed and the coefficients will be
further improved. This will indeed be the subject of our future work.

\section{Gravitational-Wave Emission}
\label{sec:GravitationalWaves}

\begin{figure*}[ht] 
\begin{center}
{\label{fig:q22-1:1} \includegraphics[width=.425\textwidth]{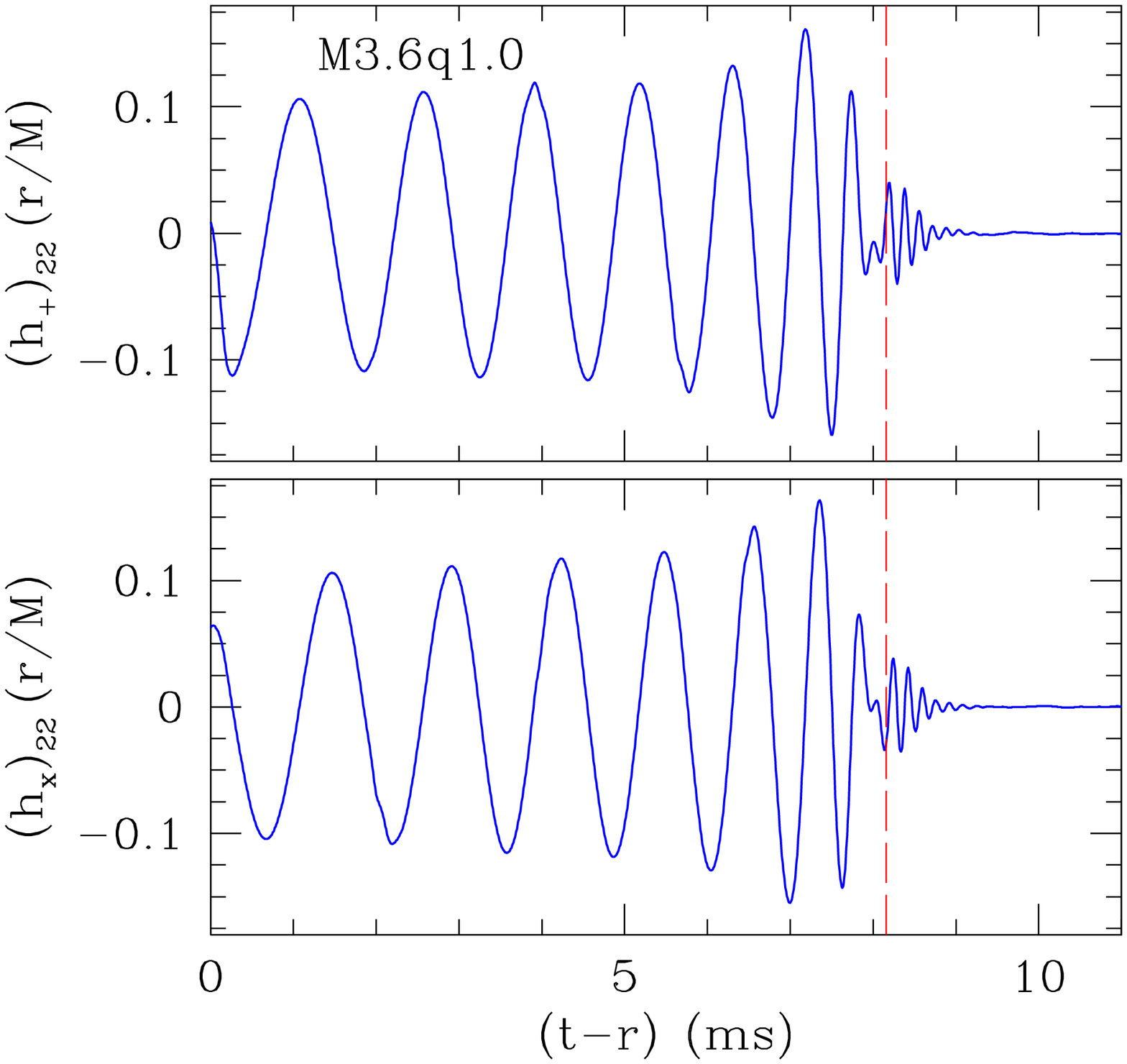}}
{\label{fig:q22-1:4} \includegraphics[width=.425\textwidth]{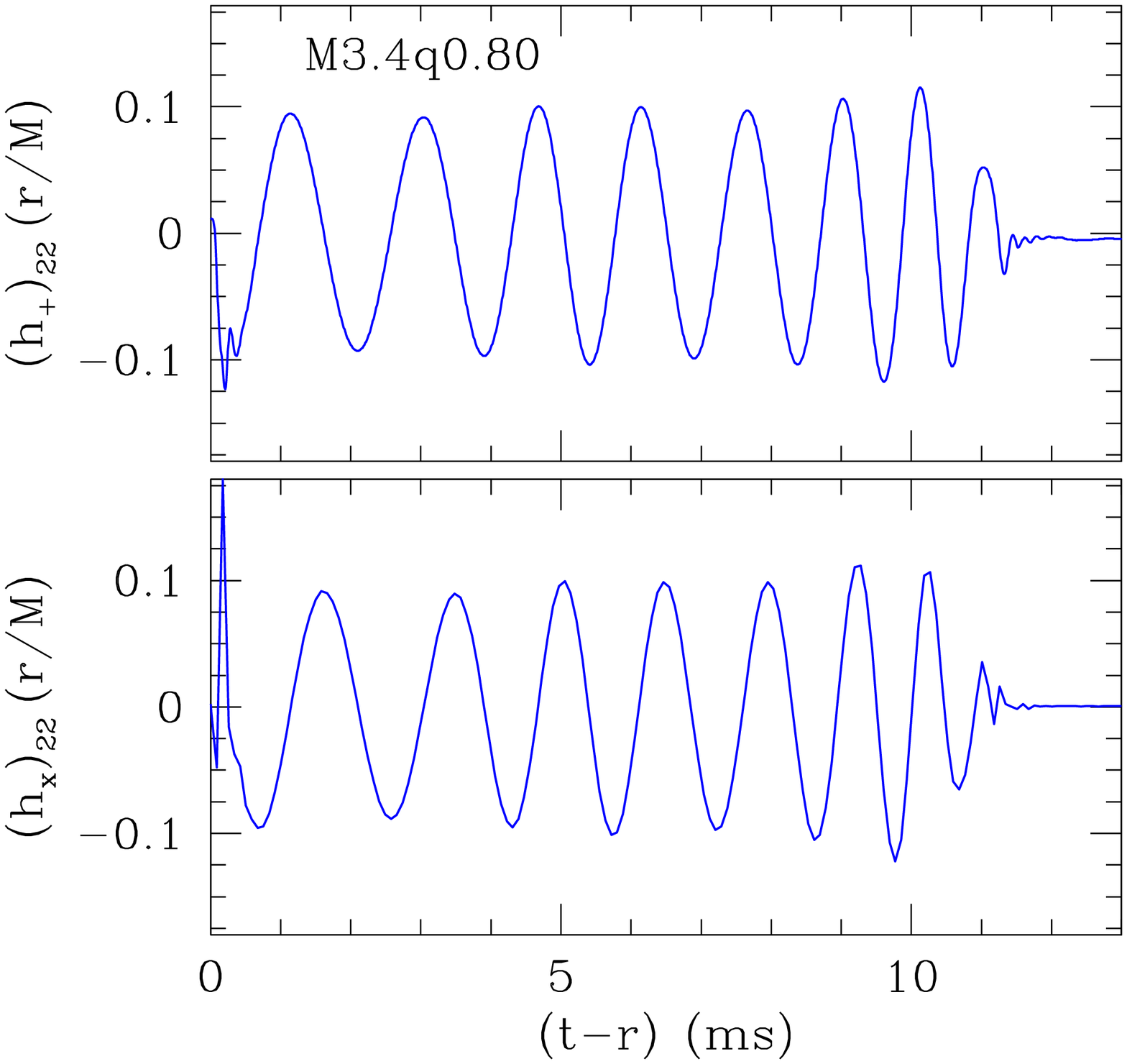}}
\vspace{-0.5 cm} 
{\label{fig:q22-1:2} \includegraphics[width=.425\textwidth]{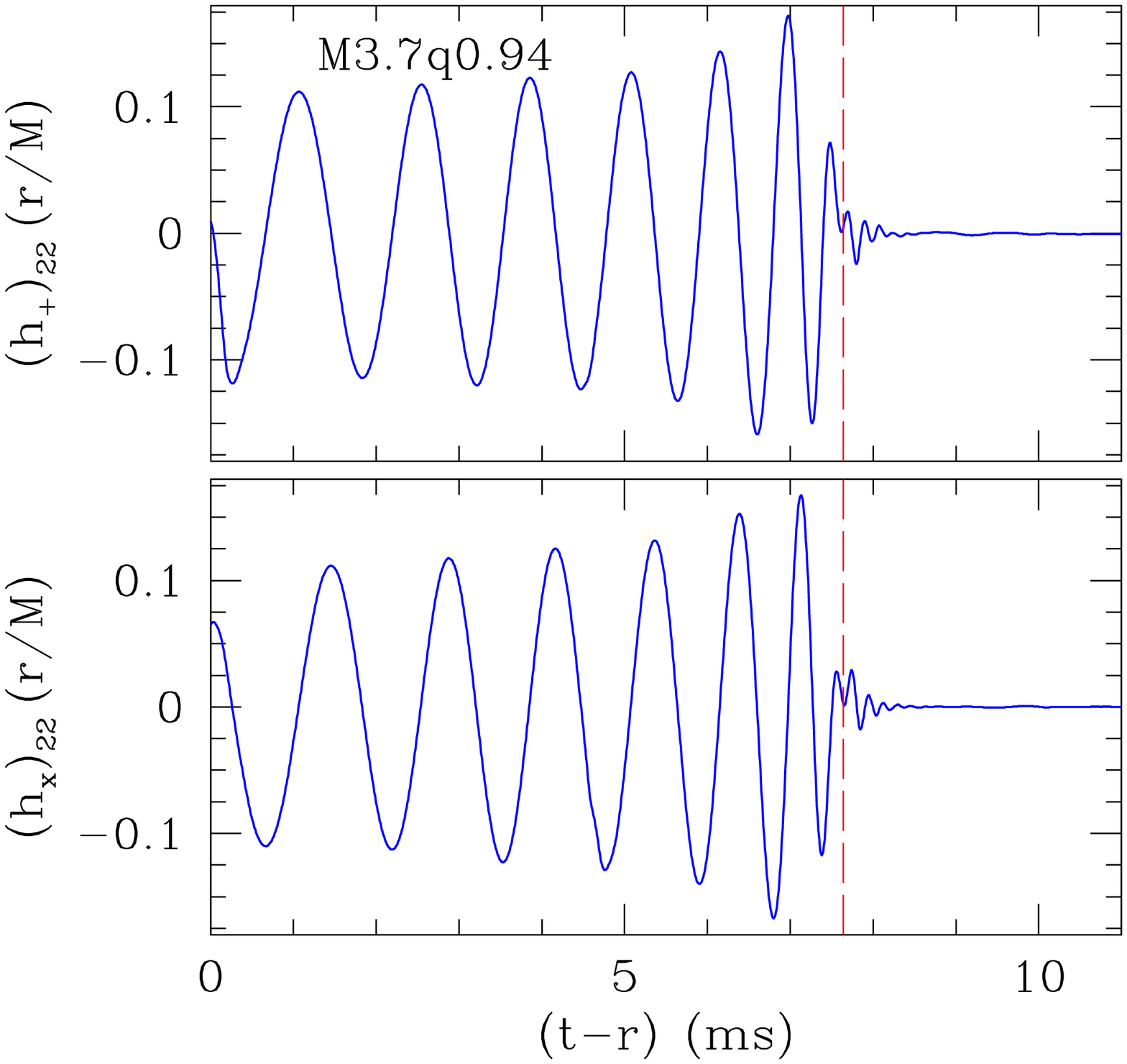}}
{\label{fig:q22-1:5} \includegraphics[width=.425\textwidth]{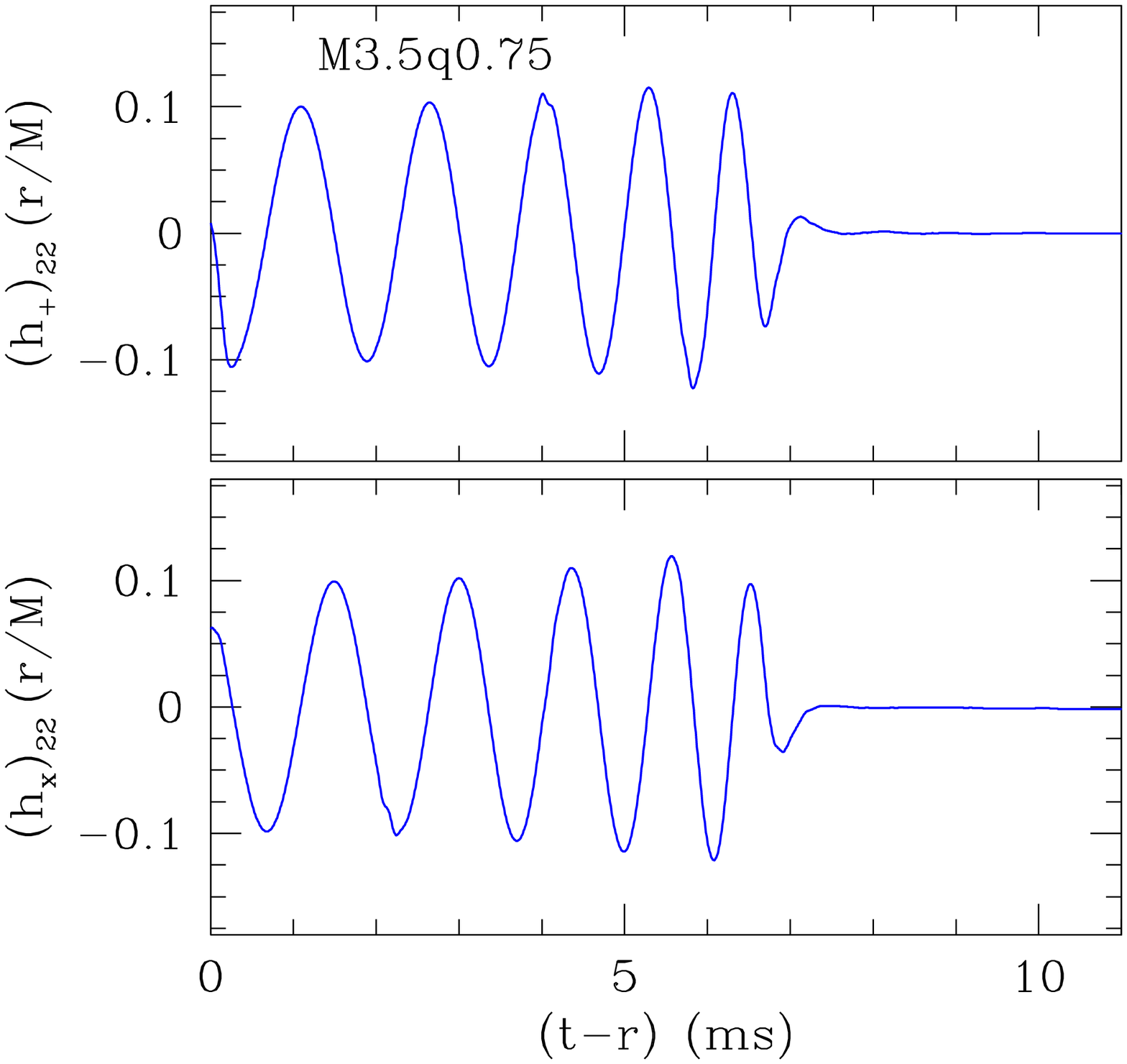}}
\vspace{-0.5 cm} 
{\label{fig:q22-1:3} \includegraphics[width=.425\textwidth]{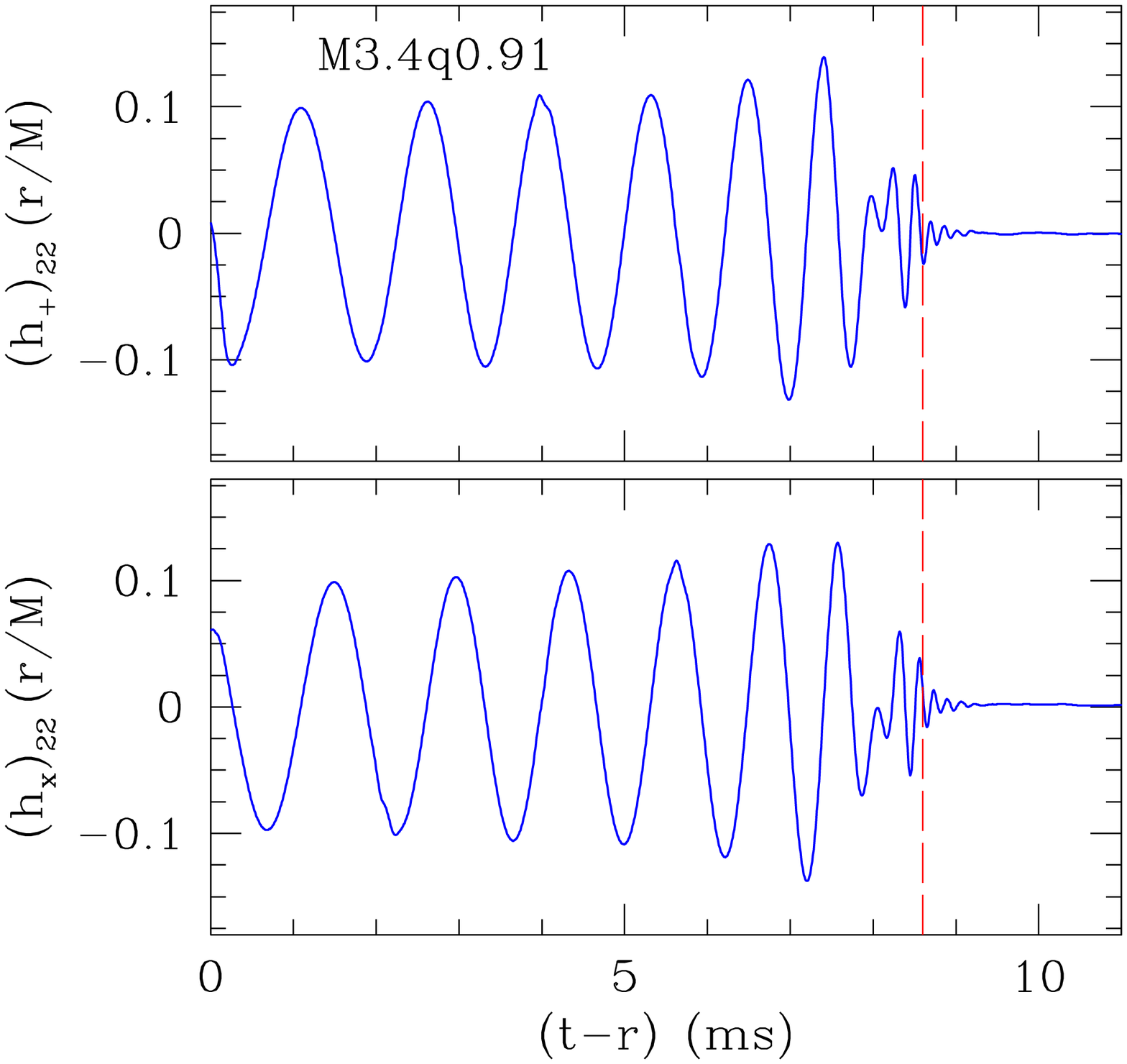}}
{\label{fig:q22-1:6} \includegraphics[width=.425\textwidth]{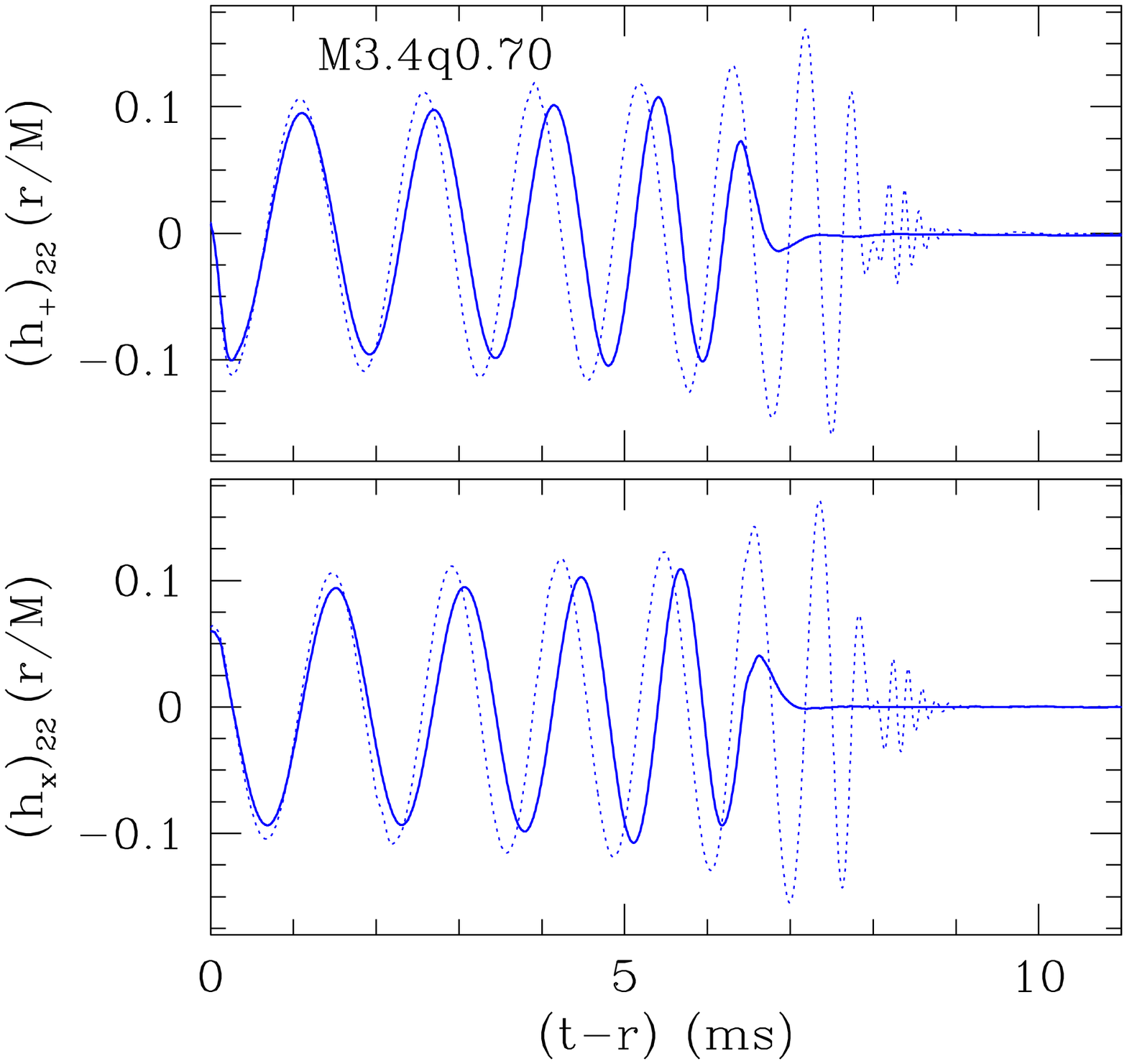}}
\end{center}
\vspace{-0.5 cm} 
\caption{Gravitational waveforms in the two polarizations ${h_+}$
  (upper panels) and ${h_\times}$ (lower panels) as computed from the
  lowest $\ell=m=2$ multipole for all the binaries considered. For
  those models where it was found, the vertical dashed lines mark the
  time of the first detection of the apparent horizon. Note that as
  the mass ratio $q$ decreases, the ringdown part of the signal starts
  earlier but it is also less evident because of the increasingly large
  accretion after the formation of the apparent horizon. Finally,
  shown as an aid to comparison, the panel of the binary
  \texttt{M3.4q070} also reports with dotted lines the waveforms for
  the equal-mass binary \texttt{M3.6q1.00}.}
\label{fig:q22-1}
\end{figure*}

Figure~\ref{fig:q22-1} shows the waveforms in the two polarizations of
the gravitational-wave amplitude \ensuremath{(h_+)_{22}} (upper
panels) and \ensuremath{(h_\times)_{22}} (lower panels) for all the
models considered and as computed from the gauge-invariant
perturbations of a Schwarzschild spacetime. As predicted by the
post-Newtonian approximation~\cite{Blanchet02}, the inspiral phase is
characterized by harmonic oscillations at roughly twice the orbital
frequency but that show an increase both in amplitude and frequency as
the merger approaches. We note that the initial part of the inspiral of the
binary \texttt{M3.4q0.80} shows a comparatively larger contamination
from the initial spurious burst of radiation. This is simply due to
the fact that such a binary has been constructed with a comparatively
larger initial violation of the constraints (\ie the violation of the
${\rm L}_2$ norm of the Hamiltonian constraint is $\sim 3\times
10^{-6}$ and about $50\%$ larger than the violation measured in other
binaries). We believe that this larger initial error is also the one
responsible for a longer time spent by this binary before the merger.

As already discussed before, because of the very high total mass of
the systems no transient HMNS forms, whose dynamics would have been
dramatically imprinted on the waveforms (\cf the detailed comparison
of the HMNS dynamics for different EOSs presented
in~\cite{Baiotti08}). As a result, the post-merger waveform is
essentially the one corresponding to the collapse of the HMNS to a
BH. Indeed, as noted above, in the low-$q$ cases
\texttt{M3.5q0.75} and \texttt{M3.4q0.70}, for which a common apparent
horizon is found almost simultaneously with the merger, the part of
the waveform produced by the newly formed BH starts
essentially together with the end of the one coming from the inspiral.

The ringdown part of the waveform starts increasingly early for
binaries with smaller mass ratios and its signature in the waveform is
also less evident. More specifically, while the ringdown of the BH
created after the merger can be clearly identified in the waveform of
the equal-mass model \texttt{M3.6q1.00}, it becomes much less clear as
one scrolls down in the different panels of figure~\ref{fig:q22-1} and
it seems almost absent in model \texttt{M3.4q0.70}. Indeed it is
necessary to examine \texttt{M3.4q0.70} on a logarithmic scale in
order to appreciate the presence of an exponential ringdown. We
believe that this behaviour is mostly likely due to the copious mass
accretion after the formation of the apparent horizon that becomes
increasingly large as the mass ratio decreases. We recall, in fact,
that the mass accretion rate following the BH formation is highly
sensitive on the mass ratio and inversely proportional to it (see
figure~\ref{fig:m-total} where this is very apparent). Under these
conditions of very intense mass accretion, the BH is continuously
``hit'' by generically nonspherical flows of matter which prevent its
natural ringdown, essentially ``chocking'' it. A detailed analysis on
the role played by mass accretion on the properties of the ringdown
has already been investigated in~\cite{Papadopoulos01}, where however
the BH ringdown was always observed because of the intrinsically
perturbative nature of the approach. The rather different accretion
regime reached in these simulations suggests therefore that the
dynamics observed in figure~\ref{fig:q22-1} reflects a nonlinear
response of the BH that was not accessible in the work
of~\cite{Papadopoulos01}. Additional work is needed to clarify the
relation between hypercritical accretion and BH ringdown and will be
the subject of future investigations.

A more systematic analysis of the waveforms as a function of the mass
ratio is beyond the scope of this paper and will be considered elsewhere
using more realistic or parametrised EOSs. Here, however, as an aid 
to comparison, the panel relative to
the binary \texttt{M3.4q070} in figure~\ref{fig:q22-1} also reports with
dotted lines the waveforms for the equal-mass binary \texttt{M3.6q1.00}
and highlights that besides the different amplitude evolution, the
mass asymmetry also results into a different phase evolution which is
likely to provide important information on the EOS. 

\begin{figure*}[t] 
{\label{fig:h-total:1} \includegraphics[width=.495\textwidth]{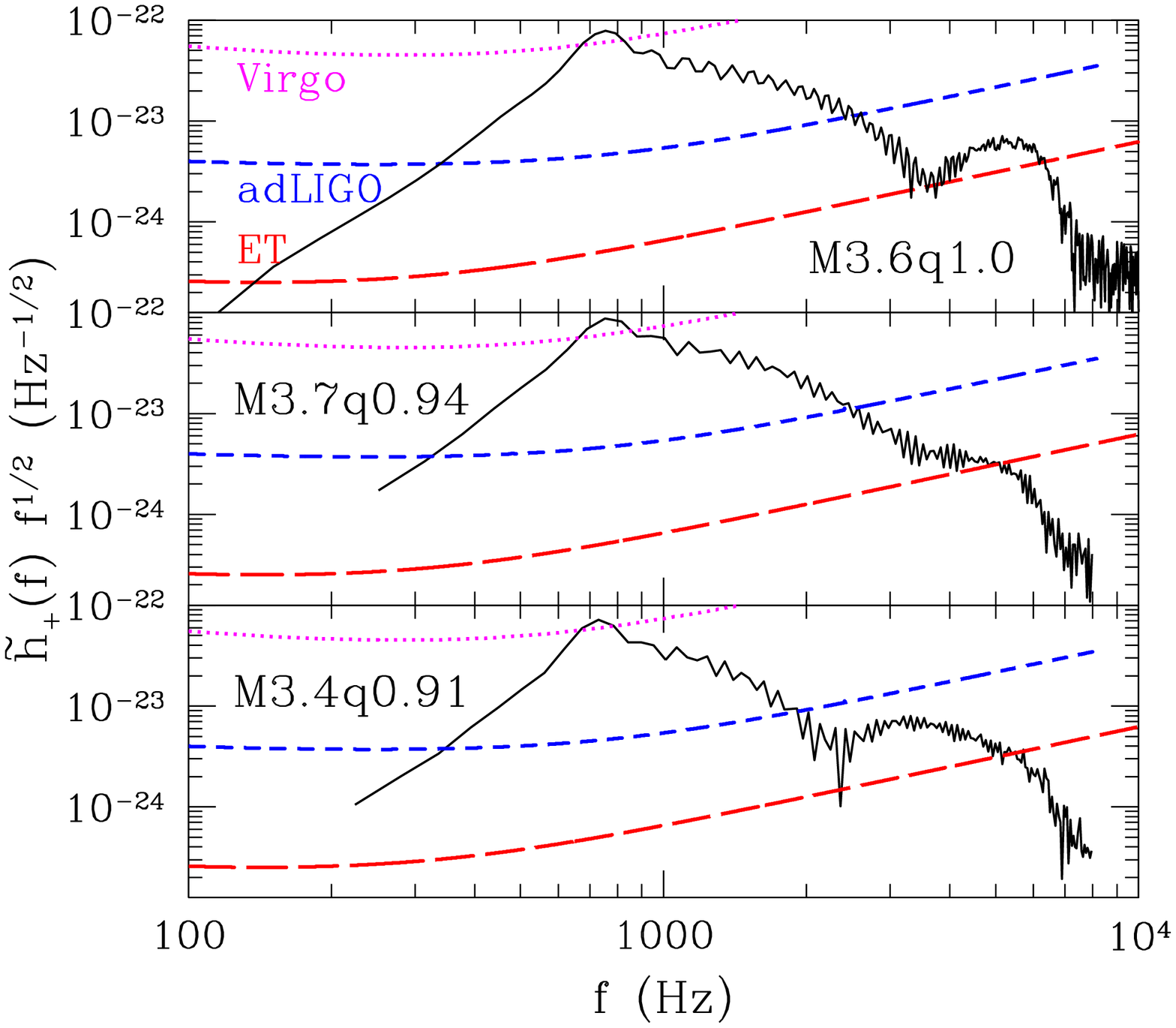}}
{\label{fig:h-total:2} \includegraphics[width=.495\textwidth]{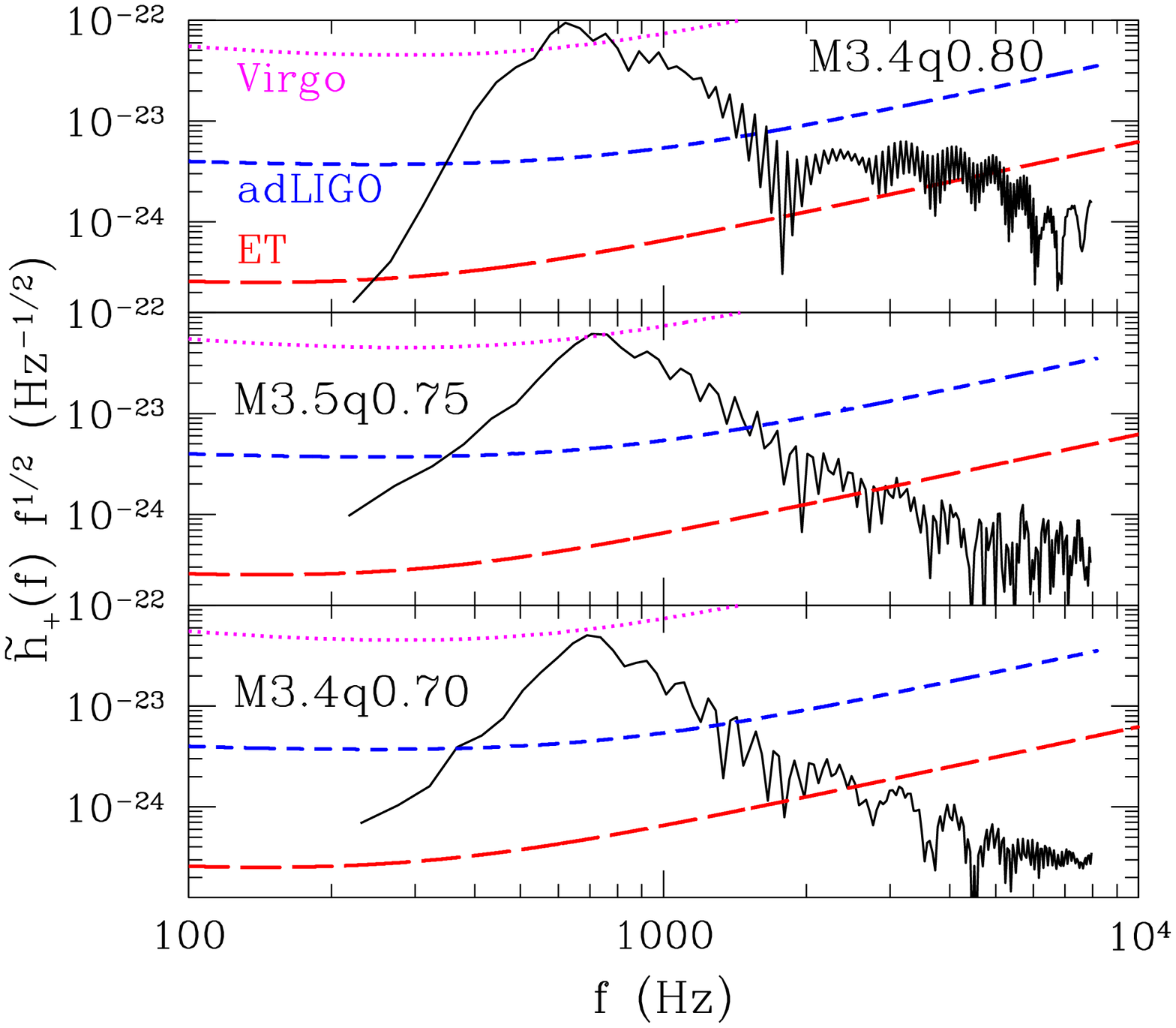}}
\caption{Scaled power spectral densities ${\tilde h}_+(f) f^{1/2}$,
  for all the binaries considered when placed at a distance of
  $100\mpc$. Shown also are the noise curves of the Virgo detector
  (dotted magenta), of the advanced LIGO detector (short-dashed blue) and of
  the planned Einstein Telescope (long-dashed red).}
\label{fig:FFT}
\end{figure*}

In addition, we show with black continuous lines in
figure~\ref{fig:FFT} the scaled power spectral densities (PSD) of
$h_+$, \ie ${\tilde h}_+(f) f^{1/2}$, for all the binaries considered
when placed at a distance of $100\mpc$ (see~\cite{Baiotti07} for a
definition of ${\tilde h}_+(f)$). Shown also here are the noise curves
of the Virgo detector (dotted magenta), of the advanced LIGO
detector~\cite{AdLIGO_noise} (short-dashed blue), and of the planned Einstein
Telescope (ET)~\cite{ET_noise} (long-dashed red).  Since the number of
cycles computed is very small, the peak emission is the one
corresponding to the last stages of the inspiral, around $0.6-0.7$ kHz
for all models considered. The amplitude, however, depends sensitively
on the mass ratio, being maximal for the high-$q$ binaries and above
the noise curve for Virgo in these cases. As the mass ratio is
decreased, in fact, the peak values of the PSD decrease and the
binaries at the distances considered become then undetectable by an
interferometer like Virgo (we recall that the binary
\texttt{M3.4q0.80} has an extended inspiral induced by the larger
initial violation of the constraints; hence its PSD amplitude is
spuriously increased in figure~\ref{fig:FFT}). New-generation
detectors such as advanced LIGO will instead be able to reveal the
inspiral signal in the frequency interval $\sim 0.3-2.0$ kHz, while
essentially all of the late-inspiral and merger signal would be
measured by the Einstein Telescope (see ref.~\cite{Punturo:2010} for
an introduction to the science reach of third-generation detectors such as
ET, and ref~\cite{Andersson:2009yt} for a detailed discussion of the
impact that gravitational waves from NSs may have on such
detectors). Table~\ref{tab:SNRs} summarizes this by reporting the
signal-to-noise ratios (SNRs) for the different detectors and clearly
highlights that present detectors are unlikely to detect any of the
binaries considered here if at a distance of $100\,\mpc$ and observed
only during the final part of the inspiral. On the other hand,
advanced detectors will be able to reveal these sources even at
such large distances (the only ones that can provide an interesting
event rate) and, in the case of third-generation detectors such as ET,
even measure them with significant SNRs.

\begin{table}[ht]
\centering
\small{\begin{tabular}[t]{lccc}
\br
Model & SNR for Virgo & SNR for adLIGO & SNR for ET  \\
\mr
\texttt{M3.6q1.00} & $0.41$  & $2.56$  & $47.47$  \\
\texttt{M3.7q0.94} & $0.41$  & $2.59$  & $48.33$  \\
\texttt{M3.4q0.91} & $0.38$  & $2.48$  & $45.40$  \\
\texttt{M3.4q0.80} & $0.46$  & $3.29$  & $55.68$  \\
\texttt{M3.5q0.75} & $0.36$  & $2.38$  & $42.56$  \\
\texttt{M3.4q0.70} & $0.34$  & $2.29$  & $40.48$  \\ 
\br
\end{tabular}}
\caption{\label{tab:SNRs} SNR for the different binaries
  considered as computed when placed at a distance of $100\mpc$ for a
  presently operating detector such as Virgo, as well as for detectors
  of second and third generation, such as advanced LIGO and ET.}
\end{table}

\section{Conclusions}
\label{sec:Conclusions}

Numerical-relativity simulations of non-vacuum spacetimes have now
reached a sufficient stability and accuracy to be able to describe in
a complete manner all of the stages of the inspiral, merger and
post-merger of binary NSs. Determining the properties of the
black-hole--torus system produced by the merger represents a key
aspect in the modelling of the central engine of SGRBs. Of the many
different properties characterizing the torus, the total rest-mass
clearly represents the most important one, since it is the torus'
binding energy which can be tapped to extract the large energies
necessary to power the SGRB emission. However, the rest-mass density
and angular momentum distributions in the torus also represent
important elements which determine its secular evolution and need to
be computed equally accurately for any satisfactory modelling of SGRB
engine.

As a first step towards modelling \textit{ab-initio} the central
engine of SGRBs, we have here presented new results from accurate and
fully general-relativistic simulations of the coalescence of
unmagnetized binary NSs with unequal masses as these are the
ones expected to yield the largest tori. The evolution of the stars
has been followed through the inspiral phase, the merger and prompt
collapse to a BH, up until the appearance of a thick accretion
disk, which was studied as it enters and remains in a regime of
quasi-steady accretion. Although we have employed a simple ideal-fluid
equation of state, we have performed a systematic study of the
properties of the black-hole--torus obtaining a number of results
that can be summarized as follows:
\begin{itemize}

\item The mass of the torus increases considerably with the mass
  asymmetry and equal-mass binaries do not produce significant tori if
  they have a total baryonic mass $M_{\rm tot} \gtrsim 3.7\,
  \Msun$. Those produced have masses $M_{\rm tor} \sim 10^{-3}\, \Msun$
  and a radial extension of $\sim 30\,\km$.

\item Tori with masses as large as $\sim 0.2\,M_{\odot}$ have been
  measured with binaries having $M_{\rm tot} \sim 3.4\, \Msun$ and
  mass ratios $q \sim 0.75-0.85$. The tori in these cases are much
  more extended with typical sizes $\gtrsim 120\,\km$.

\item The mass of the torus can be described accurately by the simple
  expression $\widetilde M_{\rm tor}(q,M_{\mathrm{tot}}) =
  \left[c_3(1+q)M_{*}-M_{\mathrm{tot}}\right] \left[c_1 (1-q) + c_2
    \right]$, involving the maximum mass for the binaries and
  some coefficients, both of which can be constrained from the simulations.

\item Using the phenomenological expression we conclude
  that tori with masses as large as ${\widetilde M}_{\rm tor} \sim
  0.35\,M_{\odot}$ can be produced for binaries with total masses
  $M_{\rm tot} \sim 2.8\, \Msun$ and $q \sim 0.75-0.85$.

\item Tori from equal-mass binaries exhibit a \textit{quasi-periodic}
  form of accretion associated with the radial epicyclic oscillations of
  the tori, while those from equal-mass binaries exhibit a
  \textit{quasi-steady} form of accretion.

\item When analyzing the evolution of the angular-momentum
  distribution in the tori, we find no evidence for the onset of
  non-axisymmetric instabilities, that angular momentum is
  transported outwards more efficiently for smaller values of $q$
  thus yielding Keplerian angular-velocity distributions, and that
  very little of the mass of the tori is unbound.

\item Present gravitational-wave detectors are unlikely to detect any of the
binaries considered here if at a distance of $100\,\mpc$ and observed
only during the final part of the inspiral. 

\item Advanced detectors will be able to reveal these sources
  even at large distances and measure them with significant SNRs in
  the case of third-generation detectors such as ET.

\end{itemize}

Overall, these results indicate that large-scale tori with large masses
and quasi-stationary evolutions can be produced as the result of the
inspiral and merger of binary NSs with unequal masses. Hence,
they may provide the energy reservoir needed to power short
GRBs. Although complete and accurate, our results are also far from
being realistic. Much remains to be done to improve them either by
considering physically-motivated EOSs, or by including the effect of
magnetic fields, or by taking into account the modifications
introduced by a self-consistent treatment of the radiation
transfer. All and each of these improvements will be the subject of
our future research.

\ack It is a pleasure to thank the group in Meudon (Paris) for
producing and making available the initial data used in these
calculations and also in particular to Dorota Gondek-Rosi\'nska for
essential help in building some of the initial models. We are also
grateful to E.~Schnetter and all the \texttt{Carpet/Cactus}
developers. We finally thank Gian Mario Manca for help in producing
some of the figures. The computations were performed on the Damiana
cluster at the AEI, on the MareNostrum cluster at the Barcelona
Supercomputing Center and on the Ranger cluster at the Texas Advanced
Computing Center through TERAGRID allocation TG-MCA02N014. This work
was supported in part by the DFG grant SFB/Transregio~7, by
``CompStar'', a Research Networking Programme of the European Science
Foundation, by the JSPS Postdoctoral Fellowship For Foreign
Researchers, Grant-in-Aid for Scientific Research (19-07803), and by
the Spanish Ministerio de Educaci\'on y Ciencia (AYA
2007-67626-C03-01).

\appendix

\section{On our accuracy: conservation of mass and angular momentum}

In a recent work~\cite{Baiotti:2009gk} we have discussed in detail the
convergence properties of our numerical simulations and, in
particular, the deterioration of the convergence rate at the merger
and during the survival of the merged object, when strong shocks are
formed and turbulence develops. In particular, in figure $3$ of that
work we have shown a very stringent measure of the overall
conservation properties of our simulations by reporting the time
evolution of the energy and angular momentum which are partially
radiated during the simulation. In order to reduce the computational
costs associated with the measurements made in~\cite{Baiotti:2009gk},
we had limited ourselves to a single configuration and in particular
one that, because of the rather high mass, formed a BH soon after the
merger. In particular, we had considered an equal-mass binary with a
total baryonic mass of $M_b=3.56\,M_{\odot}$ and a total ADM mass of
$M_{_{\rm ADM}}=3.23\,M_{\odot}$ as evolved from an initial
(coordinate) separation of $\sim 45$ km. As a result, we were able to
show an overall conservation of both mass and angular momentum to a
precision of $\sim 1\%$ over a timescale of $\sim 10\mss$.

\begin{figure*}[ht] 
\begin{center}
\includegraphics[width=0.475\textwidth]{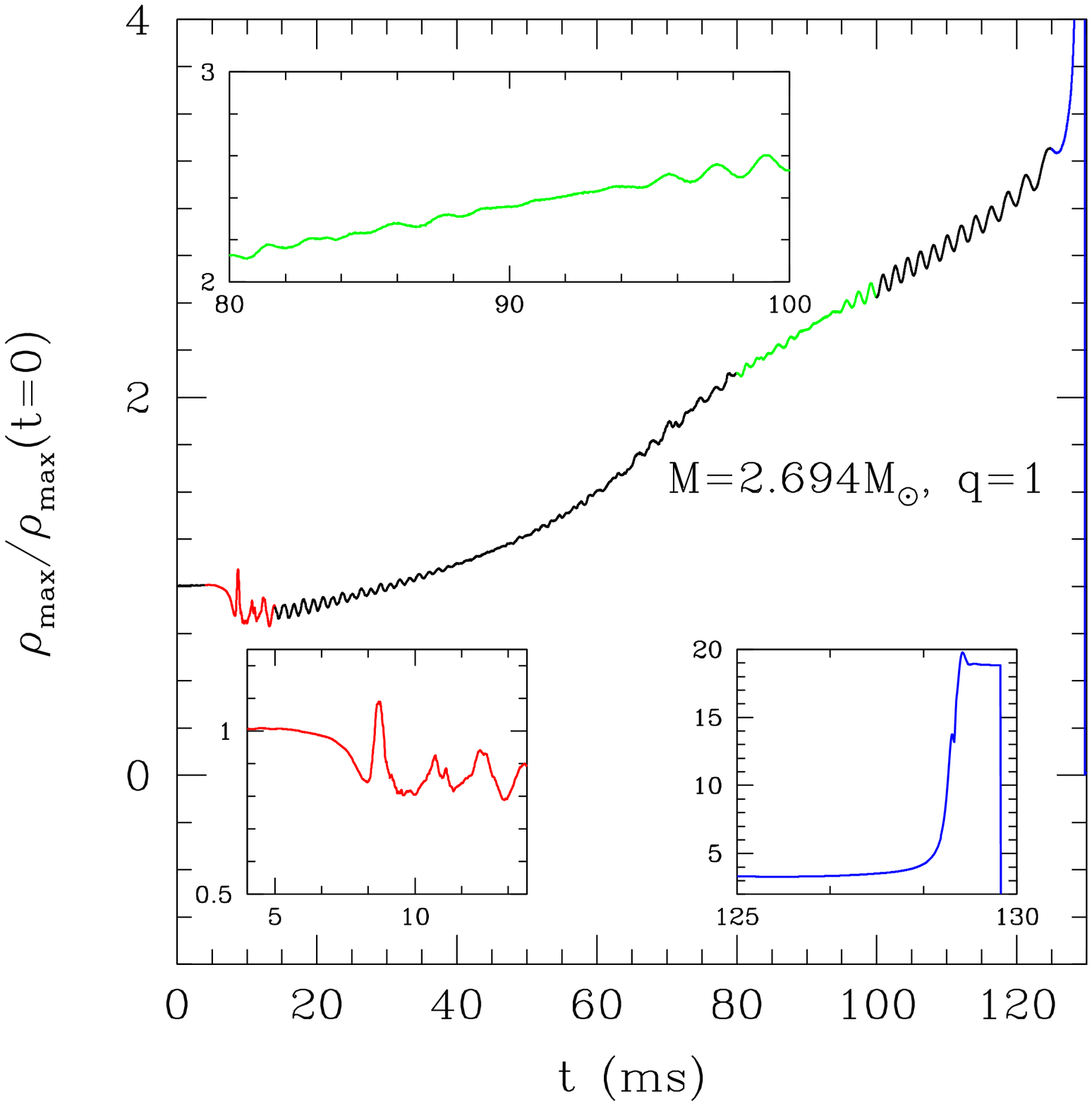}
\hskip 0.5cm
\includegraphics[width=0.475\textwidth]{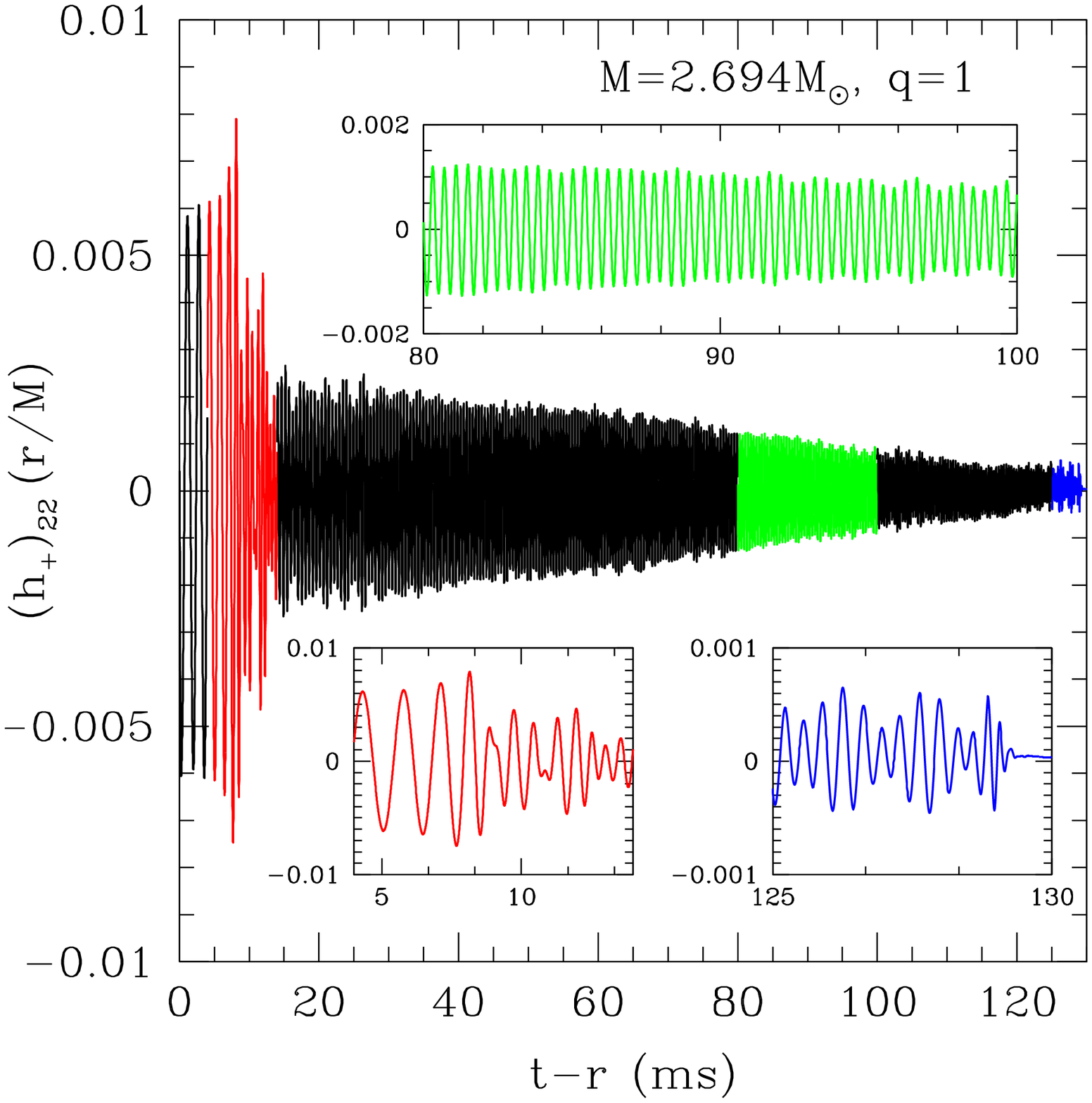}
\end{center}
\caption{Left panel: Evolution of the maximum rest-mass density
  normalized to its initial value. Shown with different colours are
  the different parts of the evolution which are then also magnified
  in the three insets (\cf the timescale to associate the insets to
  the different parts of the evolution). Right panel: The same as in
  the left panel but in terms of the $\ell=m=2$ mode of the $h_+$
  polarization amplitude.}
\label{fig:low-mass_1} 
\end{figure*}

We here reconsider the same assessment of the conservation properties
but for a far more challenging case of a binary with a small total
mass and for which the HMNS survives a considerably larger time before
collapsing to a BH. In particular, we examine the evolution of an
equal-mass binary with a total baryonic mass of $M_b=2.912\,M_{\odot}$
and a total ADM mass of $M_{_{\rm ADM}}=2.694\,M_{\odot}$ as evolved
from an initial (coordinate) separation of $\sim 45$ km; this same
binary was indicated as 1.46-45-IF in~\cite{Baiotti08} and evolved
there only up to $25\,\mss$. As representative information about the
binary inspiral and merger we report in the left panel of
figure~\ref{fig:low-mass_1} the evolution of the maximum rest-mass
density normalized to its initial value (\cf figure $15$
in~\cite{Baiotti08}). Shown with different colours are the different
parts of the evolution and which are also magnified in the different
insets (\cf the timescale to associate the insets to the three parts
of the evolution). They refer to the immediate formation of the HMNS
(red line), to the secular evolution of the HMNS as a contracting
bar-deformed object (green line) and to the exponential growth when
the threshold to black-hole formation has been crossed (blue
line). The right panel of the same figure shows instead the same
stages of the evolution but in terms of the $\ell=m=2$ amplitude of
the $+$ polarization. Note that the timescale over which the evolution
is reported is $\sim 140\,\mss$ and thus a factor $\sim 5$ larger than
the one discussed in~\cite{Baiotti08}. (As for the evolutions
in~\cite{Baiotti:2009gk}, here too we have used a rotational symmetry
around the $z$-axis to reduce computational costs.)

\begin{figure*}[ht] 
\begin{center}
\includegraphics[width=0.475\textwidth]{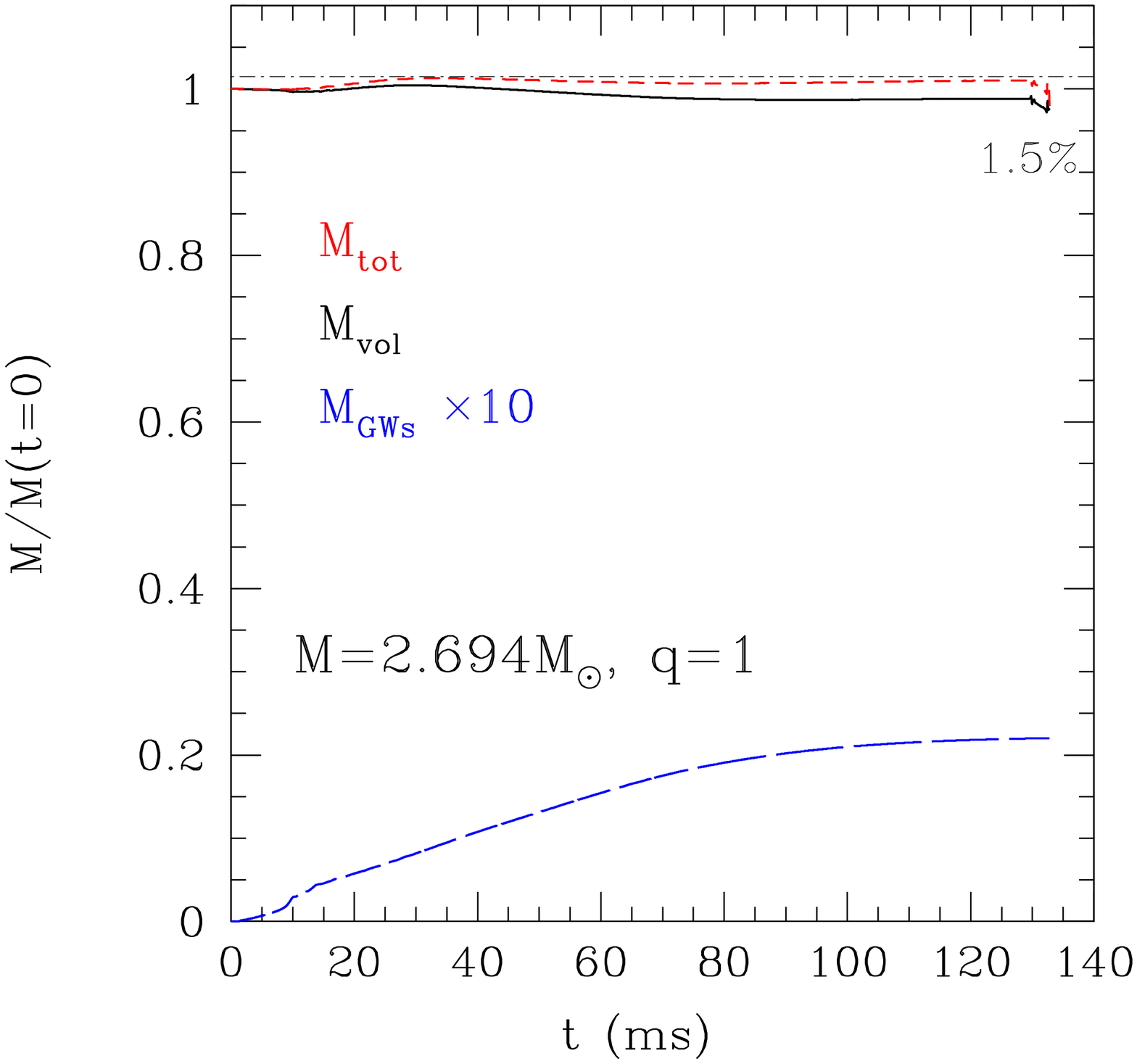}
\hskip 0.5cm
\includegraphics[width=0.475\textwidth]{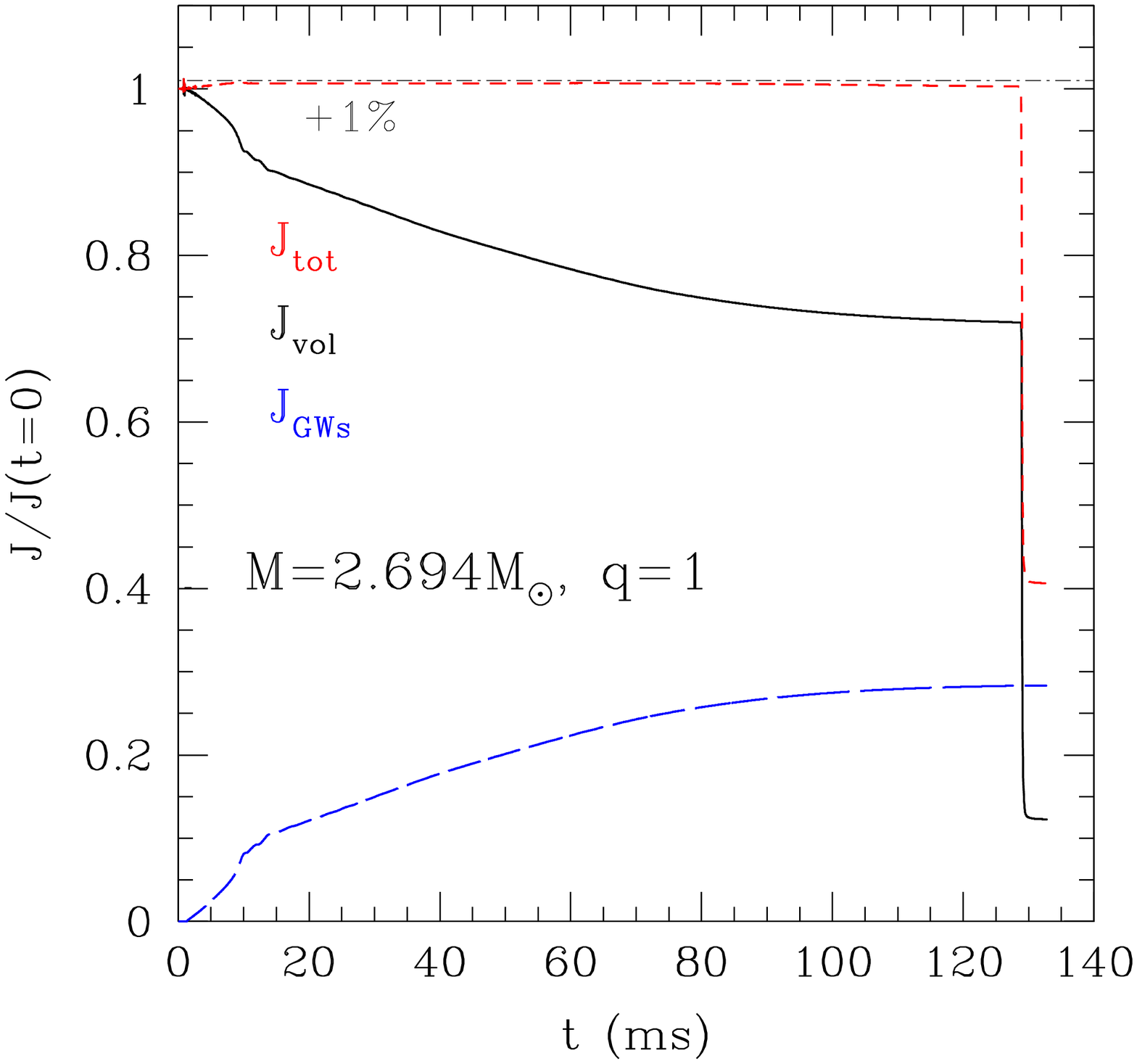}
\end{center}
\caption{Left panel: conservation of energy. The continuous black line
  is the ADM mass computed as an integral over the whole grid, while
  the long-dashed blue line is the energy carried from gravitational
  waves outside the grid and magnified by a factor of $10$; the
  short-dashed red line is the sum of the two and it should be
  conserved. The numerical violation is at most $1.5\%$ (\cf
  dot-dashed line). Right panel: the same as in the left panel but for
  the conservation of the angular momentum. Also in this case the
  violation is at most $1\%$.}
\label{fig:low-mass_2} 
\end{figure*}

In analogy with figure $3$ of~\cite{Baiotti:2009gk}, we show in the left
panel of figure~\ref{fig:low-mass_2} the evolution of the total mass
as normalized to the initial value (\cf left panel of figure $3$
of~\cite{Baiotti:2009gk}). Indicated with different lines are the
volume-integrated values of the ADM mass (solid black line), of the
energy lost to gravitational waves magnified of a factor $10$
(long-dashed blue line), and of their sum (short-dashed red line). The
last quantity should be strictly constant and this is the case to a
precision of $\sim 0.5\%$ during the inspiral, but with a secular
decrease that brings the total error to be $\sim 1.5\%$ at the end of
the simulation (as an aid to comparison the value at $1.015$ is shown
with a dot-dashed line). Similar considerations apply also to the
conservation of the angular momentum as shown in the right panel of
figure~\ref{fig:low-mass_2} (\cf right panel of figure $3$
of~\cite{Baiotti:2009gk}) which uses the same conventions as the left panel
(here $J_{\rm vol}$ is computed with the integral (15)
in~\cite{Baiotti08}). In this case the radiative losses are much
larger (almost $15\%$ of the available angular momentum is lost to
gravitational waves) but the overall conservation is still accurate to
$\sim 1\%$. Once again it is worth nothing that the timescale
over which we can show accurate conservation of mass and angular
momentum is a factor $\sim 14$ larger than the one discussed
in~\cite{Baiotti:2009gk} and provides us with great confidence over the
numerical accuracy of our results. Of course this does not provide us
with any measure of whether such results are indeed realistic.

\section*{References}
\bibliographystyle{iopart-num} 
\bibliography{}

\end{document}